\documentclass[journal]{IEEEtran}
\usepackage{amsmath,amsfonts}
\usepackage{bm}
\usepackage{array}
\usepackage{tabularx}
\usepackage{multirow}
\usepackage{cite}
\usepackage{placeins}
\usepackage{flushend}

\usepackage{tikz}
\usepackage{pgfplots}
\pgfplotsset{compat=newest,width=0.97\columnwidth,/tikz/font=\small}
\extrafloats{100}
\newenvironment{algindent}
  {\begin{list}{}{\setlength{\leftmargin}{1em}\setlength{\topsep}{0pt}}%
   \item[]}
  {\end{list}}
\newcommand{\arrayrule}{\rule[-0.9ex]{0em}{3.2ex}}
\newcommand{\arrayruleb}{\rule[-2.3ex]{0em}{5.8ex}}
\newcommand{\arrayrulec}{\rule[-3.6ex]{0em}{8.4ex}}
\newenvironment{codearray}
  {\begin{array}{|c|>{\arrayrule}l|}}
  {\end{array}}
\newcommand{\mlr}[1]{\multirow{5}{*}{$ #1 $}}
\newcommand{\mlrr}[1]{\multirow{6}{*}{$ #1 $}}
\newcommand{\mlrrr}[1]{\multirow{6.8}{*}{$ #1 $}}
\newcommand{\mlrf}[1]{\multirow{2.4}{*}{$ #1 $}}
\newcommand{\acs}{\hspace{\arraycolsep}}
\newenvironment{mat2}
  {\left[ \begin{array}{c@{\acs}c}}
  {\end{array} \right]}
\newenvironment{mat3}
  {\left[ \begin{array}{c@{\acs}c@{\acs}c}}
  {\end{array} \right]}
\newenvironment{mat32}
  {\arrayruleb \! \left[ \begin{array}{c@{\acs}c@{\acs}c}}
  {\end{array} \right]}
\newenvironment{mat4}
  {\left[ \begin{array}{c@{\acs}c@{\acs}c@{\acs}c}}
  {\end{array} \right]}
\newenvironment{mat42}
  {\arrayruleb \! \left[ \begin{array}{c@{\acs}c@{\acs}c@{\acs}c}}
  {\end{array} \right]}
\newenvironment{mat43}
  {\arrayrulec \! \left[ \begin{array}{c@{\acs}c@{\acs}c@{\acs}c}}
  {\end{array} \right]}
\newcommand{\dfreesep}{\hfill\hspace{1em}}
\newcounter{tablepart}
\newcommand{\tcfirst}
  {\setcounter{tablepart}{1}%
   \renewcommand{\thetable}{\Roman{table}\alph{tablepart}}}
\newcommand{\tcnext}
  {\addtocounter{table}{-1}%
   \stepcounter{tablepart}%
   \renewcommand{\thetable}{\Roman{table}\alph{tablepart}}}
\makeatletter
\newcommand{\tcrevert}
  {\def\@currentlabel{\Roman{table}}}
\makeatother
\newcommand{\lss}[2]{\begin{array}{@{}l@{}}\\[-2ex]#1\\#2\end{array}}
\newenvironment{captionaligned}
  {\begin{tabular}[t]{@{}l@{}}}
  {\end{tabular}}
\newcommand{\dsmax}{15}
\newcommand{\codeaa}{C_{1/2,6}^{\text{ODS}}}
\newcommand{\codeab}{C_{1/2,12}^{\text{ODS}}}
\newcommand{\codeac}{C_{1/2,31}^{\text{ODP}}}
\newcommand{\codead}{C_{1/2,31}^{\text{OBDP}}}
\newcommand{\codeba}{C_{1/3,6}^{\text{ODS}}}
\newcommand{\codebb}{C_{1/3,12}^{\text{ODS}}}
\newcommand{\codebc}{C_{1/3,19}^{\text{ODP}}}
\newcommand{\codebd}{C_{1/3,19}^{\text{OBDP}}}
\newcommand{\codebe}{C_{1/3,31}^{\text{OBDP}}}
\newcommand{\codeca}{C_{1/4,6}^{\text{ODS}}}
\newcommand{\codecb}{C_{1/4,12}^{\text{OFD}}}
\newcommand{\codecc}{C_{1/4,21}^{\text{ODP}}}
\newcommand{\codecd}{C_{1/4,21}^{\text{OBDP}}}
\newcommand{\codece}{C_{1/4,27}^{\text{OBDP}}}
\newcommand{\codeda}{C_{2/3,3}^{\text{OFD}}}
\newcommand{\codedb}{C_{2/3,6}^{\text{OFD}}}
\newcommand{\codedc}{C_{2/3,9}^{\text{ODP}}}
\newcommand{\codedd}{C_{2/3,9}^{\text{OBDP}}}
\begin{document}
\title{Convolutional Codes with Optimum Bidirectional Distance Profile}
\author
  {Ivan Stanojevi\'c and Vojin \v Senk
   \thanks
     {This research has been supported by the
      Ministry of Education, Science and Technological Development
      of the Republic of Serbia,
      through the project no.\ 451-03-68/2020-14/200156:
      ``Innovative scientific and artistic research from the FTS domain.''

      The authors are with the Department of Electrical Engineering,
      Faculty of Engineering (a.k.a. Faculty of Technical Sciences),
      University of Novi Sad, Serbia.
      E-mail: cet\_ivan@uns.ac.rs}}
\maketitle
\begin{abstract}
We define the bidirectional distance profile (BDP) of a convolutional code
as the minimum of the distance profiles of the code and its corresponding
``reverse'' code. We present tables of codes with the optimum BDP (OBDP),
which minimize the average complexity of
bidirectional sequential decoding algorithms.
The computer search is accelerated by the facts that
optimum distance profile (ODP) codes of larger memory must have
ODP codes of smaller memory as their ``prefixes'',
and that OBDP codes can be obtained by ``concatenating'' ODP
and reverse ODP codes of smaller memory.
We compare the performance of OBDP codes and other codes by simulation.
\end{abstract}
\section{Introduction}
Let
\begin{equation}
G(D) =
\left[
\begin{array}{ccc}
g_{11} (D) & \cdots & g_{1n} (D) \\
\vdots     &        & \vdots     \\
g_{k1} (D) & \cdots & g_{kn} (D)
\end{array}
\right]
\end{equation}
be the generator matrix of a rate $ R = k/n $ convolutional encoder,
where $ g_{ij} (D) \in \mathbb{F}_{2} [D] $ are binary generator polynomials
and let $ m = \max_{i,j} \deg ( g_{ij} (D) ) $ be its memory.

The $ l $th value of the column distance function (CDF) of the code
\cite{fundamentals} is
\begin{equation}
  d_{l}
= \min_{\bm{u} : \bm{u}_{0} \neq \bm{0}}
  w_{\text{H}} ( \bm{v}_{[0,l]} )
\text{,}
\end{equation}
where $ w_{\text{H}} (\cdot) $ is the Hamming weight of a sequence,
$ \bm{u} = ( \bm{u}_{0}, \bm{u}_{1}, \ldots ) $ and
$ \bm{v} = ( \bm{v}_{0}, \bm{v}_{1}, \ldots ) $ are the
sequences of $ k $-dimensional information and $ n $-dimensional code
binary vectors, respectively, and for a sequence $ \bm{x} $,
$ x_{[a,b]} = ( x_{a}, x_{a+1}, \ldots, x_{b} ) $.
If $ \bm{d} = ( d_{0}, d_{1}, \ldots ) $ and
$ \bm{d}' = ( d'_{0}, d'_{1}, \ldots ) $ are two CDFs and
$ d_{0} = d'_{0}, \ldots, d_{l-1} = d'_{l-1}, d_{l} > d'_{l} $ for some $ l $,
then we say that $ \bm{d} $ is better than $ \bm{d}' $,
i.e., $ \bm{d} > \bm{d}' $.
A code with the optimum CDF, $ \bm{d}^{*} $,
for given parameters $ k $, $ n $ and $ m $, is one for which
there is no other code with the same parameters whose CDF
$ \bm{d} $ is better than $ \bm{d}^{*} $.
The distance profile (DP) of a code \cite{fundamentals}
is its truncated CDF, $ d_{[0,m]} $, with comparison and optimality
defined in the same way. In order to allow more flexibility in the
choice of optimized code properties, we also define the shortened
distance profile, $ \text{DP}^{(s)} $, as $ d_{[0,m-s]} $, for $ s < m $,
and for notational convenience we define CDF as $ \text{DP}^{(-\infty)} $.

For most often used sequential decoding algorithms,
such as the stack algorithm \cite{zigangirov-stack,jelinek-stack},
and the Fano algorithm \cite{fano}, using ODP codes
minimizes the average number of code tree node extensions.
The influence of the CDF and DP on the decoding complexity
has been studied for the binary symmetric channel \cite{chevillat-costello}
and additive white Gaussian noise (AWGN) channel
\cite{narayanaswamy-et-al}.
In both cases, it is desirable that values $ d_{l} $ grow as quickly
as possible at the beginning of the CDF, hence the definition
of its optimality, with its earlier values having a more profound influence.
ODP codes and codes optimized according to other criteria have been
investigated extensively, e.g.,
\cite{robustly-optimal-one-half,long-one-half,some-1-3-and-1-4,high-rate,
      further-results,new-1-2-1-3-1-4,optimum-distance,optimum-spectrum}.

Although the complexity of standard unidirectional decoding algorithms
is minimized when ODP codes are used, that is not the case with
bidirectional decoding algorithms,
e.g., \cite{senk-radivojac,kallel-li,beast,parallel-fano}.
The number of visited nodes in the forward code tree depends on the DP
of the original code (forward DP), but in the backward code tree
it depends on the DP of the ``reverse'' code (backward DP).

The majority of cases when a sequential decoding algorithm is unable
to output a correct sequence is due to a too high number of visited nodes
in its code tree(s), which results in memory overflow
or excessively long decoding time.
In section \ref{bidirectional-distance-profile}
we describe a procedure for efficient search for codes which are
convenient for decoding by bidirectional sequential decoding algorithms.
The codes with optimized forward and backward DP aim to minimize
the number of nodes in both of decoder code trees.
In section \ref{code-search-results} we present codes
found by that procedure for some of the selected
sets of parameters, along with additional relevant data.
In section \ref{code-performance-results} we compare the performance
of those codes to other optimized codes used with maximum likelihood
(Viterbi) decoding and unidirectional and bidirectional stack decoding.
We evaluate the improvements in terms of the frame error rate (FER)
and overall decoding complexity by simulation.
\section{Bidirectional Distance Profile}
\label{bidirectional-distance-profile}
Let $ \bar{G}(D) = D^{m} G(D^{-1}) $ be the generator matrix of the
reverse code of the code defined by $ G(D) $, with elements
$ \bar{g}_{ij} (D) $.
Let $ \bar{\bm{d}} = ( \bar{d}_{0}, \bar{d}_{1}, \ldots ) $
be the CDF of the reverse code.
In order to take into account both decoding directions,
we define the bidirectional CDF (BCDF), $ \hat{\bm{d}} $,
as the sequence of values $ \hat{d}_{l} = \min ( d_{l}, \bar{d}_{l} ) $,
and the bidirectional $ \text{DP}^{(s)} $ ($ \text{BDP}^{(s)} $)
as $ \hat{d}_{[0,m-s]} $,
as well as the corresponding optimum BCDF and $ \text{BDP}^{(s)} $
(OBCDF and $ \text{OBDP}^{(s)} $) codes.

Let $ g_{ij}^{(l)} $ denote the $ l $th coefficient of the generator
polynomial $ g_{ij} (D) $, and let
\begin{equation}
G^{(l)} =
\left[
\begin{array}{ccc}
g_{11}^{(l)} & \cdots & g_{1n}^{(l)} \\
\vdots       &        & \vdots       \\
g_{k1}^{(l)} & \cdots & g_{kn}^{(l)}
\end{array}
\right]
\text{,}
\end{equation}
and
\begin{align}
G^{[a,b]} (D)       & = \sum_{l=a}^{b} G^{(l)}   D^{l}
\text{,}
\\
\bar{G}^{[a,b]} (D) & = \sum_{l=a}^{b} G^{(m-l)} D^{l}
\text{,}
\end{align}
so $ G(D) = G^{[0,m]} (D) $ and $ \bar{G}(D) = \bar{G}^{[0,m]} (D) $.
Also for $ G(D) = G^{[0,p]} (D) + G^{[p+1,m]} (D) $, we say that
$ G^{[0,p]} (D) $ is a prefix of $ G(D) $, and that $ G(D) $
is obtained by concatenation of $ G^{[0,p]} (D) $ and $ G^{[p+1,m]} (D) $.
It is easy to see that if the code defined by $ G(D) $ is ODP,
then the code defined by $ G^{[0,p]} (D) $ must also be ODP.

If the code memory, $ m $, is odd, we can similarly decompose
the generator matrix as
\begin{equation}
G(D) = G^{\text{F}} (D) + D^m G^{\text{B}} (D^{-1}) \text{,}
\end{equation}
with
\begin{align}
G^{\text{F}} (D) & = G^{[0,(m-1)/2]} (D)
\text{,}
\\
G^{\text{B}} (D) & = \bar{G}^{[0,(m-1)/2]} (D)
\text{.}
\end{align}
Here we call $ G^{\text{F}} (D) $ the forward half and
$ G^{\text{B}} (D) $ the backward half of $ G(D) $,
and we can express the reverse code generator matrix
as $ \bar{G}(D) = G^{\text{B}} (D) + D^m G^{\text{F}} (D^{-1}) $.
From the definitions of the corresponding CDFs, we see that
$ d_{[0,(m-1)/2]} $ depends only on $ G^{\text{F}} (D) $,
and $ \bar{d}_{[0,(m-1)/2]} $ depends only on $ G^{\text{B}} (D) $.
Since we can independently choose
$ G^{\text{F}} (D) $ and $ G^{\text{B}} (D) $,
we can make the first half of the BDP, $ \hat{d}_{[0,(m-1)/2]} $,
equal to that of an ODP code, $ d^{*}_{[0,(m-1)/2]} $,
if and only if both $ G^{\text{F}} (D) $ and $ G^{\text{B}} (D) $
define ODP codes of memory $ (m-1)/2 $.

If the code memory, $ m $, is even, we can decompose the generator matrix as
\begin{equation}
G(D) = G^{\text{F}} (D) + G^{(m/2)} D^{m/2} + D^m G^{\text{B}} (D^{-1})
\text{,}
\end{equation}
where now
\begin{align}
G^{\text{F}} (D) & = G^{[0,m/2-1]} (D)
\text{,}
\\
G^{\text{B}} (D) & = \bar{G}^{[0,m/2-1]} (D)
\text{.}
\end{align}
In this case we can optimize $ \hat{d}_{[0,m/2-1]} $ in the same way.

In order to describe the procedure for finding OBCDF and
$ \text{OBDP}^{(s)} $ codes, a few definitions are necessary.
In $ \mathbb{F}_{2} $, let $ 0 < 1 $. Binary polynomials are compared
lexicographically, i.e., for $ g(D), h(D) \in \mathbb{F}_{2} [D] $,
let $ g(D) < h(D) $ if
$ g^{(0)} = h^{(0)}, \ldots, g^{(l-1)} = h^{(l-1)}, g^{(l)} < h^{(l)} $
for some $ l $. Vectors of binary polynomials are compared lexicographically,
i.e., for $ \bm{x}, \bm{y} \in ( \mathbb{F}_{2} [D] )^{r} $,
let $ \bm{x} < \bm{y} $ if
$ x_{1} = y_{1}, \ldots, x_{l-1} = y_{l-1}, x_{l} < y_{l} $, for some $ l $.
If $ \bm{g}_{i*} $ ($ \bm{g}_{*j} $) is the
$ i $th row ($ j $th column) of a polynomial $ k \times n $ matrix $ G(D) $,
we say that it has sorted rows (columns) if
$ \bm{g}_{1*} \leq \cdots \leq \bm{g}_{k*} $
($ \bm{g}_{*1} \leq \cdots \leq \bm{g}_{*n} $).
Let $ \Pi_{l} $ denote the set of all permutations of elements
$ \{ 1, \ldots, l \} $.

For given $ k $ and $ n $, let $ {\cal U}_{m} $ denote the set of
ODP codes of memory $ m $, and let $ {\cal B}_{m} $ denote the set of
OBCDF ($ \text{OBDP}^{(s)} $) codes of memory $ m $. We can obtain the OBCDF
($ \text{OBDP}^{(s)} $) codes using the following two-stage procedure:
\begin{enumerate}
\renewcommand{\theenumi}{\Roman{enumi}}
\item
  {\bf Find all ODP codes for the desired values of $ m $.}
  \begin{itemize}
  \item[1.]
    Set $ {\cal U}_{-1} = \{ \bm{0}_{k \times n} \} $. \\
    Set $ m = 0 $.
  \item[2.]
    For all prefixes $ G'(D) \in {\cal U}_{m-1} $:
    \begin{algindent}
      For all binary $ k \times n $ matrices $ G^{(m)} $:
      \begin{algindent}
        Form $ G(D) = G'(D) + G^{(m)} D^{m} $. \\
        If $ G(D) $ has sorted rows and columns, calculate its DP. \\
        Retain in $ {\cal U}_{m} $ only the matrices whose DPs
        are the best among the calculated ones.
      \end{algindent}
    \end{algindent}
  \item[3.]
    If more sets are needed, set $ m = m + 1 $ and go to step 2.
  \end{itemize}
\item
  {\bf Find all OBCDF ($ \text{OBDP}^{(s)} $) codes for the desired
   values of $ m $.}
  \begin{itemize}
  \item[1.]
    Set $ m = \max ( 1, 2s-1 ) $.
  \item[2.]
    Set $ {\cal B}_{m} = \{ \} $. \\
    Set $ p = \lfloor (m-1)/2 \rfloor $.
  \item[3.]
    For all $ G^{\text{F}} (D) \in {\cal U}_{p} $
    and for all $ G^{\text{B}} (D) \in {\cal U}_{p} $:
    \begin{algindent}
      [Only when $ m $ is even] \\{}
      [For all binary $ k \times n $ matrices $ G^{(m/2)} $:]
      \begin{algindent}
        For all $ \pi \in \Pi_{k} $ and for all $ \rho \in \Pi_{n} $:
        \begin{algindent}
          Form
          $ G(D) =   G^{\text{F}} (D)
                   + [ G^{(m/2)} D^{m/2} ] +
                     D^{m} \underline{G}^{\text{B}} (D^{-1}) $,
          with
          $ \underline{g}^{\text{B}}_{i,j} (D) =
              g^{\text{B}}_{\pi(i),\rho(j)} (D) $. \\
          If $ G(D) $ does not define a catastrophic encoder
          \cite{fundamentals}, calculate its BCDF ($ \text{BDP}^{(s)} $). \\
          Retain in $ {\cal B}_{m} $ only the matrices whose BCDFs
          ($ \text{BDP}^{(s)} $s) are the best among the calculated ones.
        \end{algindent}
      \end{algindent}
    \end{algindent}
  \item[4.]
    For all pairs of matrices $ G'(D), G''(D) \in {\cal B}_{m} $,
    $ G'(D) \neq G''(D) $:
    \begin{algindent}
      For all $ \pi \in \Pi_{k} $ and for all $ \rho \in \Pi_{n} $:
      \begin{algindent}
        If $ G'(D) $ and $ G''(D) $ are trivially equivalent, i.e.,
        if for all $ i $, $ j $,
        $ g'_{i,j} (D) = g''_{\pi(i),\rho(j)} (D) $,
        or for all $ i $, $ j $,
        $ g'_{i,j} (D) = \bar{g}''_{\pi(i),\rho(j)} (D) $,
        eliminate from $ {\cal B}_{m} $ the lexicographically higher one of
        $ G'(D) $ and $ G''(D) $.
      \end{algindent}
    \end{algindent}
  \item[5.]
    If more sets are needed, set $ m = m + 1 $ and go to step 2.
  \end{itemize}
\end{enumerate}

In the first stage of the procedure, we test whether a generator matrix has
sorted rows and columns in order to avoid repeated calculation
of identical DPs of equivalent encoders, which differ only in the order
of inputs or outputs.

In the second stage of the procedure, the loops over permutations
of rows and columns of $ G^{\text{B}} (D) $ are necessary, since
$ d_{[0,p]} = \bar{d}_{[0,p]} = \hat{d}_{[0,p]} = d^{*}_{[0,p]} $
for all such codes, but the second part of their
BCDFs, $ \hat{d}_{[p+1,\infty]} $,
($ \text{BDP}^{(s)} $s, $ \hat{d}_{[p+1,m-s]} $), varies and must be optimized.

In most cases, we can accelerate the second stage
by omitting the loop over binary matrices $ G^{(m/2)} $
(denoted in square brackets and executed only for even $ m $).
The idea is to define $ G^{\text{F}} (D) = G^{[0,m/2]} (D) $
and take these matrices as all elements of $ {\cal U}_{m/2} $ instead of
$ {\cal U}_{m/2-1} $,
and calculate the BCDFs ($ \text{BDP}^{(s)} $s) of all the concatenations.
This implies that $ d_{m/2} = d^{*}_{m/2} $, but not necessarily that
$ \bar{d}_{m/2} = d^{*}_{m/2} $.
However, if the latter condition is satisfied,
which can be checked after the set $ {\cal B}_{m} $
has been fully constructed,
the reduced search does not affect the optimality of
$ \hat{d}_{[0,m/2]} $.

In the search for $ \text{OBDP}^{(s)} $ codes with $ s \geq 2 $,
we restrict it to the case $ m \geq 2s - 1 $,
in order to simplify the procedure.
When $ m < 2s - 1 $, concatenating
$ G^{\text{F}} (D), G^{\text{B}} (D) \in {\cal U}_{(m-1)/2} $ for
odd $ m $ or $ G^{\text{F}} (D) \in {\cal U}_{m/2} $ and
$ G^{\text{B}} (D) \in {\cal U}_{m/2-1} $ for even $ m $
may not be justified, since only optimality of $ \hat{d}_{[0,m-s]} $ is
required, and the search must include the loop over all ``middle''
matrices $ G^{[m+1-s,s-1]} (D) $.

For some values of $ k $ and $ n $, one must analyze the special case
$ m = 1 $ separately, and perform an exhaustive search of all codes,
since concatenation of $ G^{\text{F}} (D), G^{\text{B}} (D) \in {\cal U}_{0} $
may yield only catastrophic code generators.
\section{Code Search Results}
\label{code-search-results}
In general, for given $ k $, $ n $, and $ m $, OBCDF and
$ \text{OBDP}^{(s)} $ codes are not unique. In order to optimize their
error rate performance, we made an additional selection according to
their information distance spectra, $ \bm{c} = ( c_{1}, c_{2}, \ldots ) $.
Let an error event denote a path in the code trellis which starts in the
zero state, departs from it at the beginning, and returns to the zero state
only at its termination. Then $ c_{d} $ is the sum of Hamming weights
of information sequences of all error events whose codewords
have Hamming weight $ d $.
As before, we define $ \bm{c} < \bm{c}' $ if
$ c_{1} = c'_{1}, \ldots, c_{l-1} = c'_{l-1}, c_{l} < c'_{l} $
for some $ l $, and we search for
the lowest $ \bm{c} $ \cite{optimum-spectrum}.

In tables \ref{code-first}--\ref{code-last} generator matrices of OBCDF and
$ \text{OBDP}^{(s)} $ codes for rates $ 1/2 $, $ 1/3 $, $ 1/4 $,
$ 2/3 $, $ 2/4 $, and $ 3/4 $ are given in octal notation
(using the convention in \cite{robustly-optimal-one-half}, i.e.,
left-aligned). For each code the following additional information
is provided:
\begin{itemize}
\item
  its BDP, $ \hat{d}_{[0,m]} $,
\item
  its free distance, $ d_{\infty} $,
\item
  its code distance spectrum
  (also called the Viterbi spectrum in \cite{fundamentals}),
  $ \bm{a} = ( a_{1}, a_{2}, \ldots ) $,
  where $ a_{d} $ is the number of error events
  whose codewords have Hamming weight $ d $,
\item
  its information distance spectrum, $ \bm{c} = ( c_{1}, c_{2}, \ldots ) $.
\end{itemize}

In tables of $ \text{OBDP}^{(s)} $ codes we have only listed improved codes.
Here by ``improved'' we mean codes with better information distance spectra
than $ \text{OBDP}^{(s-1)} $ codes (or OBCDF codes for $ s = 0 $)
with the same $ k $, $ n $, and $ m $.
For $ R = 3/4 $ and $ m = 3 $ there are no improved
$ \text{OBDP}^{(2)} $ codes.

All the codes have optimized information distance spectra.
The matrices are unique up to their reversal and permutations
of their rows or columns, except in the cases
\begin{itemize}
\item
  $ R = 2/4 $, OBCDF, $ m = 6 $,
\item
  $ R = 2/4 $, $ \text{OBDP}^{(0)} $, $ m = 4 $,
\item
  $ R = 2/4 $, $ \text{OBDP}^{(2)} $, $ m = 6 $,
\end{itemize}
where the codes have identical code and information distance spectra,
but their generators are nontrivially different.

In tables \ref{dfree-first}--\ref{dfree-last} an overview of free distances,
$ d_{\infty}^{(s)} $, of $ \text{OBDP}^{(s)} $ codes is given
(as before, only for improved codes). For each set of parameters,
$ k $, $ n $, and $ m $, the corresponding Griesmer upper bound
on the free distance, $ d_{\infty}^{\text{G}} $, is calculated as
the highest $ d_{\infty} $ which satisfies the inequality \cite{fundamentals}
\begin{equation}
\sum_{l=0}^{ki-1}
  \left\lceil \frac{d_{\infty}}{2^{l}} \right\rceil \leq ( m + i ) n
\end{equation}
for all $ i = 1, 2, \ldots $. We see that some of the listed codes,
mostly for lower values of $ m $, are also optimum free distance codes.

A similar procedure for finding codes suitable for
bidirectional decoding is presented in \cite{kallel-li} for $ R = 1/2 $.
In order to reduce the search space, the authors restrict it
to the so called symmetric codes, which have the property that
$ g_{1} (D) = \bar{g}_{\pi(1)} (D) $ and
$ g_{2} (D) = \bar{g}_{\pi(2)} (D) $,
where
$ G(D) = \left[ \begin{array}{cc} g_{1} (D) & g_{2} (D) \end{array} \right] $
and $ \pi \in \Pi_{2} $, and hence for which $ \bm{d} = \bar{\bm{d}} $.
First, the search for symmetric codes with $ d_{[0,m]} = d^{*}_{[0,m]} $
is performed, and among those codes the one with the best distance
spectrum is selected. If there are no such codes,
the search is performed again, but using the relaxed condition
that $ d_{[0,m-4]} = d^{*}_{[0,m-4]} $. Due to the imposed symmetry,
for some values of $ m $, the codes given in \cite{kallel-li}
have slightly lower free distances than our codes.
\section{Code Performance Results}
\label{code-performance-results}
In order to compare $ \text{OBDP}^{(s)} $ codes to other codes used
in practice, we performed simulations of their FER and
decoding complexity. In the simulations we used the following decoding
algorithms:
\begin{itemize}
\item
  Viterbi algorithm (VA),
\item
  Stack algorithm (SA) \cite{zigangirov-stack,jelinek-stack},
\item
  Bidirectional stack algorithm (BSA) \cite{kallel-li}.
\end{itemize}
We have listed the codes used in the simulations
in table \ref{simulation-codes},
where OFD denotes optimum free distance codes,
and ODS denotes optimum distance spectrum codes.
Some of the codes are defined implicitly by their
parity check matrices ($ H(D) $) in the original literature.

In all simulations the length of the information sequence is $ K = 224 $
binary symbols, whereas the length of the zero-terminated codeword
is $ ( K/k + m ) n $ binary symbols, depending on the code used.
The channel is binary input AWGN with symmetric bipolar mapping,
and the signal/noise ratio (SNR) is defined as
$ 10 \log_{10} ( E_{\text{b}} / N_{0} ) $,
where $ E_{\text{b}} $ is the energy used per information symbol,
and $ N_{0} $ is the one-sided noise power spectral density.

In the descriptions of all simulation results we use the notation
$ A(C) $ to denote the application of algorithm $ A $ to decoding
sequences of code $ C $, and similarly $ A(C,M) $ when
at most $ M $ nodes are inserted in the code tree(s).
In the latter case, $ M $ can also be a set of values,
when their effects are jointly analyzed.
\subsection{Comparison to Viterbi Decoding}
Figures \ref{fer-mlc-first}--\ref{fer-mlc-last} show the comparison
of simulated FERs of BSA decoding of selected OBDP codes and VA
(maximum likelihood) decoding of OFD and ODS codes of lower memory,
for code rates $ 1/2 $, $ 1/3 $, $ 1/4 $, and $ 2/3 $.
The OBDP codes we used for this comparison are the ones with the highest
memory we found.

The results for BSA decoding are given for varying maximum allowed
number of nodes inserted in the code trees (maximum total tree size),
and FER curves from top to bottom correspond to its increasing values.
Both the situation when the decoder reaches the maximum total tree size
without a result (decoding failure) and the situation when it outputs
an incorrect decoded sequence are considered as errors.
We see that for all code rates at some point this dependence saturates,
when the FER starts to be determined only by the code properties
(free distance and distance spectrum).

Depending on the maximum total tree sizes, BSA decoding exhibits
lower FERs than VA decoding starting at certain SNR values.
The crossing points where this occurs are given in tables
\ref{fer-cross-mlc-first}--\ref{fer-cross-mlc-last}.

As the elementary operations performed in BSA and VA decoding
are different, there is not a unique way of comparing the computational
efficiency of the two algorithms, particularly since various optimizations
of both algorithms can significantly affect their overall
time and space complexities. The (SNR, FER) crossing points and the
corresponding maximum tree sizes of the BSA can, however,
serve as a rough estimate of whether BSA decoding is advantageous
to VA decoding for a certain application, i.e., when the expected
channel SNR and the hardware/software resource limitations of the decoder
are known.
\subsection{Comparison of Sequential Decoding}
In figures \ref{fer-sc-first}--\ref{fer-sc-last} simulated FERs are shown
for the following cases:
\begin{itemize}
\item
  SA decoding of ODP codes,
\item
  BSA decoding of ODP codes,
\item
  BSA decoding of OBDP codes.
\end{itemize}
The same code rates are used as in the former comparison.
In all simulations, codes in the compared systems have the same memory.
Again, the maximum total tree size, $ M $, is varied as a parameter,
and both decoding failures and incorrect results are considered as errors.

For $ R = 1/2 $ and $ R = 2/3 $, BSA decoding of ODP and
BSA decoding of OBDP codes exhibit no noticeable difference in performance.
For $ R = 1/3 $, OBDP codes perform
$ \approx 0.05 \div 0.2 \text{ dB} $ better than ODP codes
for different values of $ M $, whereas for $ R = 1/4 $, that
difference is $ \approx 0.2 \div 0.4 \text{ dB} $.
It is interesting to note that in most cases SA decoding of ODP codes
is inferior to BSA decoding of either ODP or OBDP codes,
except for ($ R = 1/3 $, $ M = 2^{24} $) and ($ R = 1/4 $, $ M = 2^{24} $),
where they are very close, and for ($ R = 2/3 $, $ M = 2^{24} $),
where the SA outperforms the BSA by $ \approx 0.25 \text{ dB} $.

Figures \ref{complexity-first}--\ref{complexity-last} show
the distributions of $ X $, the number of extended nodes
in the code tree(s) when the decoded sequence is correct.
When the number of extended nodes becomes excessively high,
and we terminate the decoding of a particular received sequence
due to practical limitations, we cannot determine whether the
decoded sequence is correct. Hence, in some results for the SA,
we only state bounds on the distribution of $ X $.
Similarly as before, for $ R = 1/2 $ and $ R = 2/3 $,
the complexity of the BSA when used with ODP and OBDP codes
is practically the same,
but for $ R = 1/3 $ and $ R = 1/4 $, we see the improvements
with OBDP codes. We also see that the BSA has lower
average complexity than the SA in most cases.
The average numbers of extended nodes are summarized in tables
\ref{average-complexity-first}--\ref{average-complexity-last},
where for the SA we give the lower bounds (LB).
\flushcolsend
\begin{table*}
\tcfirst
\caption{OBCDF codes, $ R = 1/2 $}
\tcrevert
\label{code-first}
\centering
$
\begin{codearray}
\hline
\mlr{m}   & G(D) \\
\cline{2-2}
          & \hat{d}_{[0,m]} \dfreesep d_{\infty} \\
\cline{2-2}
          & a_{d_{\infty}},\ldots,a_{d_{\infty}+\dsmax} \\
\cline{2-2}
          & c_{d_{\infty}},\ldots,c_{d_{\infty}+\dsmax} \\
\hline
\hline
\mlr{1}   & \begin{mat2} 2 & 6 \end{mat2} \\
\cline{2-2}
          & 1,2 \dfreesep 3 \\
\cline{2-2}
          & 1,1,1,1,1,1,1,1,1,1,1,1,1,1,1,1 \\
\cline{2-2}
          & 1,2,3,4,5,6,7,8,9,10,11,12,13,14,15,16 \\
\hline
\mlr{2}   & \begin{mat2} 5 & 7 \end{mat2} \\
\cline{2-2}
          & 2,3,3 \dfreesep 5 \\
\cline{2-2}
          & 1,2,4,8,16,32,64,128,256,512,1024,2048,4096,8192,16384,32768 \\
\cline{2-2}
          & 1,4,12,32,80,192,448,1024,2304,5120,11264,24576,53248,114688,245760,524288 \\
\hline
\mlr{3}   & \begin{mat2} 54 & 64 \end{mat2} \\
\cline{2-2}
          & 2,3,3,3 \dfreesep 6 \\
\cline{2-2}
          & 2,0,10,0,49,0,241,0,1185,0,5827,0,28653,0,140895,0 \\
\cline{2-2}
          & 4,0,38,0,277,0,1806,0,11063,0,65132,0,373045,0,2093866,0 \\
\hline
\mlr{4}   & \begin{mat2} 46 & 62 \end{mat2} \\
\cline{2-2}
          & 2,3,3,4,4 \dfreesep 6 \\
\cline{2-2}
          & 1,0,4,0,22,0,124,0,682,0,3729,0,20390,0,111534,0 \\
\cline{2-2}
          & 1,0,10,0,96,0,778,0,5616,0,38050,0,248136,0,1576298,0 \\
\hline
\mlr{5}   & \begin{mat2} 57 & 75 \end{mat2} \\
\cline{2-2}
          & 2,3,3,4,4,4 \dfreesep 8 \\
\cline{2-2}
          & 3,0,12,0,70,0,397,0,2223,0,12497,0,70093,0,393300,0 \\
\cline{2-2}
          & 8,0,46,0,400,0,2925,0,20446,0,137233,0,895302,0,5726816,0 \\
\hline
\mlr{6}   & \begin{mat2} 564 & 774 \end{mat2} \\
\cline{2-2}
          & 2,3,3,4,4,5,5 \dfreesep 8 \\
\cline{2-2}
          & 2,0,3,0,34,0,173,0,1057,0,5963,0,34349,0,196670,0 \\
\cline{2-2}
          & 4,0,12,0,151,0,1089,0,8440,0,58557,0,397997,0,2628445,0 \\
\hline
\mlr{7}   & \begin{mat2} 452 & 756 \end{mat2} \\
\cline{2-2}
          & 2,3,3,4,4,5,5,5 \dfreesep 10 \\
\cline{2-2}
          & 2,5,5,23,48,97,303,657,1498,3861,8982,21552,52738,125131,300651,725462 \\
\cline{2-2}
          & 6,17,20,119,298,689,2384,5745,14336,40237,101550,262476,686508,1735401,4431952,11313276 \\
\hline
\mlr{8}   & \begin{mat2} 477 & 635 \end{mat2} \\
\cline{2-2}
          & 2,3,3,4,4,5,5,6,6 \dfreesep 10 \\
\cline{2-2}
          & 1,4,2,7,20,49,135,338,824,1795,4420,10749,25592,62748,150604,361309 \\
\cline{2-2}
          & 2,12,8,31,116,315,908,2574,7120,17205,46620,122557,311912,816840,2088376,5324621 \\
\hline
\mlr{9}   & \begin{mat2} 5414 & 6064 \end{mat2} \\
\cline{2-2}
          & 2,3,3,4,4,5,5,6,6,6 \dfreesep 10 \\
\cline{2-2}
          & 2,0,7,0,21,0,135,0,816,0,4543,0,26250,0,152693,0 \\
\cline{2-2}
          & 4,0,28,0,113,0,936,0,6830,0,45980,0,310526,0,2062608,0 \\
\hline
\mlr{10}  & \begin{mat2} 4522 & 6006 \end{mat2} \\
\cline{2-2}
          & 2,3,3,4,4,5,5,6,6,6,7 \dfreesep 9 \\
\cline{2-2}
          & 1,0,1,1,1,18,21,27,110,256,549,1299,3089,7357,17339,40952 \\
\cline{2-2}
          & 1,0,5,4,9,88,123,186,796,2060,5043,13538,34287,86780,220001,555790 \\
\hline
\mlr{11}  & \begin{mat2} 4417 & 6171 \end{mat2} \\
\cline{2-2}
          & 2,3,3,4,4,5,5,6,6,6,7,7 \dfreesep 12 \\
\cline{2-2}
          & 1,2,2,7,10,46,112,218,538,1342,3409,7916,19026,46731,111971,269808 \\
\cline{2-2}
          & 2,4,8,37,52,284,754,1684,4846,13292,35988,90606,235752,616633,1570748,4016286 \\
\hline
\mlr{12}  & \begin{mat2} 54464 & 60014 \end{mat2} \\
\cline{2-2}
          & 2,3,3,4,4,5,5,6,6,6,7,7,7 \dfreesep 11 \\
\cline{2-2}
          & 1,1,1,3,7,20,34,97,188,428,974,2185,5230,12271,28423,66606 \\
\cline{2-2}
          & 1,4,3,12,49,138,242,790,1628,4204,10082,24954,63534,159920,394365,977640 \\
\hline
\mlr{13}  & \begin{mat2} 57276 & 76572 \end{mat2} \\
\cline{2-2}
          & 2,3,3,4,4,5,5,6,6,6,7,7,8,8 \dfreesep 12 \\
\cline{2-2}
          & 1,0,2,0,11,0,54,0,367,0,1823,0,10305,0,58236,0 \\
\cline{2-2}
          & 2,0,6,0,68,0,352,0,3278,0,18043,0,120794,0,777972,0 \\
\hline
\end{codearray}
$
\end{table*}
\begin{table*}
\tcnext
\caption{OBCDF codes, $ R = 1/2 $}
\centering
$
\begin{codearray}
\hline
\mlr{m}   & G(D) \\
\cline{2-2}
          & \hat{d}_{[0,m]} \dfreesep d_{\infty} \\
\cline{2-2}
          & a_{d_{\infty}},\ldots,a_{d_{\infty}+\dsmax} \\
\cline{2-2}
          & c_{d_{\infty}},\ldots,c_{d_{\infty}+\dsmax} \\
\hline
\hline
\mlr{14}  & \begin{mat2} 40375 & 71637 \end{mat2} \\
\cline{2-2}
          & 2,3,3,4,4,5,5,6,6,6,7,7,8,8,8 \dfreesep 12 \\
\cline{2-2}
          & 1,0,0,1,3,4,12,29,65,167,379,979,2429,5780,13963,33262 \\
\cline{2-2}
          & 2,0,0,3,12,20,68,185,424,1319,3314,9195,24950,64440,167430,427434 \\
\hline
\mlr{15}  & \begin{mat2} 406564 & 710774 \end{mat2} \\
\cline{2-2}
          & 2,3,3,4,4,5,5,6,6,6,7,7,8,8,8,8 \dfreesep 16 \\
\cline{2-2}
          & 2,3,2,19,31,77,200,486,1190,2901,6955,16463,40333,97493,234166,565259 \\
\cline{2-2}
          & 6,15,14,97,210,599,1626,4432,12008,31819,81850,207805,544228,1399255,3560950,9072937 \\
\hline
\mlr{16}  & \begin{mat2} 445222 & 611106 \end{mat2} \\
\cline{2-2}
          & 2,3,3,4,4,5,5,6,6,6,7,7,8,8,8,8,9 \dfreesep 14 \\
\cline{2-2}
          & 2,0,3,0,7,0,59,0,237,0,1513,0,8046,0,43697,0 \\
\cline{2-2}
          & 5,0,13,0,36,0,455,0,1966,0,15509,0,94840,0,583969,0 \\
\hline
\mlr{17}  & \begin{mat2} 563477 & 771635 \end{mat2} \\
\cline{2-2}
          & 2,3,3,4,4,5,5,6,6,6,7,7,8,8,8,8,9,9 \dfreesep 16 \\
\cline{2-2}
          & 2,0,6,0,25,0,130,0,756,0,3951,0,21952,0,122884,0 \\
\cline{2-2}
          & 6,0,22,0,176,0,1106,0,7724,0,45065,0,286066,0,1806236,0 \\
\hline
\mlr{18}  & \begin{mat2} 5632374 & 7713164 \end{mat2} \\
\cline{2-2}
          & 2,3,3,4,4,5,5,6,6,6,7,7,8,8,8,8,9,9,9 \dfreesep 18 \\
\cline{2-2}
          & 1,0,12,0,51,0,302,0,1801,0,10081,0,58219,0,339656,0 \\
\cline{2-2}
          & 5,0,70,0,365,0,2720,0,18953,0,123304,0,808989,0,5295340,0 \\
\hline
\mlr{19}  & \begin{mat2} 5574376 & 7513272 \end{mat2} \\
\cline{2-2}
          & 2,3,3,4,4,5,5,6,6,6,7,7,8,8,8,8,9,9,9,9 \dfreesep 20 \\
\cline{2-2}
          & 2,0,28,0,145,0,909,0,4901,0,29005,0,169697,0,989942,0 \\
\cline{2-2}
          & 7,0,166,0,1276,0,9351,0,58621,0,393493,0,2598366,0,16855945,0 \\
\hline
\mlr{20}  & \begin{mat2} 5736137 & 7643675 \end{mat2} \\
\cline{2-2}
          & 2,3,3,4,4,5,5,6,6,6,7,7,8,8,8,8,9,9,9,10,10 \dfreesep 18 \\
\cline{2-2}
          & 1,0,4,0,11,0,97,0,492,0,2600,0,14585,0,85009,0 \\
\cline{2-2}
          & 5,0,12,0,71,0,808,0,4778,0,29978,0,190165,0,1249104,0 \\
\hline
\mlr{21}  & \begin{mat2} 51265734 & 66263624 \end{mat2} \\
\cline{2-2}
          & 2,3,3,4,4,5,5,6,6,6,7,7,8,8,8,8,9,9,9,10,10,10 \dfreesep 21 \\
\cline{2-2}
          & 3,1,7,21,38,100,279,629,1470,3703,8763,21160,51141,123620,298830,721086 \\
\cline{2-2}
          & 17,4,59,164,342,942,2845,7020,17818,48096,121625,311128,794077,2028626,5153362,13057780 \\
\hline
\mlr{22}  & \begin{mat2} 46701662 & 62153046 \end{mat2} \\
\cline{2-2}
          & 2,3,3,4,4,5,5,6,6,6,7,7,8,8,8,8,9,9,9,10,10,10,10 \dfreesep 21 \\
\cline{2-2}
          & 1,5,7,18,30,74,175,341,863,2097,4643,11093,26454,63353,151033,361289 \\
\cline{2-2}
          & 5,36,37,168,298,670,1721,3686,9883,25366,59729,151066,382474,971554,2439115,6155432 \\
\hline
\mlr{23}  & \begin{mat2} 42544077 & 72625251 \end{mat2} \\
\cline{2-2}
          & 2,3,3,4,4,5,5,6,6,6,7,7,8,8,8,8,9,9,9,10,10,10,10,11 \dfreesep 22 \\
\cline{2-2}
          & 2,2,7,13,23,64,137,417,889,2149,5280,12735,30998,74835,179905,434926 \\
\cline{2-2}
          & 8,12,38,91,184,594,1344,4455,10286,26405,69438,179085,460606,1174367,2978060,7578150 \\
\hline
\mlr{24}  & \begin{mat2} 543567064 & 742331074 \end{mat2} \\
\cline{2-2}
          & 2,3,3,4,4,5,5,6,6,6,7,7,8,8,8,8,9,9,9,10,10,10,10,11,11 \dfreesep 23 \\
\cline{2-2}
          & 2,0,2,17,24,95,170,515,1074,2906,5823,16355,35486,94005,210721,541082 \\
\cline{2-2}
          & 10,0,22,130,214,928,1682,5714,12494,37634,78761,239180,541618,1531474,3573699,9713312 \\
\hline
\mlr{25}  & \begin{mat2} 407402732 & 675600306 \end{mat2} \\
\cline{2-2}
          & 2,3,3,4,4,5,5,6,6,6,7,7,8,8,8,8,9,9,9,10,10,10,10,11,11,11 \dfreesep 21 \\
\cline{2-2}
          & 1,1,0,1,4,8,11,29,111,227,544,1322,3099,7615,18711,44843 \\
\cline{2-2}
          & 3,2,0,10,20,68,77,248,1071,2270,6154,16072,40309,105818,274715,697696 \\
\hline
\mlr{26}  & \begin{mat2} 473063215 & 632451617 \end{mat2} \\
\cline{2-2}
          & 2,3,3,4,4,5,5,6,6,6,7,7,8,8,8,8,9,9,9,10,10,10,10,11,11,11,11 \dfreesep 24 \\
\cline{2-2}
          & 1,2,0,8,15,44,119,254,686,1627,3891,9393,22178,54338,131630,316774 \\
\cline{2-2}
          & 12,10,0,48,116,372,1224,2776,8020,20559,52266,134579,335886,871654,2224004,5617126 \\
\hline
\end{codearray}
$
\end{table*}
\begin{table*}
\tcnext
\caption{OBCDF codes, $ R = 1/2 $}
\centering
$
\begin{codearray}
\hline
\mlr{m}   & G(D) \\
\cline{2-2}
          & \hat{d}_{[0,m]} \dfreesep d_{\infty} \\
\cline{2-2}
          & a_{d_{\infty}},\ldots,a_{d_{\infty}+\dsmax} \\
\cline{2-2}
          & c_{d_{\infty}},\ldots,c_{d_{\infty}+\dsmax} \\
\hline
\hline
\mlr{27}  & \begin{mat2} 4624275754 & 6261156024 \end{mat2} \\
\cline{2-2}
          & 2,3,3,4,4,5,5,6,6,6,7,7,8,8,8,8,9,9,9,10,10,10,10,11,11,11,11,12 \dfreesep 26 \\
\cline{2-2}
          & 2,0,15,0,108,0,692,0,3843,0,22569,0,131245,0,763067,0 \\
\cline{2-2}
          & 11,0,126,0,979,0,7756,0,50209,0,330125,0,2144106,0,13772451,0 \\
\hline
\mlr{28}  & \begin{mat2} 5665532672 & 6364242746 \end{mat2} \\
\cline{2-2}
          & 2,3,3,4,4,5,5,6,6,6,7,7,8,8,8,8,9,9,9,10,10,10,10,11,11,11,11,12,12 \dfreesep 22 \\
\cline{2-2}
          & 1,0,1,0,5,2,20,15,65,153,350,619,1647,3351,8702,19362 \\
\cline{2-2}
          & 6,0,16,0,44,10,174,155,730,1791,4254,7921,23078,48701,133696,310462 \\
\hline
\mlr{29}  & \begin{mat2} 4644230357 & 7560621131 \end{mat2} \\
\cline{2-2}
          & 2,3,3,4,4,5,5,6,6,6,7,7,8,8,8,8,9,9,9,10,10,10,10,11,11,11,11,12,12,12 \dfreesep 24 \\
\cline{2-2}
          & 1,0,1,0,6,0,37,0,171,0,1021,0,5619,0,32612,0 \\
\cline{2-2}
          & 4,0,5,0,44,0,379,0,1876,0,12195,0,76292,0,496460,0 \\
\hline
\mlr{30}  & \begin{mat2} 50160016024 & 67671635754 \end{mat2} \\
\cline{2-2}
          & 2,3,3,4,4,5,5,6,6,6,7,7,8,8,8,8,9,9,9,10,10,10,10,11,11,11,11,12,12,12,13 \dfreesep 26 \\
\cline{2-2}
          & 2,0,0,11,7,29,48,91,256,587,1437,3368,8319,19620,47299,114997 \\
\cline{2-2}
          & 8,0,0,59,72,307,472,1007,3102,7333,19848,48908,127678,318196,803932,2048711 \\
\hline
\mlr{31}  & \begin{mat2} 50107314766 & 67631561012 \end{mat2} \\
\cline{2-2}
          & 2,3,3,4,4,5,5,6,6,6,7,7,8,8,8,8,9,9,9,10,10,10,10,11,11,11,11,12,12,12,13,13 \dfreesep 26 \\
\cline{2-2}
          & 1,0,3,0,5,0,47,0,241,0,1556,0,8122,0,48241,0 \\
\cline{2-2}
          & 7,0,22,0,41,0,518,0,2817,0,20778,0,121824,0,803844,0 \\
\hline
\end{codearray}
$
\end{table*}
\begin{table*}
\tcfirst
\caption{Improved $ \text{OBDP}^{(0)} $ codes, $ R = 1/2 $}
\centering
$
\begin{codearray}
\hline
\mlr{m}   & G(D) \\
\cline{2-2}
          & \hat{d}_{[0,m]} \dfreesep d_{\infty} \\
\cline{2-2}
          & a_{d_{\infty}},\ldots,a_{d_{\infty}+\dsmax} \\
\cline{2-2}
          & c_{d_{\infty}},\ldots,c_{d_{\infty}+\dsmax} \\
\hline
\hline
\mlr{6}   & \begin{mat2} 534 & 724 \end{mat2} \\
\cline{2-2}
          & 2,3,3,4,4,5,5 \dfreesep 8 \\
\cline{2-2}
          & 1,0,5,0,35,0,187,0,1074,0,6150,0,35219,0,201519,0 \\
\cline{2-2}
          & 2,0,15,0,188,0,1275,0,9350,0,63958,0,427990,0,2799113,0 \\
\hline
\mlr{9}   & \begin{mat2} 4674 & 7544 \end{mat2} \\
\cline{2-2}
          & 2,3,3,4,4,5,5,6,6,6 \dfreesep 12 \\
\cline{2-2}
          & 8,0,16,0,159,0,741,0,5027,0,26512,0,162403,0,920155,0 \\
\cline{2-2}
          & 32,0,78,0,1150,0,6309,0,52470,0,319464,0,2245020,0,14245009,0 \\
\hline
\mlr{11}  & \begin{mat2} 4173 & 6605 \end{mat2} \\
\cline{2-2}
          & 2,3,3,4,4,5,5,6,6,6,7,7 \dfreesep 13 \\
\cline{2-2}
          & 2,3,3,19,44,80,252,594,1348,3360,7950,19094,46574,112583,270590,652046 \\
\cline{2-2}
          & 4,14,13,102,280,624,2152,5510,13892,36800,93960,242868,631352,1624706,4135928,10519542 \\
\hline
\mlr{12}  & \begin{mat2} 52274 & 71664 \end{mat2} \\
\cline{2-2}
          & 2,3,3,4,4,5,5,6,6,6,7,7,7 \dfreesep 15 \\
\cline{2-2}
          & 4,9,19,56,130,282,678,1652,3980,9818,23499,56170,136279,328772,795142,1918633 \\
\cline{2-2}
          & 14,52,105,376,1064,2560,6502,17586,45590,120568,309661,786856,2026339,5168048,13173984,33424410 \\
\hline
\mlr{15}  & \begin{mat2} 440564 & 616714 \end{mat2} \\
\cline{2-2}
          & 2,3,3,4,4,5,5,6,6,6,7,7,8,8,8,8 \dfreesep 17 \\
\cline{2-2}
          & 1,8,20,33,80,226,482,1213,2920,7028,16734,40733,98071,237946,573035,1388178 \\
\cline{2-2}
          & 1,56,128,264,652,2108,4886,13108,34276,88728,224854,583484,1486739,3815700,9669453,24620504 \\
\hline
\mlr{16}  & \begin{mat2} 445362 & 611036 \end{mat2} \\
\cline{2-2}
          & 2,3,3,4,4,5,5,6,6,6,7,7,8,8,8,8,9 \dfreesep 17 \\
\cline{2-2}
          & 2,3,7,25,34,102,241,612,1470,3387,8484,20123,48652,117882,284036,686012 \\
\cline{2-2}
          & 8,16,35,154,252,828,2103,6054,15824,39010,104642,265504,682316,1758370,4477036,11392906 \\
\hline
\mlr{17}  & \begin{mat2} 401177 & 670535 \end{mat2} \\
\cline{2-2}
          & 2,3,3,4,4,5,5,6,6,6,7,7,8,8,8,8,9,9 \dfreesep 18 \\
\cline{2-2}
          & 2,0,18,0,91,0,604,0,3538,0,19822,0,117513,0,683985,0 \\
\cline{2-2}
          & 4,0,87,0,636,0,5181,0,36765,0,239530,0,1619524,0,10597857,0 \\
\hline
\mlr{18}  & \begin{mat2} 4440574 & 6107364 \end{mat2} \\
\cline{2-2}
          & 2,3,3,4,4,5,5,6,6,6,7,7,8,8,8,8,9,9,9 \dfreesep 20 \\
\cline{2-2}
          & 7,0,58,0,289,0,1721,0,10138,0,58330,0,340617,0,1989323,0 \\
\cline{2-2}
          & 38,0,395,0,2512,0,17809,0,123623,0,810906,0,5310432,0,34430210,0 \\
\hline
\mlr{19}  & \begin{mat2} 4560552 & 6062726 \end{mat2} \\
\cline{2-2}
          & 2,3,3,4,4,5,5,6,6,6,7,7,8,8,8,8,9,9,9,9 \dfreesep 20 \\
\cline{2-2}
          & 1,6,12,36,81,187,464,1038,2470,6047,14686,35507,85293,206065,499088,1202417 \\
\cline{2-2}
          & 4,20,72,282,668,1807,4832,11736,30210,79065,203656,520103,1326504,3378337,8608464,21769383 \\
\hline
\mlr{20}  & \begin{mat2} 4775443 & 6344455 \end{mat2} \\
\cline{2-2}
          & 2,3,3,4,4,5,5,6,6,6,7,7,8,8,8,8,9,9,9,10,10 \dfreesep 20 \\
\cline{2-2}
          & 1,0,15,0,70,0,411,0,2543,0,14608,0,84761,0,494579,0 \\
\cline{2-2}
          & 2,0,96,0,541,0,3967,0,28600,0,190168,0,1249440,0,8132898,0 \\
\hline
\mlr{21}  & \begin{mat2} 50265354 & 67453424 \end{mat2} \\
\cline{2-2}
          & 2,3,3,4,4,5,5,6,6,6,7,7,8,8,8,8,9,9,9,10,10,10 \dfreesep 21 \\
\cline{2-2}
          & 1,2,14,14,43,110,243,625,1530,3587,8750,21327,51095,123591,298740,721114 \\
\cline{2-2}
          & 5,6,102,106,353,1010,2601,7200,18748,46388,121322,313082,792621,2026094,5156420,13066470 \\
\hline
\mlr{22}  & \begin{mat2} 46176562 & 62453546 \end{mat2} \\
\cline{2-2}
          & 2,3,3,4,4,5,5,6,6,6,7,7,8,8,8,8,9,9,9,10,10,10,10 \dfreesep 23 \\
\cline{2-2}
          & 3,10,25,54,142,316,756,1885,4310,10476,25718,61652,149210,360445,870632,2101995 \\
\cline{2-2}
          & 13,74,181,478,1380,3226,8562,22800,55484,143438,373748,949018,2426128,6164790,15628314,39541942 \\
\hline
\mlr{23}  & \begin{mat2} 41732353 & 70767461 \end{mat2} \\
\cline{2-2}
          & 2,3,3,4,4,5,5,6,6,6,7,7,8,8,8,8,9,9,9,10,10,10,10,11 \dfreesep 24 \\
\cline{2-2}
          & 8,10,26,67,164,378,939,2219,5235,12581,31065,74467,180365,435642,1049311,2535173 \\
\cline{2-2}
          & 36,74,196,581,1510,4030,10480,26367,67880,170681,450754,1146139,2927348,7445082,18825902,47649257 \\
\hline
\end{codearray}
$
\end{table*}
\begin{table*}
\tcnext
\caption{Improved $ \text{OBDP}^{(0)} $ codes, $ R = 1/2 $}
\centering
$
\begin{codearray}
\hline
\mlr{m}   & G(D) \\
\cline{2-2}
          & \hat{d}_{[0,m]} \dfreesep d_{\infty} \\
\cline{2-2}
          & a_{d_{\infty}},\ldots,a_{d_{\infty}+\dsmax} \\
\cline{2-2}
          & c_{d_{\infty}},\ldots,c_{d_{\infty}+\dsmax} \\
\hline
\hline
\mlr{24}  & \begin{mat2} 415755274 & 704555144 \end{mat2} \\
\cline{2-2}
          & 2,3,3,4,4,5,5,6,6,6,7,7,8,8,8,8,9,9,9,10,10,10,10,11,11 \dfreesep 24 \\
\cline{2-2}
          & 2,6,10,43,71,185,471,1113,2687,6414,15450,37093,89955,217024,525537,1268771 \\
\cline{2-2}
          & 8,42,72,369,636,1825,5090,12903,33616,85298,219384,558859,1428740,3630140,9252378,23407229 \\
\hline
\mlr{25}  & \begin{mat2} 416262512 & 706141156 \end{mat2} \\
\cline{2-2}
          & 2,3,3,4,4,5,5,6,6,6,7,7,8,8,8,8,9,9,9,10,10,10,10,11,11,11 \dfreesep 25 \\
\cline{2-2}
          & 1,5,15,37,101,222,552,1302,3247,7798,18585,45060,108888,262757,633828,1529804 \\
\cline{2-2}
          & 1,46,109,310,911,2248,6124,15750,42111,106944,272899,698782,1778636,4518654,11438870,28915486 \\
\hline
\mlr{26}  & \begin{mat2} 512043315 & 731676517 \end{mat2} \\
\cline{2-2}
          & 2,3,3,4,4,5,5,6,6,6,7,7,8,8,8,8,9,9,9,10,10,10,10,11,11,11,11 \dfreesep 27 \\
\cline{2-2}
          & 12,21,42,132,265,661,1586,3824,9326,22529,53988,131359,317267,764690,1847681,4456194 \\
\cline{2-2}
          & 86,182,412,1332,2859,7914,20108,52266,135126,345828,874878,2241076,5684179,14346982,36243111,91258732 \\
\hline
\mlr{27}  & \begin{mat2} 4147233534 & 7053240024 \end{mat2} \\
\cline{2-2}
          & 2,3,3,4,4,5,5,6,6,6,7,7,8,8,8,8,9,9,9,10,10,10,10,11,11,11,11,12 \dfreesep 27 \\
\cline{2-2}
          & 5,9,21,51,143,348,789,1985,4686,11338,27390,65330,158325,382935,923390,2228788 \\
\cline{2-2}
          & 21,58,177,504,1459,3678,9253,25354,64144,165034,420008,1056750,2696715,6848208,17309872,43687776 \\
\hline
\mlr{28}  & \begin{mat2} 4075047376 & 6757475322 \end{mat2} \\
\cline{2-2}
          & 2,3,3,4,4,5,5,6,6,6,7,7,8,8,8,8,9,9,9,10,10,10,10,11,11,11,11,12,12 \dfreesep 28 \\
\cline{2-2}
          & 8,13,29,77,153,406,948,2282,5641,13615,32869,79341,191796,461580,1114771,2690510 \\
\cline{2-2}
          & 44,97,248,755,1610,4520,11428,29390,77412,197663,506202,1285577,3276444,8275324,20940684,52835864 \\
\hline
\mlr{29}  & \begin{mat2} 4317713123 & 6562116075 \end{mat2} \\
\cline{2-2}
          & 2,3,3,4,4,5,5,6,6,6,7,7,8,8,8,8,9,9,9,10,10,10,10,11,11,11,11,12,12,12 \dfreesep 29 \\
\cline{2-2}
          & 3,14,34,93,216,474,1191,2793,6788,16367,39474,95537,230254,557179,1345191,3248804 \\
\cline{2-2}
          & 21,116,308,926,2362,5802,15487,38700,99984,253852,647586,1645592,4162206,10555198,26627409,67077488 \\
\hline
\mlr{30}  & \begin{mat2} 46747737324 & 62131433034 \end{mat2} \\
\cline{2-2}
          & 2,3,3,4,4,5,5,6,6,6,7,7,8,8,8,8,9,9,9,10,10,10,10,11,11,11,11,12,12,12,13 \dfreesep 28 \\
\cline{2-2}
          & 1,2,8,27,38,98,266,595,1432,3400,8166,19795,47549,116011,278337,672296 \\
\cline{2-2}
          & 10,20,72,231,362,1114,3052,7219,18970,48058,122092,311879,790970,2029865,5105014,12900446 \\
\hline
\mlr{31}  & \begin{mat2} 46026512472 & 75150113146 \end{mat2} \\
\cline{2-2}
          & 2,3,3,4,4,5,5,6,6,6,7,7,8,8,8,8,9,9,9,10,10,10,10,11,11,11,11,12,12,12,13,13 \dfreesep 30 \\
\cline{2-2}
          & 2,13,29,46,114,273,695,1656,4061,10139,23841,57951,138928,335182,811329,1956453 \\
\cline{2-2}
          & 8,109,264,458,1226,3337,9214,22800,59318,156527,388026,997057,2505802,6329020,16017520,40291965 \\
\hline
\end{codearray}
$
\end{table*}
\begin{table*}
\caption{Improved $ \text{OBDP}^{(1)} $ codes, $ R = 1/2 $}
\centering
$
\begin{codearray}
\hline
\mlr{m}   & G(D) \\
\cline{2-2}
          & \hat{d}_{[0,m]} \dfreesep d_{\infty} \\
\cline{2-2}
          & a_{d_{\infty}},\ldots,a_{d_{\infty}+\dsmax} \\
\cline{2-2}
          & c_{d_{\infty}},\ldots,c_{d_{\infty}+\dsmax} \\
\hline
\hline
\mlr{10}  & \begin{mat2} 4046 & 6772 \end{mat2} \\
\cline{2-2}
          & 2,3,3,4,4,5,5,6,6,6,6 \dfreesep 12 \\
\cline{2-2}
          & 2,2,4,19,29,92,216,478,1217,2876,6997,16974,40753,97819,236441,571250 \\
\cline{2-2}
          & 4,10,16,95,204,672,1698,4352,12106,30408,80082,208828,538758,1376111,3519696,9000972 \\
\hline
\mlr{16}  & \begin{mat2} 406072 & 674546 \end{mat2} \\
\cline{2-2}
          & 2,3,3,4,4,5,5,6,6,6,7,7,8,8,8,8,8 \dfreesep 18 \\
\cline{2-2}
          & 4,0,36,0,215,0,1193,0,6876,0,40019,0,234643,0,1368655,0 \\
\cline{2-2}
          & 10,0,218,0,1753,0,11572,0,78359,0,522463,0,3469359,0,22598988,0 \\
\hline
\mlr{23}  & \begin{mat2} 50635563 & 72252135 \end{mat2} \\
\cline{2-2}
          & 2,3,3,4,4,5,5,6,6,6,7,7,8,8,8,8,9,9,9,10,10,10,10,10 \dfreesep 24 \\
\cline{2-2}
          & 3,12,32,75,157,383,930,2166,5272,12752,30939,74494,180401,435873,1050818,2539839 \\
\cline{2-2}
          & 18,84,254,711,1600,4299,11206,27838,72724,185216,475650,1211558,3092764,7831391,19793920,50005199 \\
\hline
\mlr{27}  & \begin{mat2} 4165155544 & 7064264114 \end{mat2} \\
\cline{2-2}
          & 2,3,3,4,4,5,5,6,6,6,7,7,8,8,8,8,9,9,9,10,10,10,10,11,11,11,11,11 \dfreesep 28 \\
\cline{2-2}
          & 18,0,124,0,656,0,3925,0,22350,0,131237,0,764142,0,4461030,0 \\
\cline{2-2}
          & 111,0,1149,0,7081,0,48667,0,315934,0,2078191,0,13413083,0,85882061,0 \\
\hline
\mlr{30}  & \begin{mat2} 51477667274 & 73460644664 \end{mat2} \\
\cline{2-2}
          & 2,3,3,4,4,5,5,6,6,6,7,7,8,8,8,8,9,9,9,10,10,10,10,11,11,11,11,12,12,12,12 \dfreesep 30 \\
\cline{2-2}
          & 4,24,46,100,237,584,1357,3452,8253,19669,47667,115791,278516,672645,1624094,3920469 \\
\cline{2-2}
          & 36,220,508,1062,2726,7406,17920,48868,124436,313079,799302,2042921,5149004,13009757,32788850,82509447 \\
\hline
\end{codearray}
$
\end{table*}
\begin{table*}
\caption{Improved $ \text{OBDP}^{(2)} $ codes, $ R = 1/2 $, $ m \geq 3 $}
\centering
$
\begin{codearray}
\hline
\mlr{m}   & G(D) \\
\cline{2-2}
          & \hat{d}_{[0,m]} \dfreesep d_{\infty} \\
\cline{2-2}
          & a_{d_{\infty}},\ldots,a_{d_{\infty}+\dsmax} \\
\cline{2-2}
          & c_{d_{\infty}},\ldots,c_{d_{\infty}+\dsmax} \\
\hline
\hline
\mlr{4}   & \begin{mat2} 46 & 72 \end{mat2} \\
\cline{2-2}
          & 2,3,3,3,3 \dfreesep 7 \\
\cline{2-2}
          & 2,3,4,16,37,68,176,432,925,2156,5153,11696,26868,62885,145085,334024 \\
\cline{2-2}
          & 4,12,20,72,225,500,1324,3680,8967,22270,57403,142234,348830,867106,2134239,5205290 \\
\hline
\mlr{6}   & \begin{mat2} 434 & 724 \end{mat2} \\
\cline{2-2}
          & 2,3,3,4,4,4,4 \dfreesep 9 \\
\cline{2-2}
          & 2,5,8,23,47,107,278,660,1611,3813,8944,21450,51351,122611,292979,698573 \\
\cline{2-2}
          & 4,18,40,142,305,780,2238,5960,15967,41042,103748,266606,682057,1732618,4389827,11061590 \\
\hline
\mlr{8}   & \begin{mat2} 435 & 657 \end{mat2} \\
\cline{2-2}
          & 2,3,3,4,4,5,5,5,6 \dfreesep 12 \\
\cline{2-2}
          & 11,0,50,0,286,0,1630,0,9639,0,55152,0,320782,0,1859184,0 \\
\cline{2-2}
          & 33,0,281,0,2179,0,15035,0,105166,0,692330,0,4580007,0,29692894,0 \\
\hline
\mlr{11}  & \begin{mat2} 4363 & 7335 \end{mat2} \\
\cline{2-2}
          & 2,3,3,4,4,5,5,6,6,6,6,6 \dfreesep 14 \\
\cline{2-2}
          & 4,0,49,0,193,0,1251,0,7273,0,42114,0,243240,0,1426044,0 \\
\cline{2-2}
          & 12,0,337,0,1533,0,12384,0,84317,0,563311,0,3654007,0,23880276,0 \\
\hline
\mlr{13}  & \begin{mat2} 42756 & 64712 \end{mat2} \\
\cline{2-2}
          & 2,3,3,4,4,5,5,6,6,6,7,7,7,7 \dfreesep 16 \\
\cline{2-2}
          & 5,16,23,63,134,359,861,1988,4841,11791,28522,68365,165514,399273,965805,2331416 \\
\cline{2-2}
          & 22,102,134,477,1098,3363,8524,21488,56556,148217,384586,974407,2499990,6375885,16253918,41227076 \\
\hline
\mlr{17}  & \begin{mat2} 434323 & 731231 \end{mat2} \\
\cline{2-2}
          & 2,3,3,4,4,5,5,6,6,6,7,7,8,8,8,8,8,8 \dfreesep 19 \\
\cline{2-2}
          & 3,11,24,53,110,310,749,1742,4193,10113,24550,58735,142140,342902,829084,2004765 \\
\cline{2-2}
          & 9,68,154,382,886,2844,7443,18790,49543,127486,328320,837908,2148344,5476172,13951296,35423220 \\
\hline
\mlr{20}  & \begin{mat2} 4147563 & 6624135 \end{mat2} \\
\cline{2-2}
          & 2,3,3,4,4,5,5,6,6,6,7,7,8,8,8,8,9,9,9,9,9 \dfreesep 22 \\
\cline{2-2}
          & 7,15,47,95,203,549,1258,3077,7181,17523,43029,103401,249374,602288,1452102,3505799 \\
\cline{2-2}
          & 50,99,342,845,2016,5655,14334,38009,93346,244425,635378,1618353,4115592,10447428,26426548,66802643 \\
\hline
\mlr{24}  & \begin{mat2} 417136244 & 707535014 \end{mat2} \\
\cline{2-2}
          & 2,3,3,4,4,5,5,6,6,6,7,7,8,8,8,8,9,9,9,10,10,10,10,10,10 \dfreesep 25 \\
\cline{2-2}
          & 5,11,30,90,190,458,1087,2653,6430,15462,37244,89881,217545,525906,1266182,3063805 \\
\cline{2-2}
          & 25,70,230,788,1864,4860,12453,33174,85460,218746,555936,1421972,3625131,9216014,23275660,58924666 \\
\hline
\mlr{28}  & \begin{mat2} 4151623456 & 7041374442 \end{mat2} \\
\cline{2-2}
          & 2,3,3,4,4,5,5,6,6,6,7,7,8,8,8,8,9,9,9,10,10,10,10,11,11,11,11,11,11 \dfreesep 28 \\
\cline{2-2}
          & 1,9,36,58,187,400,928,2368,5659,13576,32867,79062,190826,462068,1112801,2689950 \\
\cline{2-2}
          & 4,57,324,518,1988,4430,11412,30434,77644,198170,508278,1287946,3271686,8319600,20992388,53044336 \\
\hline
\mlr{31}  & \begin{mat2} 55701344176 & 75157532472 \end{mat2} \\
\cline{2-2}
          & 2,3,3,4,4,5,5,6,6,6,7,7,8,8,8,8,9,9,9,10,10,10,10,11,11,11,11,12,12,12,12,12 \dfreesep 31 \\
\cline{2-2}
          & 8,23,58,125,281,733,1713,4143,9991,23926,57845,139572,336440,810781,1959183,4728747 \\
\cline{2-2}
          & 54,226,612,1306,3401,9536,22905,59712,151701,386082,980125,2482196,6277316,15808396,39853965,100246860 \\
\hline
\end{codearray}
$
\end{table*}
\begin{table*}
\caption{Improved $ \text{OBDP}^{(3)} $ codes, $ R = 1/2 $, $ m \geq 5 $}
\centering
$
\begin{codearray}
\hline
\mlr{m}   & G(D) \\
\cline{2-2}
          & \hat{d}_{[0,m]} \dfreesep d_{\infty} \\
\cline{2-2}
          & a_{d_{\infty}},\ldots,a_{d_{\infty}+\dsmax} \\
\cline{2-2}
          & c_{d_{\infty}},\ldots,c_{d_{\infty}+\dsmax} \\
\hline
\hline
\mlr{5}   & \begin{mat2} 53 & 75 \end{mat2} \\
\cline{2-2}
          & 2,3,3,3,3,3 \dfreesep 8 \\
\cline{2-2}
          & 1,8,7,12,48,95,281,605,1272,3334,7615,18131,43197,99210,237248,559238 \\
\cline{2-2}
          & 2,36,32,62,332,701,2342,5503,12506,36234,88576,225685,574994,1400192,3554210,8845154 \\
\hline
\mlr{7}   & \begin{mat2} 452 & 766 \end{mat2} \\
\cline{2-2}
          & 2,3,3,4,4,4,4,5 \dfreesep 10 \\
\cline{2-2}
          & 1,5,13,29,56,120,309,754,1801,4431,10648,25517,61205,146167,350623,841939 \\
\cline{2-2}
          & 2,23,62,165,404,932,2704,7166,18442,49069,127184,327463,838316,2129221,5406334,13695355 \\
\hline
\mlr{9}   & \begin{mat2} 4554 & 7524 \end{mat2} \\
\cline{2-2}
          & 2,3,3,4,4,5,5,5,5,5 \dfreesep 12 \\
\cline{2-2}
          & 1,7,14,32,68,177,437,956,2334,5664,13883,33778,80662,193575,467213,1129345 \\
\cline{2-2}
          & 2,25,70,230,510,1433,3926,9414,25334,66010,172488,448964,1143806,2912083,7423470,18899417 \\
\hline
\mlr{12}  & \begin{mat2} 46554 & 75624 \end{mat2} \\
\cline{2-2}
          & 2,3,3,4,4,5,5,6,6,6,6,6,6 \dfreesep 15 \\
\cline{2-2}
          & 2,13,21,45,145,285,690,1744,4064,9863,23861,57245,139041,336883,811332,1956191 \\
\cline{2-2}
          & 8,72,129,308,1147,2588,6778,18928,47454,123114,318697,810512,2090693,5349342,13581004,34424188 \\
\hline
\mlr{14}  & \begin{mat2} 52635 & 71077 \end{mat2} \\
\cline{2-2}
          & 2,3,3,4,4,5,5,6,6,6,7,7,7,7,7 \dfreesep 17 \\
\cline{2-2}
          & 5,13,40,85,161,418,1005,2403,5912,14402,34552,83138,200335,484926,1170885,2825117 \\
\cline{2-2}
          & 29,64,256,650,1505,3964,10171,26758,70714,185294,474668,1211154,3087807,7889166,20042891,50767166 \\
\hline
\mlr{18}  & \begin{mat2} 4066164 & 7105374 \end{mat2} \\
\cline{2-2}
          & 2,3,3,4,4,5,5,6,6,6,7,7,8,8,8,8,8,8,8 \dfreesep 20 \\
\cline{2-2}
          & 2,10,29,67,169,367,894,2107,5082,12154,29396,72137,172983,417642,1009082,2435146 \\
\cline{2-2}
          & 8,44,182,521,1426,3551,9302,23477,60952,155706,401204,1046089,2661056,6784618,17248868,43690606 \\
\hline
\mlr{21}  & \begin{mat2} 44632424 & 76170134 \end{mat2} \\
\cline{2-2}
          & 2,3,3,4,4,5,5,6,6,6,7,7,8,8,8,8,9,9,9,9,9,9 \dfreesep 23 \\
\cline{2-2}
          & 10,9,41,115,277,659,1494,3667,8789,21303,51503,123968,299152,722189,1747531,4216229 \\
\cline{2-2}
          & 64,62,341,1012,2845,7214,17382,46390,118617,306218,786307,1994170,5064272,12846060,32579919,82184500 \\
\hline
\mlr{25}  & \begin{mat2} 415744152 & 704555326 \end{mat2} \\
\cline{2-2}
          & 2,3,3,4,4,5,5,6,6,6,7,7,8,8,8,8,9,9,9,10,10,10,10,10,10,10 \dfreesep 26 \\
\cline{2-2}
          & 5,0,83,0,448,0,2663,0,15521,0,90420,0,525540,0,3064270,0 \\
\cline{2-2}
          & 24,0,691,0,4524,0,31818,0,212706,0,1392288,0,9001380,0,57699795,0 \\
\hline
\mlr{29}  & \begin{mat2} 4027363533 & 7175037345 \end{mat2} \\
\cline{2-2}
          & 2,3,3,4,4,5,5,6,6,6,7,7,8,8,8,8,9,9,9,10,10,10,10,11,11,11,11,11,11,11 \dfreesep 30 \\
\cline{2-2}
          & 23,0,161,0,964,0,5633,0,32876,0,191684,0,1114754,0,6504467,0 \\
\cline{2-2}
          & 159,0,1550,0,11196,0,75291,0,496307,0,3223737,0,20637017,0,131587711,0 \\
\hline
\end{codearray}
$
\end{table*}
\begin{table*}
\caption{Improved $ \text{OBDP}^{(4)} $ codes, $ R = 1/2 $, $ m \geq 7 $}
\centering
$
\begin{codearray}
\hline
\mlr{m}   & G(D) \\
\cline{2-2}
          & \hat{d}_{[0,m]} \dfreesep d_{\infty} \\
\cline{2-2}
          & a_{d_{\infty}},\ldots,a_{d_{\infty}+\dsmax} \\
\cline{2-2}
          & c_{d_{\infty}},\ldots,c_{d_{\infty}+\dsmax} \\
\hline
\hline
\mlr{10}  & \begin{mat2} 4752 & 6166 \end{mat2} \\
\cline{2-2}
          & 2,3,3,4,4,5,5,5,5,5,5 \dfreesep 14 \\
\cline{2-2}
          & 14,0,80,0,397,0,2385,0,13962,0,80558,0,469959,0,2736038,0 \\
\cline{2-2}
          & 65,0,540,0,3451,0,24749,0,168341,0,1109898,0,7276046,0,47024669,0 \\
\hline
\mlr{13}  & \begin{mat2} 43372 & 65446 \end{mat2} \\
\cline{2-2}
          & 2,3,3,4,4,5,5,6,6,6,6,6,6,6 \dfreesep 16 \\
\cline{2-2}
          & 3,13,28,62,139,349,860,2188,4942,11706,28782,69737,168870,405984,979609,2365761 \\
\cline{2-2}
          & 18,57,160,460,1154,3311,8706,23912,58360,148350,388210,1002441,2570496,6529100,16591680,42062277 \\
\hline
\mlr{15}  & \begin{mat2} 412764 & 667114 \end{mat2} \\
\cline{2-2}
          & 2,3,3,4,4,5,5,6,6,6,7,7,7,7,7,7 \dfreesep 18 \\
\cline{2-2}
          & 5,19,40,108,208,461,1223,2887,7047,17140,41241,99750,239505,579028,1400448,3376495 \\
\cline{2-2}
          & 26,109,294,862,1898,4531,12920,33269,87530,226758,580132,1487822,3780132,9636484,24488010,61947071 \\
\hline
\mlr{19}  & \begin{mat2} 5327632 & 7026236 \end{mat2} \\
\cline{2-2}
          & 2,3,3,4,4,5,5,6,6,6,7,7,8,8,8,8,8,8,9,9 \dfreesep 22 \\
\cline{2-2}
          & 31,0,151,0,879,0,5179,0,29716,0,173543,0,1012113,0,5898333,0 \\
\cline{2-2}
          & 218,0,1227,0,9280,0,63228,0,414407,0,2714184,0,17553879,0,112379079,0 \\
\hline
\mlr{22}  & \begin{mat2} 42431526 & 72713352 \end{mat2} \\
\cline{2-2}
          & 2,3,3,4,4,5,5,6,6,6,7,7,8,8,8,8,9,9,9,9,9,9,9 \dfreesep 24 \\
\cline{2-2}
          & 16,0,108,0,639,0,3727,0,21298,0,124547,0,728158,0,4239458,0 \\
\cline{2-2}
          & 82,0,868,0,6696,0,45536,0,295575,0,1940887,0,12591660,0,80540832,0 \\
\hline
\mlr{26}  & \begin{mat2} 506477557 & 722102055 \end{mat2} \\
\cline{2-2}
          & 2,3,3,4,4,5,5,6,6,6,7,7,8,8,8,8,9,9,9,10,10,10,10,10,10,10,10 \dfreesep 27 \\
\cline{2-2}
          & 7,25,47,116,271,686,1585,3856,9457,22211,54716,131399,317591,768660,1853483,4470440 \\
\cline{2-2}
          & 39,204,463,1190,3039,8106,20215,52456,137671,341522,887948,2246280,5700767,14450312,36447601,91678560 \\
\hline
\mlr{30}  & \begin{mat2} 52762256724 & 64643055434 \end{mat2} \\
\cline{2-2}
          & 2,3,3,4,4,5,5,6,6,6,7,7,8,8,8,8,9,9,9,10,10,10,10,11,11,11,11,11,11,12,12 \dfreesep 30 \\
\cline{2-2}
          & 2,17,42,112,258,578,1459,3357,8132,20299,47868,115612,279189,672286,1626470,3922381 \\
\cline{2-2}
          & 16,137,398,1244,3068,7248,19562,47851,122486,324251,803744,2041702,5167536,13012190,32874642,82620557 \\
\hline
\end{codearray}
$
\end{table*}
\begin{table*}
\caption{Improved $ \text{OBDP}^{(5)} $ codes, $ R = 1/2 $, $ m \geq 9 $}
\centering
$
\begin{codearray}
\hline
\mlr{m}   & G(D) \\
\cline{2-2}
          & \hat{d}_{[0,m]} \dfreesep d_{\infty} \\
\cline{2-2}
          & a_{d_{\infty}},\ldots,a_{d_{\infty}+\dsmax} \\
\cline{2-2}
          & c_{d_{\infty}},\ldots,c_{d_{\infty}+\dsmax} \\
\hline
\hline
\mlr{11}  & \begin{mat2} 4325 & 6747 \end{mat2} \\
\cline{2-2}
          & 2,3,3,4,4,5,5,5,5,6,6,6 \dfreesep 15 \\
\cline{2-2}
          & 14,21,34,101,249,597,1373,3317,8014,19559,47302,113723,274266,662666,1600334,3857052 \\
\cline{2-2}
          & 66,98,220,788,2083,5424,13771,35966,93970,246720,635694,1623432,4149988,10599534,26964102,68271886 \\
\hline
\mlr{14}  & \begin{mat2} 42437 & 72711 \end{mat2} \\
\cline{2-2}
          & 2,3,3,4,4,5,5,6,6,6,6,6,6,7,7 \dfreesep 17 \\
\cline{2-2}
          & 2,23,30,71,204,393,1024,2593,6000,14473,35592,85300,205943,495822,1198985,2893119 \\
\cline{2-2}
          & 4,142,210,520,1752,3600,10516,28850,72086,186244,486840,1238964,3174023,8058126,20514015,51960096 \\
\hline
\mlr{16}  & \begin{mat2} 551576 & 755072 \end{mat2} \\
\cline{2-2}
          & 2,3,3,4,4,5,5,6,6,6,7,7,7,7,7,8,8 \dfreesep 19 \\
\cline{2-2}
          & 9,23,44,114,248,603,1471,3460,8530,20549,49631,119792,288771,696786,1682350,4066959 \\
\cline{2-2}
          & 41,148,352,974,2290,6080,16313,41338,108486,279190,717443,1830546,4661987,11853186,30053680,76105216 \\
\hline
\mlr{20}  & \begin{mat2} 4271771 & 6465163 \end{mat2} \\
\cline{2-2}
          & 2,3,3,4,4,5,5,6,6,6,7,7,8,8,8,8,8,8,8,9,9 \dfreesep 22 \\
\cline{2-2}
          & 4,16,49,100,212,500,1254,3086,7326,17724,42885,103374,250020,603244,1452195,3507704 \\
\cline{2-2}
          & 22,94,370,808,2010,5196,14038,37156,95230,244942,626798,1599454,4082304,10367564,26198528,66276392 \\
\hline
\mlr{23}  & \begin{mat2} 40560411 & 67665743 \end{mat2} \\
\cline{2-2}
          & 2,3,3,4,4,5,5,6,6,6,7,7,8,8,8,8,9,9,9,9,9,10,10,10 \dfreesep 24 \\
\cline{2-2}
          & 1,12,34,64,151,395,937,2203,5343,12988,30974,75085,182024,437317,1059067,2555260 \\
\cline{2-2}
          & 2,72,226,518,1450,4181,10442,25989,68478,178314,449866,1156167,2956164,7475449,18998962,48042454 \\
\hline
\mlr{27}  & \begin{mat2} 4020517504 & 7170726534 \end{mat2} \\
\cline{2-2}
          & 2,3,3,4,4,5,5,6,6,6,7,7,8,8,8,8,9,9,9,10,10,10,10,10,10,11,11,11 \dfreesep 28 \\
\cline{2-2}
          & 9,17,72,131,344,793,1971,4565,11365,26950,65994,158336,383423,923113,2233769,5383067 \\
\cline{2-2}
          & 42,113,640,1197,3612,8727,23938,59257,157498,395176,1026358,2593690,6626288,16713673,42369486,106690671 \\
\hline
\mlr{31}  & \begin{mat2} 42523570626 & 64546507642 \end{mat2} \\
\cline{2-2}
          & 2,3,3,4,4,5,5,6,6,6,7,7,8,8,8,8,9,9,9,10,10,10,10,11,11,11,11,11,11,11,12,12 \dfreesep 32 \\
\cline{2-2}
          & 39,0,252,0,1379,0,8405,0,47762,0,278396,0,1627011,0,9479225,0 \\
\cline{2-2}
          & 329,0,2708,0,17089,0,119182,0,759940,0,4907375,0,31438671,0,199350291,0 \\
\hline
\end{codearray}
$
\end{table*}
\begin{table*}
\caption{Improved $ \text{OBDP}^{(6)} $ codes, $ R = 1/2 $, $ m \geq 11 $}
\centering
$
\begin{codearray}
\hline
\mlr{m}   & G(D) \\
\cline{2-2}
          & \hat{d}_{[0,m]} \dfreesep d_{\infty} \\
\cline{2-2}
          & a_{d_{\infty}},\ldots,a_{d_{\infty}+\dsmax} \\
\cline{2-2}
          & c_{d_{\infty}},\ldots,c_{d_{\infty}+\dsmax} \\
\hline
\hline
\mlr{12}  & \begin{mat2} 43134 & 73724 \end{mat2} \\
\cline{2-2}
          & 2,3,3,4,4,5,5,5,5,6,6,6,6 \dfreesep 16 \\
\cline{2-2}
          & 14,38,35,108,342,724,1604,4020,9825,23899,57724,138584,335272,808236,1952588,4714596 \\
\cline{2-2}
          & 60,188,288,952,2754,6628,16606,44640,116712,304987,785180,2002076,5132496,13067080,33236580,84264272 \\
\hline
\mlr{17}  & \begin{mat2} 461253 & 751565 \end{mat2} \\
\cline{2-2}
          & 2,3,3,4,4,5,5,6,6,6,7,7,7,7,7,7,8,8 \dfreesep 20 \\
\cline{2-2}
          & 9,28,55,113,330,747,1812,4299,10117,24904,60321,145294,350099,846621,2043057,4930302 \\
\cline{2-2}
          & 46,170,414,975,3278,7975,20570,52437,132678,346830,891562,2271608,5770478,14698027,37181052,93955814 \\
\hline
\mlr{21}  & \begin{mat2} 46124774 & 62422564 \end{mat2} \\
\cline{2-2}
          & 2,3,3,4,4,5,5,6,6,6,7,7,8,8,8,8,8,9,9,9,9,9 \dfreesep 24 \\
\cline{2-2}
          & 45,0,205,0,1267,0,7451,0,42602,0,250539,0,1456088,0,8476055,0 \\
\cline{2-2}
          & 302,0,1847,0,13595,0,92696,0,604413,0,3982980,0,25624480,0,163618242,0 \\
\hline
\mlr{24}  & \begin{mat2} 513541374 & 663212164 \end{mat2} \\
\cline{2-2}
          & 2,3,3,4,4,5,5,6,6,6,7,7,8,8,8,8,9,9,9,9,9,9,9,9,9 \dfreesep 26 \\
\cline{2-2}
          & 24,0,170,0,924,0,5316,0,31345,0,182781,0,1059865,0,6179147,0 \\
\cline{2-2}
          & 149,0,1630,0,10554,0,68762,0,462827,0,3001342,0,19227629,0,122654694,0 \\
\hline
\mlr{28}  & \begin{mat2} 5621541772 & 6312710476 \end{mat2} \\
\cline{2-2}
          & 2,3,3,4,4,5,5,6,6,6,7,7,8,8,8,8,9,9,9,10,10,10,10,10,10,10,10,10,10 \dfreesep 29 \\
\cline{2-2}
          & 10,29,85,158,408,969,2257,5744,13589,33041,79462,191109,463485,1117630,2697269,6511035 \\
\cline{2-2}
          & 74,252,819,1640,4794,12384,30519,82608,207251,531920,1346046,3400202,8645743,21814508,54927249,138132926 \\
\hline
\end{codearray}
$
\end{table*}
\begin{table*}
\caption{Improved $ \text{OBDP}^{(7)} $ codes, $ R = 1/2 $, $ m \geq 13 $}
\centering
$
\begin{codearray}
\hline
\mlr{m}   & G(D) \\
\cline{2-2}
          & \hat{d}_{[0,m]} \dfreesep d_{\infty} \\
\cline{2-2}
          & a_{d_{\infty}},\ldots,a_{d_{\infty}+\dsmax} \\
\cline{2-2}
          & c_{d_{\infty}},\ldots,c_{d_{\infty}+\dsmax} \\
\hline
\hline
\mlr{13}  & \begin{mat2} 54232 & 76676 \end{mat2} \\
\cline{2-2}
          & 2,3,3,4,4,5,5,5,5,6,6,6,7,7 \dfreesep 16 \\
\cline{2-2}
          & 2,12,29,70,158,292,883,2000,4883,11709,28348,68907,164552,398298,964176,2322775 \\
\cline{2-2}
          & 4,60,184,502,1414,2662,8908,21462,56990,147211,380780,982795,2484436,6342994,16184098,40921627 \\
\hline
\mlr{16}  & \begin{mat2} 441612 & 764566 \end{mat2} \\
\cline{2-2}
          & 2,3,3,4,4,5,5,6,6,6,6,6,6,7,7,7,8 \dfreesep 19 \\
\cline{2-2}
          & 6,22,46,102,262,633,1579,3746,8845,21196,51274,123650,298940,723634,1745518,4210802 \\
\cline{2-2}
          & 28,140,318,858,2500,6644,17857,45444,115165,294036,755602,1922784,4902904,12490556,31621014,79865860 \\
\hline
\mlr{18}  & \begin{mat2} 4625434 & 7534724 \end{mat2} \\
\cline{2-2}
          & 2,3,3,4,4,5,5,6,6,6,7,7,7,7,7,8,8,8,8 \dfreesep 21 \\
\cline{2-2}
          & 10,29,69,165,347,901,2180,5041,12399,29522,71856,174244,418629,1011550,2444837,5899055 \\
\cline{2-2}
          & 60,204,561,1518,3527,9914,25616,63406,168071,424978,1093098,2801834,7092835,17993692,45582743,115014128 \\
\hline
\mlr{22}  & \begin{mat2} 54733632 & 60302236 \end{mat2} \\
\cline{2-2}
          & 2,3,3,4,4,5,5,6,6,6,7,7,8,8,8,8,8,8,8,8,9,9,9 \dfreesep 24 \\
\cline{2-2}
          & 6,29,62,135,325,777,1853,4430,10662,26012,62665,151620,365898,882440,2129925,5146568 \\
\cline{2-2}
          & 34,215,552,1299,3464,8841,22836,58240,149086,386220,983616,2512418,6367826,16117418,40715392,102778092 \\
\hline
\mlr{25}  & \begin{mat2} 505443406 & 671573632 \end{mat2} \\
\cline{2-2}
          & 2,3,3,4,4,5,5,6,6,6,7,7,8,8,8,8,9,9,9,9,9,9,9,9,9,10 \dfreesep 26 \\
\cline{2-2}
          & 1,16,41,104,236,569,1370,3229,7688,19010,45550,109213,264497,640102,1545054,3727715 \\
\cline{2-2}
          & 4,132,366,986,2478,6645,17482,43737,110372,289344,731264,1846511,4698444,11915342,30079166,75764209 \\
\hline
\mlr{29}  & \begin{mat2} 4124636521 & 6672055513 \end{mat2} \\
\cline{2-2}
          & 2,3,3,4,4,5,5,6,6,6,7,7,8,8,8,8,9,9,9,10,10,10,10,10,10,10,10,10,10,11 \dfreesep 30 \\
\cline{2-2}
          & 12,46,80,232,463,1148,2823,6716,16371,39516,96254,231334,557147,1347521,3255460,7855376 \\
\cline{2-2}
          & 112,380,742,2518,5354,14514,37798,96086,246894,629678,1619976,4089802,10320348,26111155,65884494,165655522 \\
\hline
\end{codearray}
$
\end{table*}
\begin{table*}
\caption{Improved $ \text{OBDP}^{(8)} $ codes, $ R = 1/2 $, $ m \geq 15 $}
\centering
$
\begin{codearray}
\hline
\mlr{m}   & G(D) \\
\cline{2-2}
          & \hat{d}_{[0,m]} \dfreesep d_{\infty} \\
\cline{2-2}
          & a_{d_{\infty}},\ldots,a_{d_{\infty}+\dsmax} \\
\cline{2-2}
          & c_{d_{\infty}},\ldots,c_{d_{\infty}+\dsmax} \\
\hline
\hline
\mlr{17}  & \begin{mat2} 422767 & 723745 \end{mat2} \\
\cline{2-2}
          & 2,3,3,4,4,5,5,6,6,6,6,6,6,6,7,7,7,7 \dfreesep 20 \\
\cline{2-2}
          & 8,35,61,139,318,800,1870,4577,11240,26781,64637,156119,376873,906954,2191764,5299061 \\
\cline{2-2}
          & 44,225,486,1261,3150,8804,21896,57561,150812,382075,976340,2491469,6338808,16023400,40596468,102679609 \\
\hline
\mlr{19}  & \begin{mat2} 4307646 & 6562572 \end{mat2} \\
\cline{2-2}
          & 2,3,3,4,4,5,5,6,6,6,7,7,7,7,7,7,8,8,8,8 \dfreesep 22 \\
\cline{2-2}
          & 22,0,171,0,909,0,5197,0,30434,0,177034,0,1034402,0,6019927,0 \\
\cline{2-2}
          & 123,0,1367,0,9600,0,63447,0,423511,0,2766631,0,17934581,0,114643198,0 \\
\hline
\mlr{23}  & \begin{mat2} 42547561 & 64515053 \end{mat2} \\
\cline{2-2}
          & 2,3,3,4,4,5,5,6,6,6,7,7,8,8,8,8,8,9,9,9,9,9,9,9 \dfreesep 25 \\
\cline{2-2}
          & 11,33,63,155,392,894,2279,5413,12921,31180,75366,182252,439619,1060837,2561211,6179005 \\
\cline{2-2}
          & 61,244,557,1430,4142,10222,28163,71466,182017,465264,1189506,3031960,7684541,19444324,49126387,123802696 \\
\hline
\mlr{26}  & \begin{mat2} 463375615 & 627035103 \end{mat2} \\
\cline{2-2}
          & 2,3,3,4,4,5,5,6,6,6,7,7,8,8,8,8,9,9,9,9,9,9,9,9,10,10,10 \dfreesep 28 \\
\cline{2-2}
          & 32,0,251,0,1388,0,7863,0,45506,0,265148,0,1542960,0,8996318,0 \\
\cline{2-2}
          & 230,0,2496,0,16477,0,107000,0,690760,0,4491289,0,28746811,0,182938866,0 \\
\hline
\mlr{30}  & \begin{mat2} 53754445264 & 65055123214 \end{mat2} \\
\cline{2-2}
          & 2,3,3,4,4,5,5,6,6,6,7,7,8,8,8,8,9,9,9,10,10,10,10,10,10,10,11,11,11,11,12 \dfreesep 31 \\
\cline{2-2}
          & 11,46,125,240,590,1435,3402,8177,19812,47401,115482,279371,673700,1627130,3925407,9472951 \\
\cline{2-2}
          & 109,464,1335,2814,7446,19410,48672,124378,319324,802686,2056136,5213210,13150436,33147834,83319039,209113130 \\
\hline
\end{codearray}
$
\end{table*}
\begin{table*}
\tcfirst
\caption{OBCDF codes, $ R = 1/3 $}
\centering
$
\begin{codearray}
\hline
\mlr{m}   & G(D) \\
\cline{2-2}
          & \hat{d}_{[0,m]} \dfreesep d_{\infty} \\
\cline{2-2}
          & a_{d_{\infty}},\ldots,a_{d_{\infty}+\dsmax} \\
\cline{2-2}
          & c_{d_{\infty}},\ldots,c_{d_{\infty}+\dsmax} \\
\hline
\hline
\mlr{1}   & \begin{mat3} 2 & 4 & 6 \end{mat3} \\
\cline{2-2}
          & 2,4 \dfreesep 4 \\
\cline{2-2}
          & 1,0,1,0,1,0,1,0,1,0,1,0,1,0,1,0 \\
\cline{2-2}
          & 1,0,2,0,3,0,4,0,5,0,6,0,7,0,8,0 \\
\hline
\mlr{2}   & \begin{mat3} 5 & 7 & 7 \end{mat3} \\
\cline{2-2}
          & 3,4,5 \dfreesep 8 \\
\cline{2-2}
          & 2,0,5,0,13,0,34,0,89,0,233,0,610,0,1597,0 \\
\cline{2-2}
          & 3,0,15,0,58,0,201,0,655,0,2052,0,6255,0,18687,0 \\
\hline
\mlr{3}   & \begin{mat3} 44 & 54 & 74 \end{mat3} \\
\cline{2-2}
          & 3,4,5,6 \dfreesep 9 \\
\cline{2-2}
          & 1,1,1,2,3,7,10,11,25,45,58,95,173,272,421,688 \\
\cline{2-2}
          & 1,2,3,4,9,28,40,50,129,250,358,622,1201,2064,3389,5792 \\
\hline
\mlr{4}   & \begin{mat3} 52 & 66 & 76 \end{mat3} \\
\cline{2-2}
          & 3,4,5,6,7 \dfreesep 12 \\
\cline{2-2}
          & 5,0,3,0,13,0,62,0,108,0,328,0,1051,0,2544,0 \\
\cline{2-2}
          & 12,0,12,0,56,0,320,0,693,0,2324,0,8380,0,23009,0 \\
\hline
\mlr{5}   & \begin{mat3} 45 & 51 & 77 \end{mat3} \\
\cline{2-2}
          & 3,4,5,6,7,8 \dfreesep 12 \\
\cline{2-2}
          & 2,0,2,0,10,0,17,0,67,0,189,0,480,0,1483,0 \\
\cline{2-2}
          & 3,0,4,0,36,0,74,0,344,0,1152,0,3477,0,11996,0 \\
\hline
\mlr{6}   & \begin{mat3} 434 & 564 & 704 \end{mat3} \\
\cline{2-2}
          & 3,4,5,6,7,8,9 \dfreesep 13 \\
\cline{2-2}
          & 1,1,0,4,5,3,11,14,21,50,87,133,211,378,649,1075 \\
\cline{2-2}
          & 1,2,0,14,19,12,45,64,105,302,539,828,1479,2848,5165,9130 \\
\hline
\mlr{7}   & \begin{mat3} 446 & 616 & 722 \end{mat3} \\
\cline{2-2}
          & 3,4,5,6,7,8,9,9 \dfreesep 14 \\
\cline{2-2}
          & 1,0,3,0,3,0,18,0,47,0,135,0,376,0,1090,0 \\
\cline{2-2}
          & 1,0,9,0,13,0,84,0,254,0,873,0,2734,0,8948,0 \\
\hline
\mlr{8}   & \begin{mat3} 533 & 575 & 665 \end{mat3} \\
\cline{2-2}
          & 3,4,5,6,7,8,9,10,10 \dfreesep 15 \\
\cline{2-2}
          & 1,1,0,1,2,7,3,11,27,42,59,92,167,294,581,947 \\
\cline{2-2}
          & 3,2,0,2,4,30,17,50,119,246,367,640,1183,2172,4575,7940 \\
\hline
\mlr{9}   & \begin{mat3} 4674 & 6754 & 7544 \end{mat3} \\
\cline{2-2}
          & 3,4,5,6,7,8,9,10,10,11 \dfreesep 18 \\
\cline{2-2}
          & 4,0,4,0,13,0,25,0,97,0,314,0,797,0,2450,0 \\
\cline{2-2}
          & 12,0,16,0,51,0,132,0,589,0,2152,0,6209,0,21214,0 \\
\hline
\mlr{10}  & \begin{mat3} 5772 & 6056 & 7206 \end{mat3} \\
\cline{2-2}
          & 3,4,5,6,7,8,9,10,10,11,12 \dfreesep 18 \\
\cline{2-2}
          & 1,0,2,3,3,1,10,25,24,16,76,111,228,333,620,955 \\
\cline{2-2}
          & 2,0,6,11,14,9,58,125,140,120,528,765,1792,2761,5344,8581 \\
\hline
\mlr{11}  & \begin{mat3} 4135 & 5057 & 7263 \end{mat3} \\
\cline{2-2}
          & 3,4,5,6,7,8,9,10,11,11,11,11 \dfreesep 21 \\
\cline{2-2}
          & 1,3,2,5,5,11,28,33,56,94,183,298,486,916,1502,2437 \\
\cline{2-2}
          & 1,10,6,22,31,46,142,192,358,652,1303,2264,3988,7874,13592,23094 \\
\hline
\mlr{12}  & \begin{mat3} 51624 & 66234 & 71154 \end{mat3} \\
\cline{2-2}
          & 3,4,5,6,7,8,9,10,11,11,12,12,13 \dfreesep 22 \\
\cline{2-2}
          & 2,3,5,1,9,7,24,18,53,76,142,251,403,776,1317,2033 \\
\cline{2-2}
          & 8,7,24,3,42,37,158,86,392,572,1084,1973,3484,6936,12454,19865 \\
\hline
\mlr{13}  & \begin{mat3} 53256 & 65126 & 72552 \end{mat3} \\
\cline{2-2}
          & 3,4,5,6,7,8,9,10,11,11,12,12,13,13 \dfreesep 20 \\
\cline{2-2}
          & 1,0,2,0,4,0,4,0,24,0,61,0,145,0,418,0 \\
\cline{2-2}
          & 2,0,7,0,16,0,19,0,134,0,389,0,1080,0,3351,0 \\
\hline
\end{codearray}
$
\end{table*}
\begin{table*}
\tcnext
\caption{OBCDF codes, $ R = 1/3 $}
\centering
$
\begin{codearray}
\hline
\mlr{m}   & G(D) \\
\cline{2-2}
          & \hat{d}_{[0,m]} \dfreesep d_{\infty} \\
\cline{2-2}
          & a_{d_{\infty}},\ldots,a_{d_{\infty}+\dsmax} \\
\cline{2-2}
          & c_{d_{\infty}},\ldots,c_{d_{\infty}+\dsmax} \\
\hline
\hline
\mlr{14}  & \begin{mat3} 40701 & 53765 & 67273 \end{mat3} \\
\cline{2-2}
          & 3,4,5,6,7,8,9,10,11,11,12,13,13,13,14 \dfreesep 22 \\
\cline{2-2}
          & 1,0,0,1,2,1,3,6,18,16,56,48,119,154,353,438 \\
\cline{2-2}
          & 2,0,0,3,4,1,18,22,102,80,324,262,764,1058,2596,3406 \\
\hline
\mlr{15}  & \begin{mat3} 520454 & 644124 & 733334 \end{mat3} \\
\cline{2-2}
          & 3,4,5,6,7,8,9,10,11,11,12,13,13,13,14,15 \dfreesep 26 \\
\cline{2-2}
          & 5,0,7,0,12,0,36,0,114,0,310,0,899,0,2706,0 \\
\cline{2-2}
          & 23,0,28,0,66,0,212,0,834,0,2512,0,8081,0,26574,0 \\
\hline
\mlr{16}  & \begin{mat3} 403402 & 517712 & 730156 \end{mat3} \\
\cline{2-2}
          & 3,4,5,6,7,8,9,10,11,11,12,13,13,14,14,15,15 \dfreesep 24 \\
\cline{2-2}
          & 1,0,1,0,2,0,5,0,20,0,45,0,170,0,437,0 \\
\cline{2-2}
          & 2,0,1,0,10,0,19,0,94,0,210,0,1084,0,3142,0 \\
\hline
\mlr{17}  & \begin{mat3} 421765 & 531607 & 706321 \end{mat3} \\
\cline{2-2}
          & 3,4,5,6,7,8,9,10,11,11,12,13,13,14,14,15,15,16 \dfreesep 29 \\
\cline{2-2}
          & 1,1,4,5,3,14,24,46,76,105,190,312,585,915,1549,2741 \\
\cline{2-2}
          & 1,2,18,22,11,64,138,270,510,786,1424,2510,4943,8114,14457,26560 \\
\hline
\mlr{18}  & \begin{mat3} 4304304 & 5060254 & 6501424 \end{mat3} \\
\cline{2-2}
          & 3,4,5,6,7,8,9,10,11,11,12,13,13,14,15,15,16,16,17 \dfreesep 23 \\
\cline{2-2}
          & 1,0,0,0,0,0,0,2,3,3,5,10,7,31,31,45 \\
\cline{2-2}
          & 1,0,0,0,0,0,0,4,15,12,19,36,39,184,167,286 \\
\hline
\mlr{19}  & \begin{mat3} 4763236 & 6146306 & 7454762 \end{mat3} \\
\cline{2-2}
          & 3,4,5,6,7,8,9,10,11,11,12,13,13,14,15,15,16,16,16,16 \dfreesep 26 \\
\cline{2-2}
          & 1,0,0,0,1,0,4,0,3,0,31,0,48,0,170,0 \\
\cline{2-2}
          & 4,0,0,0,2,0,35,0,12,0,207,0,300,0,1238,0 \\
\hline
\mlr{20}  & \begin{mat3} 5704623 & 6231075 & 7432617 \end{mat3} \\
\cline{2-2}
          & 3,4,5,6,7,8,9,10,11,11,12,13,13,14,15,15,16,16,17,17,17 \dfreesep 31 \\
\cline{2-2}
          & 2,2,1,3,4,6,10,14,12,45,74,106,205,380,645,1035 \\
\cline{2-2}
          & 6,16,5,10,16,46,66,90,86,318,530,834,1759,3512,6053,10094 \\
\hline
\mlr{21}  & \begin{mat3} 46324714 & 54425344 & 62027664 \end{mat3} \\
\cline{2-2}
          & 3,4,5,6,7,8,9,10,11,11,12,13,13,14,15,15,16,16,17,17,18,18 \dfreesep 32 \\
\cline{2-2}
          & 1,0,1,2,3,5,2,16,22,31,59,87,170,294,485,871 \\
\cline{2-2}
          & 4,0,2,4,14,23,6,100,140,191,418,647,1296,2464,4260,8009 \\
\hline
\mlr{22}  & \begin{mat3} 40666602 & 53634752 & 67377566 \end{mat3} \\
\cline{2-2}
          & 3,4,5,6,7,8,9,10,11,11,12,13,13,14,15,15,16,16,17,18,18,18,19 \dfreesep 34 \\
\cline{2-2}
          & 1,0,3,0,5,0,22,0,50,0,190,0,531,0,1451,0 \\
\cline{2-2}
          & 3,0,6,0,17,0,120,0,290,0,1404,0,4217,0,12663,0 \\
\hline
\mlr{23}  & \begin{mat3} 51275623 & 66500617 & 71746075 \end{mat3} \\
\cline{2-2}
          & 3,4,5,6,7,8,9,10,11,11,12,13,13,14,15,15,16,16,17,18,18,18,19,19 \dfreesep 37 \\
\cline{2-2}
          & 1,3,3,3,10,8,21,44,82,119,203,384,606,1042,1779,2965 \\
\cline{2-2}
          & 7,12,13,14,58,60,141,316,634,952,1809,3570,5888,10520,18863,33156 \\
\hline
\mlr{24}  & \begin{mat3} 551571614 & 616366264 & 770370374 \end{mat3} \\
\cline{2-2}
          & 3,4,5,6,7,8,9,10,11,11,12,13,13,14,15,15,16,16,17,18,18,19,19,20,20 \dfreesep 34 \\
\cline{2-2}
          & 1,0,0,0,1,0,5,3,13,10,26,28,59,136,191,307 \\
\cline{2-2}
          & 2,0,0,0,4,0,28,17,94,50,166,224,486,1098,1654,2823 \\
\hline
\mlr{25}  & \begin{mat3} 527061652 & 641577756 & 737773026 \end{mat3} \\
\cline{2-2}
          & 3,4,5,6,7,8,9,10,11,11,12,13,13,14,15,15,16,16,17,18,18,19,19,20,20,20 \dfreesep 36 \\
\cline{2-2}
          & 1,0,2,0,5,0,2,0,28,0,65,0,193,0,464,0 \\
\cline{2-2}
          & 2,0,12,0,21,0,8,0,185,0,450,0,1648,0,4182,0 \\
\hline
\mlr{26}  & \begin{mat3} 533361127 & 650525053 & 724436665 \end{mat3} \\
\cline{2-2}
          & 3,4,5,6,7,8,9,10,11,11,12,13,13,14,15,15,16,16,17,18,18,19,19,20,21,21,22 \dfreesep 38 \\
\cline{2-2}
          & 2,0,0,1,2,5,8,11,23,24,44,80,175,191,337,651 \\
\cline{2-2}
          & 6,0,0,3,18,33,54,79,170,162,388,640,1514,1691,3204,6477 \\
\hline
\end{codearray}
$
\end{table*}
\begin{table*}
\tcnext
\caption{OBCDF codes, $ R = 1/3 $}
\centering
$
\begin{codearray}
\hline
\mlr{m}   & G(D) \\
\cline{2-2}
          & \hat{d}_{[0,m]} \dfreesep d_{\infty} \\
\cline{2-2}
          & a_{d_{\infty}},\ldots,a_{d_{\infty}+\dsmax} \\
\cline{2-2}
          & c_{d_{\infty}},\ldots,c_{d_{\infty}+\dsmax} \\
\hline
\hline
\mlr{27}  & \begin{mat3} 4017417334 & 5150442624 & 7336076004 \end{mat3} \\
\cline{2-2}
          & 3,4,5,6,7,8,9,10,11,11,12,13,13,14,15,15,16,16,17,18,18,19,20,20,20,20,21,22 \dfreesep 38 \\
\cline{2-2}
          & 1,0,1,0,5,0,6,0,15,0,61,0,120,0,348,0 \\
\cline{2-2}
          & 5,0,2,0,18,0,28,0,76,0,354,0,853,0,2632,0 \\
\hline
\mlr{28}  & \begin{mat3} 4333324526 & 5054103212 & 6522555542 \end{mat3} \\
\cline{2-2}
          & 3,4,5,6,7,8,9,10,11,11,12,13,13,14,15,15,16,16,17,18,18,19,20,20,21,21,21,22,22 \dfreesep 41 \\
\cline{2-2}
          & 4,0,3,3,2,10,8,14,17,37,33,104,160,205,463,810 \\
\cline{2-2}
          & 20,0,11,18,10,58,44,102,105,264,259,774,1398,1830,4247,7760 \\
\hline
\mlr{29}  & \begin{mat3} 4446061757 & 5406076103 & 7460043031 \end{mat3} \\
\cline{2-2}
          & 3,4,5,6,7,8,9,10,11,11,12,13,13,14,15,15,16,16,17,18,18,19,20,20,20,21,22,22,22,23 \dfreesep 38 \\
\cline{2-2}
          & 1,0,0,1,0,0,1,2,3,3,5,3,14,35,38,94 \\
\cline{2-2}
          & 2,0,0,1,0,0,2,10,20,31,26,29,94,251,286,716 \\
\hline
\mlr{30}  & \begin{mat3} 44137047714 & 63744372044 & 76211621174 \end{mat3} \\
\cline{2-2}
          & 3,4,5,6,7,8,9,10,11,11,12,13,13,14,15,15,16,16,17,18,18,19,20,20,21,21,22,22,23,24,24 \dfreesep 41 \\
\cline{2-2}
          & 1,1,0,1,3,0,1,5,9,7,8,24,19,83,114,217 \\
\cline{2-2}
          & 5,2,0,4,15,0,9,34,49,54,38,184,145,648,988,2062 \\
\hline
\mlr{31}  & \begin{mat3} 40043101266 & 65362363716 & 71200036512 \end{mat3} \\
\cline{2-2}
          & 3,4,5,6,7,8,9,10,11,11,12,13,13,14,15,15,16,16,17,18,18,19,20,20,21,21,22,22,23,23,23,23 \dfreesep 45 \\
\cline{2-2}
          & 2,1,3,3,3,3,5,14,17,41,46,84,190,320,522,855 \\
\cline{2-2}
          & 4,4,9,14,17,10,31,88,119,286,350,646,1646,2964,4898,8546 \\
\hline
\end{codearray}
$
\end{table*}
\begin{table*}
\caption{Improved $ \text{OBDP}^{(0)} $ codes, $ R = 1/3 $}
\centering
$
\begin{codearray}
\hline
\mlr{m}   & G(D) \\
\cline{2-2}
          & \hat{d}_{[0,m]} \dfreesep d_{\infty} \\
\cline{2-2}
          & a_{d_{\infty}},\ldots,a_{d_{\infty}+\dsmax} \\
\cline{2-2}
          & c_{d_{\infty}},\ldots,c_{d_{\infty}+\dsmax} \\
\hline
\hline
\mlr{3}   & \begin{mat3} 54 & 64 & 74 \end{mat3} \\
\cline{2-2}
          & 3,4,5,6 \dfreesep 10 \\
\cline{2-2}
          & 3,0,2,0,15,0,24,0,87,0,188,0,557,0,1354,0 \\
\cline{2-2}
          & 6,0,6,0,58,0,118,0,507,0,1284,0,4323,0,11846,0 \\
\hline
\mlr{5}   & \begin{mat3} 43 & 57 & 71 \end{mat3} \\
\cline{2-2}
          & 3,4,5,6,7,8 \dfreesep 12 \\
\cline{2-2}
          & 2,0,1,0,8,0,31,0,55,0,193,0,544,0,1463,0 \\
\cline{2-2}
          & 3,0,3,0,29,0,147,0,322,0,1249,0,4122,0,12486,0 \\
\hline
\mlr{6}   & \begin{mat3} 474 & 514 & 764 \end{mat3} \\
\cline{2-2}
          & 3,4,5,6,7,8,9 \dfreesep 14 \\
\cline{2-2}
          & 1,2,3,3,4,13,11,28,57,78,126,223,415,666,1110,1917 \\
\cline{2-2}
          & 2,4,10,11,16,59,58,160,326,510,884,1675,3204,5534,9900,17801 \\
\hline
\mlr{7}   & \begin{mat3} 466 & 536 & 662 \end{mat3} \\
\cline{2-2}
          & 3,4,5,6,7,8,9,9 \dfreesep 16 \\
\cline{2-2}
          & 3,0,9,0,15,0,48,0,134,0,393,0,1195,0,3360,0 \\
\cline{2-2}
          & 6,0,30,0,62,0,252,0,837,0,2758,0,9363,0,29590,0 \\
\hline
\mlr{8}   & \begin{mat3} 575 & 623 & 727 \end{mat3} \\
\cline{2-2}
          & 3,4,5,6,7,8,9,10,10 \dfreesep 18 \\
\cline{2-2}
          & 1,4,10,9,17,14,39,56,85,207,258,484,917,1390,2572,4362 \\
\cline{2-2}
          & 2,10,50,37,92,92,274,402,600,1579,2130,4120,8306,13108,25524,45034 \\
\hline
\mlr{13}  & \begin{mat3} 40246 & 53322 & 67576 \end{mat3} \\
\cline{2-2}
          & 3,4,5,6,7,8,9,10,11,11,12,12,13,13 \dfreesep 22 \\
\cline{2-2}
          & 1,0,1,2,5,5,8,11,30,51,67,138,192,364,630,1003 \\
\cline{2-2}
          & 2,0,2,4,18,19,38,47,180,301,422,936,1386,2754,5042,8549 \\
\hline
\mlr{25}  & \begin{mat3} 513467436 & 664525446 & 716556172 \end{mat3} \\
\cline{2-2}
          & 3,4,5,6,7,8,9,10,11,11,12,13,13,14,15,15,16,16,17,18,18,19,19,20,20,20 \dfreesep 41 \\
\cline{2-2}
          & 2,7,9,3,19,37,48,84,154,274,445,728,1263,2135,3638,6297 \\
\cline{2-2}
          & 12,40,43,18,129,286,390,658,1328,2482,4117,7272,13039,23074,40990,73458 \\
\hline
\mlr{28}  & \begin{mat3} 4430505436 & 5463114632 & 7432121422 \end{mat3} \\
\cline{2-2}
          & 3,4,5,6,7,8,9,10,11,11,12,13,13,14,15,15,16,16,17,18,18,19,20,20,21,21,21,22,22 \dfreesep 41 \\
\cline{2-2}
          & 1,1,0,1,1,4,5,17,13,27,56,75,152,289,455,812 \\
\cline{2-2}
          & 1,2,0,4,5,14,27,116,81,168,440,580,1320,2538,4453,7640 \\
\hline
\mlr{29}  & \begin{mat3} 4156353665 & 5272053621 & 6646554507 \end{mat3} \\
\cline{2-2}
          & 3,4,5,6,7,8,9,10,11,11,12,13,13,14,15,15,16,16,17,18,18,19,20,20,20,21,22,22,22,23 \dfreesep 44 \\
\cline{2-2}
          & 1,0,4,0,9,0,36,0,94,0,292,0,773,0,2216,0 \\
\cline{2-2}
          & 5,0,21,0,51,0,224,0,666,0,2319,0,7151,0,22111,0 \\
\hline
\end{codearray}
$
\end{table*}
\begin{table*}
\caption{Improved $ \text{OBDP}^{(1)} $ codes, $ R = 1/3 $}
\centering
$
\begin{codearray}
\hline
\mlr{m}   & G(D) \\
\cline{2-2}
          & \hat{d}_{[0,m]} \dfreesep d_{\infty} \\
\cline{2-2}
          & a_{d_{\infty}},\ldots,a_{d_{\infty}+\dsmax} \\
\cline{2-2}
          & c_{d_{\infty}},\ldots,c_{d_{\infty}+\dsmax} \\
\hline
\hline
\mlr{1}   & \begin{mat3} 2 & 6 & 6 \end{mat3} \\
\cline{2-2}
          & 2,3 \dfreesep 5 \\
\cline{2-2}
          & 1,1,1,1,1,1,1,1,1,1,1,1,1,1,1,1 \\
\cline{2-2}
          & 1,2,3,4,5,6,7,8,9,10,11,12,13,14,15,16 \\
\hline
\mlr{5}   & \begin{mat3} 57 & 63 & 75 \end{mat3} \\
\cline{2-2}
          & 3,4,5,6,7,7 \dfreesep 12 \\
\cline{2-2}
          & 1,0,5,0,8,0,19,0,82,0,201,0,568,0,1646,0 \\
\cline{2-2}
          & 2,0,17,0,30,0,109,0,496,0,1453,0,4540,0,15035,0 \\
\hline
\mlr{9}   & \begin{mat3} 4464 & 5154 & 6374 \end{mat3} \\
\cline{2-2}
          & 3,4,5,6,7,8,9,10,10,10 \dfreesep 18 \\
\cline{2-2}
          & 2,1,1,5,5,12,14,31,53,64,151,263,410,703,1191,2016 \\
\cline{2-2}
          & 6,1,2,17,28,48,60,187,324,416,1072,1935,3174,5869,10568,18392 \\
\hline
\mlr{10}  & \begin{mat3} 4662 & 5646 & 6272 \end{mat3} \\
\cline{2-2}
          & 3,4,5,6,7,8,9,10,10,11,11 \dfreesep 20 \\
\cline{2-2}
          & 2,0,11,0,20,0,35,0,148,0,393,0,1200,0,3559,0 \\
\cline{2-2}
          & 3,0,39,0,104,0,185,0,944,0,2840,0,9819,0,32463,0 \\
\hline
\mlr{27}  & \begin{mat3} 5040413754 & 6772040424 & 7075467434 \end{mat3} \\
\cline{2-2}
          & 3,4,5,6,7,8,9,10,11,11,12,13,13,14,15,15,16,16,17,18,18,19,20,20,20,20,21,21 \dfreesep 40 \\
\cline{2-2}
          & 1,0,2,0,9,0,10,0,40,0,116,0,410,0,1050,0 \\
\cline{2-2}
          & 2,0,12,0,45,0,54,0,265,0,1034,0,3714,0,10486,0 \\
\hline
\end{codearray}
$
\end{table*}
\begin{table*}
\caption{Improved $ \text{OBDP}^{(2)} $ codes, $ R = 1/3 $, $ m \geq 3 $}
\centering
$
\begin{codearray}
\hline
\mlr{m}   & G(D) \\
\cline{2-2}
          & \hat{d}_{[0,m]} \dfreesep d_{\infty} \\
\cline{2-2}
          & a_{d_{\infty}},\ldots,a_{d_{\infty}+\dsmax} \\
\cline{2-2}
          & c_{d_{\infty}},\ldots,c_{d_{\infty}+\dsmax} \\
\hline
\hline
\mlr{5}   & \begin{mat3} 45 & 53 & 67 \end{mat3} \\
\cline{2-2}
          & 3,4,5,6,6,7 \dfreesep 12 \\
\cline{2-2}
          & 1,0,4,0,9,0,25,0,70,0,205,0,580,0,1675,0 \\
\cline{2-2}
          & 1,0,13,0,34,0,129,0,418,0,1400,0,4535,0,14902,0 \\
\hline
\mlr{6}   & \begin{mat3} 514 & 564 & 674 \end{mat3} \\
\cline{2-2}
          & 3,4,5,6,7,7,8 \dfreesep 15 \\
\cline{2-2}
          & 3,2,6,11,7,17,32,53,80,153,246,427,793,1247,2103,3599 \\
\cline{2-2}
          & 9,4,24,52,41,90,202,360,538,1142,1912,3520,6865,11440,20299,36600 \\
\hline
\mlr{7}   & \begin{mat3} 456 & 646 & 772 \end{mat3} \\
\cline{2-2}
          & 3,4,5,6,7,8,8,9 \dfreesep 16 \\
\cline{2-2}
          & 1,2,6,9,6,15,21,40,77,112,216,341,575,1026,1700,3045 \\
\cline{2-2}
          & 2,4,20,39,32,87,120,238,536,784,1604,2699,4778,9016,15668,29541 \\
\hline
\mlr{10}  & \begin{mat3} 4732 & 5562 & 6346 \end{mat3} \\
\cline{2-2}
          & 3,4,5,6,7,8,9,10,10,10,11 \dfreesep 21 \\
\cline{2-2}
          & 4,5,7,11,9,25,45,78,114,205,355,605,1060,1774,2989,5080 \\
\cline{2-2}
          & 10,14,33,56,55,128,247,546,768,1480,2781,5024,9218,16318,28693,50856 \\
\hline
\mlr{15}  & \begin{mat3} 431724 & 507134 & 651654 \end{mat3} \\
\cline{2-2}
          & 3,4,5,6,7,8,9,10,11,11,12,13,13,13,13,14 \dfreesep 28 \\
\cline{2-2}
          & 5,0,18,0,32,0,107,0,326,0,890,0,2518,0,7577,0 \\
\cline{2-2}
          & 21,0,75,0,178,0,688,0,2459,0,7372,0,23063,0,76974,0 \\
\hline
\mlr{16}  & \begin{mat3} 472562 & 571076 & 761172 \end{mat3} \\
\cline{2-2}
          & 3,4,5,6,7,8,9,10,11,11,12,13,13,14,14,14,14 \dfreesep 28 \\
\cline{2-2}
          & 4,0,4,0,34,0,66,0,157,0,429,0,1277,0,3793,0 \\
\cline{2-2}
          & 13,0,14,0,194,0,450,0,1052,0,3344,0,11373,0,36122,0 \\
\hline
\mlr{21}  & \begin{mat3} 43617114 & 50176144 & 65531464 \end{mat3} \\
\cline{2-2}
          & 3,4,5,6,7,8,9,10,11,11,12,13,13,14,15,15,16,16,17,17,17,17 \dfreesep 34 \\
\cline{2-2}
          & 2,0,4,0,14,0,32,0,126,0,338,0,999,0,2781,0 \\
\cline{2-2}
          & 7,0,19,0,70,0,187,0,872,0,2703,0,8869,0,27234,0 \\
\hline
\mlr{24}  & \begin{mat3} 522211124 & 647163454 & 730404334 \end{mat3} \\
\cline{2-2}
          & 3,4,5,6,7,8,9,10,11,11,12,13,13,14,15,15,16,16,17,18,18,19,19,19,20 \dfreesep 35 \\
\cline{2-2}
          & 1,1,1,1,0,1,10,12,8,23,47,67,70,227,308,469 \\
\cline{2-2}
          & 5,4,1,6,0,8,68,76,66,172,337,498,568,1994,2658,4554 \\
\hline
\mlr{28}  & \begin{mat3} 4331523516 & 5216527052 & 7134531542 \end{mat3} \\
\cline{2-2}
          & 3,4,5,6,7,8,9,10,11,11,12,13,13,14,15,15,16,16,17,18,18,19,20,20,21,21,21,21,21 \dfreesep 44 \\
\cline{2-2}
          & 5,2,7,4,14,10,42,40,116,137,326,514,764,1307,2288,3879 \\
\cline{2-2}
          & 32,14,32,22,102,50,238,290,942,1151,2876,4914,7418,13371,24116,43447 \\
\hline
\end{codearray}
$
\end{table*}
\begin{table*}
\caption{Improved $ \text{OBDP}^{(3)} $ codes, $ R = 1/3 $, $ m \geq 5 $}
\centering
$
\begin{codearray}
\hline
\mlr{m}   & G(D) \\
\cline{2-2}
          & \hat{d}_{[0,m]} \dfreesep d_{\infty} \\
\cline{2-2}
          & a_{d_{\infty}},\ldots,a_{d_{\infty}+\dsmax} \\
\cline{2-2}
          & c_{d_{\infty}},\ldots,c_{d_{\infty}+\dsmax} \\
\hline
\hline
\mlr{5}   & \begin{mat3} 47 & 53 & 75 \end{mat3} \\
\cline{2-2}
          & 3,4,5,5,5,6 \dfreesep 13 \\
\cline{2-2}
          & 1,3,6,4,5,12,14,33,66,106,179,317,513,766,1297,2251 \\
\cline{2-2}
          & 1,8,26,20,19,62,86,204,420,710,1345,2606,4343,6790,12305,22356 \\
\hline
\mlr{6}   & \begin{mat3} 474 & 534 & 664 \end{mat3} \\
\cline{2-2}
          & 3,4,5,6,6,7,7 \dfreesep 15 \\
\cline{2-2}
          & 3,3,6,9,4,18,35,45,77,153,263,436,764,1209,2046,3550 \\
\cline{2-2}
          & 7,8,22,44,22,94,219,282,531,1104,1939,3460,6538,11006,19478,35738 \\
\hline
\mlr{7}   & \begin{mat3} 452 & 662 & 756 \end{mat3} \\
\cline{2-2}
          & 3,4,5,6,7,7,8,9 \dfreesep 16 \\
\cline{2-2}
          & 1,0,8,0,24,0,51,0,133,0,405,0,1129,0,3532,0 \\
\cline{2-2}
          & 1,0,24,0,113,0,287,0,898,0,3020,0,9436,0,32644,0 \\
\hline
\mlr{9}   & \begin{mat3} 4754 & 5324 & 6744 \end{mat3} \\
\cline{2-2}
          & 3,4,5,6,7,8,9,9,10,10 \dfreesep 20 \\
\cline{2-2}
          & 5,0,17,0,42,0,83,0,282,0,758,0,2537,0,6850,0 \\
\cline{2-2}
          & 11,0,70,0,204,0,490,0,1959,0,5742,0,21621,0,64990,0 \\
\hline
\mlr{12}  & \begin{mat3} 42664 & 53714 & 70344 \end{mat3} \\
\cline{2-2}
          & 3,4,5,6,7,8,9,10,11,11,11,12,12 \dfreesep 23 \\
\cline{2-2}
          & 1,2,7,7,8,25,27,43,88,144,240,435,752,1231,2122,3641 \\
\cline{2-2}
          & 1,6,29,28,32,130,159,256,582,1048,1786,3448,6312,10952,19922,35602 \\
\hline
\mlr{17}  & \begin{mat3} 422543 & 662711 & 725135 \end{mat3} \\
\cline{2-2}
          & 3,4,5,6,7,8,9,10,11,11,12,13,13,14,14,14,15,16 \dfreesep 29 \\
\cline{2-2}
          & 1,0,3,4,8,15,29,45,63,119,188,309,567,944,1550,2707 \\
\cline{2-2}
          & 1,0,13,18,38,74,181,306,449,892,1490,2556,4993,8946,15184,27418 \\
\hline
\mlr{18}  & \begin{mat3} 5373724 & 6575134 & 7227654 \end{mat3} \\
\cline{2-2}
          & 3,4,5,6,7,8,9,10,11,11,12,13,13,14,15,15,15,15,15 \dfreesep 26 \\
\cline{2-2}
          & 1,0,3,0,1,0,6,0,19,0,44,0,106,0,380,0 \\
\cline{2-2}
          & 2,0,15,0,8,0,36,0,83,0,289,0,814,0,3045,0 \\
\hline
\mlr{20}  & \begin{mat3} 4602545 & 5611033 & 7747647 \end{mat3} \\
\cline{2-2}
          & 3,4,5,6,7,8,9,10,11,11,12,13,13,14,15,15,16,16,16,17,17 \dfreesep 33 \\
\cline{2-2}
          & 1,1,2,5,7,10,27,37,62,122,232,359,592,975,1635,2880 \\
\cline{2-2}
          & 5,2,6,26,37,46,193,252,432,950,1908,3112,5370,9302,16539,30286 \\
\hline
\mlr{26}  & \begin{mat3} 410707237 & 506133675 & 724647071 \end{mat3} \\
\cline{2-2}
          & 3,4,5,6,7,8,9,10,11,11,12,13,13,14,15,15,16,16,17,18,18,19,19,20,20,20,21 \dfreesep 41 \\
\cline{2-2}
          & 2,0,2,12,11,12,24,42,86,117,222,399,648,1090,1774,3040 \\
\cline{2-2}
          & 10,0,10,60,55,80,164,266,646,902,1858,3638,5914,10664,18170,32172 \\
\hline
\end{codearray}
$
\end{table*}
\begin{table*}
\caption{Improved $ \text{OBDP}^{(4)} $ codes, $ R = 1/3 $, $ m \geq 7 $}
\centering
$
\begin{codearray}
\hline
\mlr{m}   & G(D) \\
\cline{2-2}
          & \hat{d}_{[0,m]} \dfreesep d_{\infty} \\
\cline{2-2}
          & a_{d_{\infty}},\ldots,a_{d_{\infty}+\dsmax} \\
\cline{2-2}
          & c_{d_{\infty}},\ldots,c_{d_{\infty}+\dsmax} \\
\hline
\hline
\mlr{9}   & \begin{mat3} 4714 & 5334 & 7724 \end{mat3} \\
\cline{2-2}
          & 3,4,5,6,7,8,8,9,9,10 \dfreesep 20 \\
\cline{2-2}
          & 3,7,8,11,19,27,41,85,137,226,411,705,1146,2001,3449,5722 \\
\cline{2-2}
          & 8,23,36,57,94,167,256,561,996,1686,3298,6059,10250,18665,34092,58896 \\
\hline
\mlr{10}  & \begin{mat3} 4456 & 6546 & 7672 \end{mat3} \\
\cline{2-2}
          & 3,4,5,6,7,8,9,9,9,10,10 \dfreesep 22 \\
\cline{2-2}
          & 9,0,24,0,53,0,145,0,424,0,1244,0,3648,0,10551,0 \\
\cline{2-2}
          & 28,0,111,0,335,0,982,0,3276,0,10625,0,34911,0,109888,0 \\
\hline
\mlr{11}  & \begin{mat3} 4551 & 5157 & 7353 \end{mat3} \\
\cline{2-2}
          & 3,4,5,6,7,8,9,10,10,10,11,11 \dfreesep 22 \\
\cline{2-2}
          & 1,4,4,12,16,15,41,66,96,180,282,502,907,1478,2547,4287 \\
\cline{2-2}
          & 2,10,14,60,84,93,232,426,674,1308,2204,4184,7884,13384,24394,43097 \\
\hline
\mlr{13}  & \begin{mat3} 42162 & 51572 & 64476 \end{mat3} \\
\cline{2-2}
          & 3,4,5,6,7,8,9,10,11,11,11,12,12,13 \dfreesep 24 \\
\cline{2-2}
          & 3,0,8,0,15,0,46,0,163,0,422,0,1264,0,3603,0 \\
\cline{2-2}
          & 7,0,31,0,66,0,255,0,1058,0,3054,0,10282,0,32605,0 \\
\hline
\mlr{14}  & \begin{mat3} 45755 & 55163 & 75271 \end{mat3} \\
\cline{2-2}
          & 3,4,5,6,7,8,9,10,11,11,12,12,12,12,13 \dfreesep 27 \\
\cline{2-2}
          & 3,11,9,13,18,38,52,126,156,308,532,977,1479,2715,4193,7818 \\
\cline{2-2}
          & 9,48,45,62,100,246,330,888,1138,2420,4446,9022,13489,26632,42647,83562 \\
\hline
\mlr{16}  & \begin{mat3} 431516 & 675542 & 731052 \end{mat3} \\
\cline{2-2}
          & 3,4,5,6,7,8,9,10,11,11,12,13,13,13,13,14,15 \dfreesep 28 \\
\cline{2-2}
          & 1,4,1,1,11,26,20,41,94,134,234,374,649,1135,1854,3159 \\
\cline{2-2}
          & 2,12,4,7,50,136,126,281,624,982,1868,3094,5688,10417,17696,31781 \\
\hline
\mlr{19}  & \begin{mat3} 4340472 & 5270746 & 7165336 \end{mat3} \\
\cline{2-2}
          & 3,4,5,6,7,8,9,10,11,11,12,13,13,14,15,15,15,15,16,17 \dfreesep 32 \\
\cline{2-2}
          & 1,2,4,9,9,20,28,43,82,125,253,390,659,1165,1950,3282 \\
\cline{2-2}
          & 6,6,14,45,52,128,182,317,602,1007,2086,3352,6024,11119,19626,34402 \\
\hline
\mlr{21}  & \begin{mat3} 40555364 & 65473174 & 71462344 \end{mat3} \\
\cline{2-2}
          & 3,4,5,6,7,8,9,10,11,11,12,13,13,14,15,15,16,16,16,17,17,17 \dfreesep 35 \\
\cline{2-2}
          & 1,2,6,8,13,16,35,69,93,164,288,471,827,1496,2409,4014 \\
\cline{2-2}
          & 3,8,24,40,71,84,245,464,689,1264,2378,4178,7655,14458,24299,42102 \\
\hline
\mlr{22}  & \begin{mat3} 42105642 & 66074416 & 72756752 \end{mat3} \\
\cline{2-2}
          & 3,4,5,6,7,8,9,10,11,11,12,13,13,14,15,15,16,16,17,17,17,18,18 \dfreesep 36 \\
\cline{2-2}
          & 1,0,12,0,18,0,63,0,155,0,504,0,1438,0,4115,0 \\
\cline{2-2}
          & 3,0,41,0,105,0,436,0,1150,0,4230,0,13192,0,41548,0 \\
\hline
\mlr{24}  & \begin{mat3} 455116344 & 574065364 & 603503174 \end{mat3} \\
\cline{2-2}
          & 3,4,5,6,7,8,9,10,11,11,12,13,13,14,15,15,16,16,17,18,18,18,18,18,19 \dfreesep 39 \\
\cline{2-2}
          & 2,3,3,9,7,18,51,59,102,174,308,509,854,1549,2520,4339 \\
\cline{2-2}
          & 6,10,17,56,43,124,365,444,834,1448,2612,4630,8334,15806,26654,48072 \\
\hline
\mlr{26}  & \begin{mat3} 432166577 & 504132623 & 653567051 \end{mat3} \\
\cline{2-2}
          & 3,4,5,6,7,8,9,10,11,11,12,13,13,14,15,15,16,16,17,18,18,19,19,19,20,20,20 \dfreesep 42 \\
\cline{2-2}
          & 1,2,10,12,13,31,58,70,111,223,415,599,973,1917,3284,5321 \\
\cline{2-2}
          & 8,8,54,64,88,201,414,542,942,1983,3748,5885,9840,20053,35506,60285 \\
\hline
\mlr{29}  & \begin{mat3} 5341100365 & 6555670627 & 7205114353 \end{mat3} \\
\cline{2-2}
          & 3,4,5,6,7,8,9,10,11,11,12,13,13,14,15,15,16,16,17,18,18,19,20,20,20,21,21,21,22,23 \dfreesep 44 \\
\cline{2-2}
          & 1,2,1,3,4,9,13,33,38,65,144,251,386,675,1164,1882 \\
\cline{2-2}
          & 4,8,4,19,24,61,96,277,298,549,1230,2267,3760,6919,12316,20608 \\
\hline
\mlr{31}  & \begin{mat3} 41775322742 & 64230711252 & 70124437316 \end{mat3} \\
\cline{2-2}
          & 3,4,5,6,7,8,9,10,11,11,12,13,13,14,15,15,16,16,17,18,18,19,20,20,21,21,22,22,22,23,23,24 \dfreesep 48 \\
\cline{2-2}
          & 2,2,5,8,9,15,31,63,108,159,291,492,846,1414,2352,4028 \\
\cline{2-2}
          & 10,16,22,40,66,95,216,477,870,1383,2700,4650,8374,14690,25576,45874 \\
\hline
\end{codearray}
$
\end{table*}
\begin{table*}
\caption{Improved $ \text{OBDP}^{(5)} $ codes, $ R = 1/3 $, $ m \geq 9 $}
\centering
$
\begin{codearray}
\hline
\mlr{m}   & G(D) \\
\cline{2-2}
          & \hat{d}_{[0,m]} \dfreesep d_{\infty} \\
\cline{2-2}
          & a_{d_{\infty}},\ldots,a_{d_{\infty}+\dsmax} \\
\cline{2-2}
          & c_{d_{\infty}},\ldots,c_{d_{\infty}+\dsmax} \\
\hline
\hline
\mlr{9}   & \begin{mat3} 4764 & 5134 & 6674 \end{mat3} \\
\cline{2-2}
          & 3,4,5,6,7,7,7,8,9,9 \dfreesep 20 \\
\cline{2-2}
          & 3,10,5,13,20,30,60,85,173,281,475,829,1352,2234,3865,6677 \\
\cline{2-2}
          & 6,32,24,67,110,194,378,589,1272,2233,3932,7175,12422,21404,38742,69627 \\
\hline
\mlr{10}  & \begin{mat3} 4652 & 5662 & 7716 \end{mat3} \\
\cline{2-2}
          & 3,4,5,6,7,8,8,9,9,9,9 \dfreesep 22 \\
\cline{2-2}
          & 7,0,27,0,55,0,161,0,413,0,1200,0,3809,0,10521,0 \\
\cline{2-2}
          & 17,0,122,0,345,0,1102,0,3214,0,10215,0,36004,0,109166,0 \\
\hline
\mlr{11}  & \begin{mat3} 4671 & 6625 & 7537 \end{mat3} \\
\cline{2-2}
          & 3,4,5,6,7,8,9,9,9,10,10,10 \dfreesep 24 \\
\cline{2-2}
          & 16,0,28,0,77,0,211,0,608,0,1824,0,5145,0,15137,0 \\
\cline{2-2}
          & 56,0,133,0,482,0,1480,0,4781,0,15769,0,49606,0,159840,0 \\
\hline
\mlr{12}  & \begin{mat3} 47274 & 51454 & 63364 \end{mat3} \\
\cline{2-2}
          & 3,4,5,6,7,8,9,10,10,11,11,12,12 \dfreesep 24 \\
\cline{2-2}
          & 1,7,9,14,18,21,60,93,142,238,432,750,1225,2129,3626,6120 \\
\cline{2-2}
          & 4,21,36,74,98,117,404,651,1030,1842,3532,6476,11206,20543,36688,64372 \\
\hline
\mlr{15}  & \begin{mat3} 451464 & 556414 & 756174 \end{mat3} \\
\cline{2-2}
          & 3,4,5,6,7,8,9,10,11,11,12,12,12,12,13,14 \dfreesep 28 \\
\cline{2-2}
          & 2,7,6,11,20,33,56,88,170,264,431,779,1326,2175,3600,6462 \\
\cline{2-2}
          & 6,25,32,63,106,203,388,662,1294,2062,3678,6899,12438,21685,37226,69676 \\
\hline
\mlr{17}  & \begin{mat3} 426163 & 664435 & 722757 \end{mat3} \\
\cline{2-2}
          & 3,4,5,6,7,8,9,10,11,11,12,13,13,13,13,13,14,14 \dfreesep 32 \\
\cline{2-2}
          & 19,0,20,0,89,0,230,0,626,0,1872,0,5479,0,15988,0 \\
\cline{2-2}
          & 78,0,128,0,584,0,1867,0,5343,0,17599,0,57024,0,181138,0 \\
\hline
\mlr{18}  & \begin{mat3} 5361654 & 6561724 & 7237134 \end{mat3} \\
\cline{2-2}
          & 3,4,5,6,7,8,9,10,11,11,12,13,13,14,14,14,15,15,16 \dfreesep 31 \\
\cline{2-2}
          & 1,1,7,11,14,27,44,66,81,162,263,455,815,1367,2265,3807 \\
\cline{2-2}
          & 5,2,25,60,92,180,310,474,639,1344,2293,4068,7575,13498,23161,40822 \\
\hline
\mlr{20}  & \begin{mat3} 4610365 & 5460353 & 6233627 \end{mat3} \\
\cline{2-2}
          & 3,4,5,6,7,8,9,10,11,11,12,13,13,14,15,15,15,16,16,17,17 \dfreesep 34 \\
\cline{2-2}
          & 2,0,12,0,20,0,103,0,231,0,667,0,1908,0,5829,0 \\
\cline{2-2}
          & 8,0,57,0,115,0,686,0,1838,0,5777,0,18122,0,60955,0 \\
\hline
\mlr{22}  & \begin{mat3} 41070752 & 52326416 & 66353642 \end{mat3} \\
\cline{2-2}
          & 3,4,5,6,7,8,9,10,11,11,12,13,13,14,15,15,16,16,16,17,17,18,19 \dfreesep 37 \\
\cline{2-2}
          & 2,10,4,7,10,25,69,86,147,245,440,696,1210,2026,3532,6020 \\
\cline{2-2}
          & 4,48,16,40,52,156,469,614,1143,1956,3776,6300,11464,20148,36676,65032 \\
\hline
\mlr{23}  & \begin{mat3} 46652143 & 56623517 & 77044371 \end{mat3} \\
\cline{2-2}
          & 3,4,5,6,7,8,9,10,11,11,12,13,13,14,15,15,16,16,17,17,17,17,18,19 \dfreesep 38 \\
\cline{2-2}
          & 1,2,6,13,13,30,44,58,120,228,388,588,983,1717,3023,5167 \\
\cline{2-2}
          & 4,12,32,71,78,196,308,430,972,1916,3436,5464,9546,17773,32514,57351 \\
\hline
\mlr{29}  & \begin{mat3} 4252213203 & 6666141151 & 7206766575 \end{mat3} \\
\cline{2-2}
          & 3,4,5,6,7,8,9,10,11,11,12,13,13,14,15,15,16,16,17,18,18,19,20,20,20,20,21,21,22,22 \dfreesep 45 \\
\cline{2-2}
          & 1,3,4,6,7,19,26,39,86,136,237,399,631,1166,1880,3281 \\
\cline{2-2}
          & 3,14,24,42,47,134,172,320,678,1160,2191,3778,6301,12294,20618,37478 \\
\hline
\mlr{30}  & \begin{mat3} 41056336124 & 50665003304 & 72432267234 \end{mat3} \\
\cline{2-2}
          & 3,4,5,6,7,8,9,10,11,11,12,13,13,14,15,15,16,16,17,18,18,19,20,20,21,21,21,21,22,22,23 \dfreesep 46 \\
\cline{2-2}
          & 2,0,7,0,15,0,46,0,138,0,391,0,1130,0,3251,0 \\
\cline{2-2}
          & 4,0,33,0,96,0,268,0,1082,0,3358,0,10661,0,33813,0 \\
\hline
\end{codearray}
$
\end{table*}
\begin{table*}
\caption{Improved $ \text{OBDP}^{(6)} $ codes, $ R = 1/3 $, $ m \geq 11 $}
\centering
$
\begin{codearray}
\hline
\mlr{m}   & G(D) \\
\cline{2-2}
          & \hat{d}_{[0,m]} \dfreesep d_{\infty} \\
\cline{2-2}
          & a_{d_{\infty}},\ldots,a_{d_{\infty}+\dsmax} \\
\cline{2-2}
          & c_{d_{\infty}},\ldots,c_{d_{\infty}+\dsmax} \\
\hline
\hline
\mlr{11}  & \begin{mat3} 4713 & 5175 & 6557 \end{mat3} \\
\cline{2-2}
          & 3,4,5,6,7,8,8,9,10,10,11,11 \dfreesep 24 \\
\cline{2-2}
          & 13,0,32,0,78,0,202,0,614,0,1808,0,4971,0,15006,0 \\
\cline{2-2}
          & 43,0,162,0,507,0,1420,0,4857,0,16023,0,48501,0,160167,0 \\
\hline
\mlr{12}  & \begin{mat3} 46354 & 53344 & 77224 \end{mat3} \\
\cline{2-2}
          & 3,4,5,6,7,8,9,9,9,10,11,11,12 \dfreesep 24 \\
\cline{2-2}
          & 1,4,9,20,16,30,53,83,148,224,442,746,1282,2113,3535,6212 \\
\cline{2-2}
          & 2,12,38,92,90,172,336,577,1102,1762,3620,6522,11608,20201,35544,64818 \\
\hline
\mlr{13}  & \begin{mat3} 45572 & 65346 & 77562 \end{mat3} \\
\cline{2-2}
          & 3,4,5,6,7,8,9,10,10,10,11,11,12,12 \dfreesep 26 \\
\cline{2-2}
          & 4,10,11,21,25,39,75,124,235,384,657,1027,1811,3184,5339,9351 \\
\cline{2-2}
          & 10,38,56,107,136,263,522,908,1814,3068,5574,9203,17176,31432,55316,100927 \\
\hline
\mlr{15}  & \begin{mat3} 476524 & 614734 & 745054 \end{mat3} \\
\cline{2-2}
          & 3,4,5,6,7,8,9,10,11,11,11,12,12,12,13,13 \dfreesep 28 \\
\cline{2-2}
          & 1,6,11,18,19,25,54,91,155,266,454,784,1275,2198,3839,6512 \\
\cline{2-2}
          & 4,22,50,98,114,189,372,657,1238,2182,4002,7090,12376,22106,39874,71486 \\
\hline
\mlr{16}  & \begin{mat3} 433672 & 521662 & 713176 \end{mat3} \\
\cline{2-2}
          & 3,4,5,6,7,8,9,10,11,11,12,12,12,13,13,14,15 \dfreesep 30 \\
\cline{2-2}
          & 8,0,25,0,55,0,169,0,404,0,1314,0,3864,0,10890,0 \\
\cline{2-2}
          & 30,0,113,0,337,0,1163,0,3186,0,11275,0,36751,0,113762,0 \\
\hline
\mlr{18}  & \begin{mat3} 4277664 & 6654714 & 7235344 \end{mat3} \\
\cline{2-2}
          & 3,4,5,6,7,8,9,10,11,11,12,13,13,13,13,14,15,15,15 \dfreesep 32 \\
\cline{2-2}
          & 2,3,8,18,17,30,61,84,184,299,457,796,1385,2337,3951,6787 \\
\cline{2-2}
          & 8,11,32,100,82,184,448,588,1496,2481,4050,7214,13446,23531,41710,74793 \\
\hline
\mlr{19}  & \begin{mat3} 4547062 & 5524736 & 7507206 \end{mat3} \\
\cline{2-2}
          & 3,4,5,6,7,8,9,10,11,11,12,13,13,14,14,14,15,15,15,16 \dfreesep 34 \\
\cline{2-2}
          & 6,0,26,0,60,0,149,0,481,0,1331,0,3994,0,11484,0 \\
\cline{2-2}
          & 25,0,137,0,387,0,1077,0,3944,0,11998,0,39988,0,125290,0 \\
\hline
\mlr{21}  & \begin{mat3} 42750214 & 51233774 & 64160664 \end{mat3} \\
\cline{2-2}
          & 3,4,5,6,7,8,9,10,11,11,12,13,13,14,15,15,15,16,16,17,17,18 \dfreesep 36 \\
\cline{2-2}
          & 4,0,18,0,45,0,123,0,326,0,977,0,2883,0,8317,0 \\
\cline{2-2}
          & 9,0,103,0,293,0,853,0,2811,0,8686,0,28541,0,89634,0 \\
\hline
\mlr{23}  & \begin{mat3} 43121251 & 67516157 & 73152555 \end{mat3} \\
\cline{2-2}
          & 3,4,5,6,7,8,9,10,11,11,12,13,13,14,15,15,16,16,16,17,17,18,18,19 \dfreesep 39 \\
\cline{2-2}
          & 2,4,12,20,23,45,65,117,207,346,612,1043,1854,3038,5156,8751 \\
\cline{2-2}
          & 6,26,56,100,157,304,479,898,1729,3126,5704,10118,18768,32190,56784,100850 \\
\hline
\mlr{24}  & \begin{mat3} 462271234 & 607726154 & 754734624 \end{mat3} \\
\cline{2-2}
          & 3,4,5,6,7,8,9,10,11,11,12,13,13,14,15,15,16,16,17,17,17,17,18,19,20 \dfreesep 40 \\
\cline{2-2}
          & 3,0,9,0,42,0,150,0,328,0,1043,0,3041,0,8812,0 \\
\cline{2-2}
          & 10,0,42,0,282,0,1172,0,2835,0,10090,0,32194,0,101498,0 \\
\hline
\mlr{26}  & \begin{mat3} 426222537 & 664662375 & 722527171 \end{mat3} \\
\cline{2-2}
          & 3,4,5,6,7,8,9,10,11,11,12,13,13,14,15,15,16,16,17,18,18,18,19,19,20,20,20 \dfreesep 43 \\
\cline{2-2}
          & 1,8,16,14,28,33,69,134,247,409,632,1047,1833,3165,5394,9170 \\
\cline{2-2}
          & 7,42,78,92,180,224,565,1134,2163,3716,6004,10508,19269,34560,61492,108876 \\
\hline
\mlr{27}  & \begin{mat3} 4037232304 & 5326267124 & 6745544234 \end{mat3} \\
\cline{2-2}
          & 3,4,5,6,7,8,9,10,11,11,12,13,13,14,15,15,16,16,17,18,18,19,19,20,20,20,21,21 \dfreesep 44 \\
\cline{2-2}
          & 2,9,5,14,25,40,71,97,190,292,502,934,1612,2789,4602,7768 \\
\cline{2-2}
          & 8,37,34,82,162,248,540,769,1552,2540,4492,8806,16032,29303,50082,88022 \\
\hline
\mlr{31}  & \begin{mat3} 41464054546 & 50310262676 & 72347375622 \end{mat3} \\
\cline{2-2}
          & 3,4,5,6,7,8,9,10,11,11,12,13,13,14,15,15,16,16,17,18,18,19,20,20,21,21,21,21,21,22,23,24 \dfreesep 49 \\
\cline{2-2}
          & 3,7,12,10,21,35,65,95,183,256,453,864,1357,2438,4085,6828 \\
\cline{2-2}
          & 17,34,60,56,147,236,495,784,1633,2426,4221,8652,14247,26590,46303,80324 \\
\hline
\end{codearray}
$
\end{table*}
\begin{table*}
\caption{Improved $ \text{OBDP}^{(7)} $ codes, $ R = 1/3 $, $ m \geq 13 $}
\centering
$
\begin{codearray}
\hline
\mlr{m}   & G(D) \\
\cline{2-2}
          & \hat{d}_{[0,m]} \dfreesep d_{\infty} \\
\cline{2-2}
          & a_{d_{\infty}},\ldots,a_{d_{\infty}+\dsmax} \\
\cline{2-2}
          & c_{d_{\infty}},\ldots,c_{d_{\infty}+\dsmax} \\
\hline
\hline
\mlr{13}  & \begin{mat3} 46162 & 53132 & 77336 \end{mat3} \\
\cline{2-2}
          & 3,4,5,6,7,8,9,9,9,9,10,10,11,12 \dfreesep 26 \\
\cline{2-2}
          & 2,9,15,19,20,49,78,127,227,352,643,1103,1834,3066,5194,9108 \\
\cline{2-2}
          & 4,33,74,117,108,303,536,949,1780,2966,5710,10129,17762,30866,54594,99854 \\
\hline
\mlr{14}  & \begin{mat3} 43663 & 52457 & 71535 \end{mat3} \\
\cline{2-2}
          & 3,4,5,6,7,8,9,10,10,10,11,12,12,13,13 \dfreesep 28 \\
\cline{2-2}
          & 11,0,33,0,84,0,212,0,627,0,1788,0,5234,0,15204,0 \\
\cline{2-2}
          & 34,0,182,0,541,0,1535,0,5248,0,16413,0,52508,0,166186,0 \\
\hline
\mlr{16}  & \begin{mat3} 423546 & 533762 & 705532 \end{mat3} \\
\cline{2-2}
          & 3,4,5,6,7,8,9,10,11,11,11,12,12,12,13,13,14 \dfreesep 30 \\
\cline{2-2}
          & 2,8,12,18,33,59,66,110,216,401,656,1125,1886,3224,5567,9339 \\
\cline{2-2}
          & 6,30,50,98,204,381,442,838,1682,3333,5774,10255,18318,32678,58694,103219 \\
\hline
\mlr{17}  & \begin{mat3} 423551 & 533655 & 705457 \end{mat3} \\
\cline{2-2}
          & 3,4,5,6,7,8,9,10,11,11,12,12,12,13,13,13,14,15 \dfreesep 32 \\
\cline{2-2}
          & 12,0,38,0,72,0,208,0,676,0,1920,0,5410,0,15988,0 \\
\cline{2-2}
          & 35,0,212,0,466,0,1512,0,5430,0,17303,0,53720,0,173070,0 \\
\hline
\mlr{19}  & \begin{mat3} 4171572 & 5254476 & 6670162 \end{mat3} \\
\cline{2-2}
          & 3,4,5,6,7,8,9,10,11,11,12,13,13,13,13,14,14,15,15,16 \dfreesep 34 \\
\cline{2-2}
          & 2,10,11,16,23,53,93,134,244,402,677,1142,1968,3454,5676,9723 \\
\cline{2-2}
          & 6,42,54,84,138,341,678,974,1952,3366,5870,10652,18996,35122,60068,107667 \\
\hline
\mlr{20}  & \begin{mat3} 4435733 & 5462547 & 7432445 \end{mat3} \\
\cline{2-2}
          & 3,4,5,6,7,8,9,10,11,11,12,13,13,14,14,15,15,15,15,16,16 \dfreesep 36 \\
\cline{2-2}
          & 13,0,39,0,70,0,230,0,702,0,1865,0,5877,0,16427,0 \\
\cline{2-2}
          & 67,0,228,0,485,0,1838,0,6171,0,18080,0,62043,0,188293,0 \\
\hline
\mlr{22}  & \begin{mat3} 45631756 & 62643652 & 77105026 \end{mat3} \\
\cline{2-2}
          & 3,4,5,6,7,8,9,10,11,11,12,13,13,14,15,15,15,16,16,17,17,17,18 \dfreesep 38 \\
\cline{2-2}
          & 4,0,29,0,63,0,170,0,482,0,1429,0,4219,0,11832,0 \\
\cline{2-2}
          & 22,0,148,0,430,0,1331,0,4197,0,13739,0,44577,0,135691,0 \\
\hline
\mlr{24}  & \begin{mat3} 472505474 & 557504744 & 634537614 \end{mat3} \\
\cline{2-2}
          & 3,4,5,6,7,8,9,10,11,11,12,13,13,14,15,15,16,16,16,16,17,18,18,18,18 \dfreesep 41 \\
\cline{2-2}
          & 3,17,19,16,38,52,107,180,300,525,883,1470,2520,4441,7243,12516 \\
\cline{2-2}
          & 11,110,113,88,288,396,841,1462,2676,4878,8577,15094,26896,49608,83467,151088 \\
\hline
\mlr{25}  & \begin{mat3} 433276406 & 505564432 & 652363676 \end{mat3} \\
\cline{2-2}
          & 3,4,5,6,7,8,9,10,11,11,12,13,13,14,15,15,16,16,17,17,17,17,18,19,19,20 \dfreesep 42 \\
\cline{2-2}
          & 3,7,8,20,28,61,93,138,264,459,731,1291,2206,3683,6209,10601 \\
\cline{2-2}
          & 14,27,44,118,186,473,704,1146,2306,4259,7168,12987,23414,40559,71074,126639 \\
\hline
\mlr{27}  & \begin{mat3} 4057646264 & 6543047644 & 7144641274 \end{mat3} \\
\cline{2-2}
          & 3,4,5,6,7,8,9,10,11,11,12,13,13,14,15,15,16,16,17,18,18,18,19,19,19,20,20,21 \dfreesep 45 \\
\cline{2-2}
          & 6,10,19,24,30,56,118,191,324,556,902,1545,2686,4662,7742,13160 \\
\cline{2-2}
          & 28,56,129,156,214,404,994,1588,2796,5264,8878,15892,28742,52396,89660,159418 \\
\hline
\mlr{28}  & \begin{mat3} 4263051036 & 5136327046 & 6406375572 \end{mat3} \\
\cline{2-2}
          & 3,4,5,6,7,8,9,10,11,11,12,13,13,14,15,15,16,16,17,18,18,19,19,19,20,20,20,21,21 \dfreesep 46 \\
\cline{2-2}
          & 1,7,14,21,35,58,96,171,258,434,803,1435,2249,3840,6705,11150 \\
\cline{2-2}
          & 4,39,94,143,248,422,758,1537,2310,4170,8056,14923,24552,43386,79382,137094 \\
\hline
\mlr{30}  & \begin{mat3} 41473244174 & 50304075044 & 72354466714 \end{mat3} \\
\cline{2-2}
          & 3,4,5,6,7,8,9,10,11,11,12,13,13,14,15,15,16,16,17,18,18,19,20,20,20,21,21,22,22,22,23 \dfreesep 47 \\
\cline{2-2}
          & 2,3,8,6,12,27,44,59,117,169,330,600,962,1609,2855,4803 \\
\cline{2-2}
          & 4,18,46,32,70,202,358,432,919,1478,3102,5712,9628,16966,31527,54842 \\
\hline
\end{codearray}
$
\end{table*}
\begin{table*}
\caption{Improved $ \text{OBDP}^{(8)} $ codes, $ R = 1/3 $, $ m \geq 15 $}
\centering
$
\begin{codearray}
\hline
\mlr{m}   & G(D) \\
\cline{2-2}
          & \hat{d}_{[0,m]} \dfreesep d_{\infty} \\
\cline{2-2}
          & a_{d_{\infty}},\ldots,a_{d_{\infty}+\dsmax} \\
\cline{2-2}
          & c_{d_{\infty}},\ldots,c_{d_{\infty}+\dsmax} \\
\hline
\hline
\mlr{15}  & \begin{mat3} 454644 & 515614 & 735374 \end{mat3} \\
\cline{2-2}
          & 3,4,5,6,7,8,9,10,10,11,11,12,12,12,13,13 \dfreesep 30 \\
\cline{2-2}
          & 25,0,31,0,132,0,273,0,940,0,2593,0,7600,0,22069,0 \\
\cline{2-2}
          & 110,0,186,0,872,0,2137,0,8083,0,24624,0,79435,0,249539,0 \\
\hline
\mlr{17}  & \begin{mat3} 437551 & 524655 & 714457 \end{mat3} \\
\cline{2-2}
          & 3,4,5,6,7,8,9,10,11,11,11,12,12,12,13,14,14,15 \dfreesep 32 \\
\cline{2-2}
          & 10,0,29,0,82,0,241,0,677,0,1838,0,5470,0,15872,0 \\
\cline{2-2}
          & 33,0,141,0,514,0,1696,0,5577,0,16468,0,54199,0,170871,0 \\
\hline
\mlr{18}  & \begin{mat3} 4317654 & 6751024 & 7312704 \end{mat3} \\
\cline{2-2}
          & 3,4,5,6,7,8,9,10,11,11,12,12,12,13,13,14,15,15,16 \dfreesep 33 \\
\cline{2-2}
          & 4,14,13,14,38,62,102,142,304,474,823,1358,2311,3988,6555,11499 \\
\cline{2-2}
          & 10,62,85,80,218,420,772,1054,2490,4116,7479,12920,23077,41474,71093,129634 \\
\hline
\mlr{20}  & \begin{mat3} 4357257 & 5022735 & 6571063 \end{mat3} \\
\cline{2-2}
          & 3,4,5,6,7,8,9,10,11,11,12,13,13,13,13,14,15,15,16,16,16 \dfreesep 36 \\
\cline{2-2}
          & 6,19,16,24,47,63,110,192,360,597,955,1722,2848,4868,8431,14248 \\
\cline{2-2}
          & 28,97,92,170,312,457,866,1560,3154,5543,9190,17504,30216,53912,96950,169676 \\
\hline
\mlr{21}  & \begin{mat3} 42672474 & 53715064 & 70331444 \end{mat3} \\
\cline{2-2}
          & 3,4,5,6,7,8,9,10,11,11,12,13,13,14,14,14,15,16,16,16,17,18 \dfreesep 37 \\
\cline{2-2}
          & 3,8,13,27,39,57,91,161,289,511,861,1387,2452,4138,7157,12128 \\
\cline{2-2}
          & 9,48,63,150,237,392,679,1224,2371,4530,7955,13502,24982,43592,78943,139490 \\
\hline
\mlr{23}  & \begin{mat3} 42277535 & 66232457 & 72533663 \end{mat3} \\
\cline{2-2}
          & 3,4,5,6,7,8,9,10,11,11,12,13,13,14,15,15,15,16,16,17,17,17,18,19 \dfreesep 40 \\
\cline{2-2}
          & 6,13,13,38,41,69,147,206,356,586,1030,1757,2999,5094,8710,14897 \\
\cline{2-2}
          & 26,69,68,262,310,533,1238,1830,3300,5706,10534,18609,33396,59004,103788,185091 \\
\hline
\mlr{25}  & \begin{mat3} 520705436 & 644257446 & 733720172 \end{mat3} \\
\cline{2-2}
          & 3,4,5,6,7,8,9,10,11,11,12,13,13,14,15,15,16,16,16,16,17,18,18,18,18,19 \dfreesep 42 \\
\cline{2-2}
          & 2,7,13,23,31,48,89,150,273,446,777,1311,2116,3679,6188,10735 \\
\cline{2-2}
          & 4,35,82,139,234,362,714,1296,2548,4352,7676,13659,23250,41777,72912,131797 \\
\hline
\mlr{26}  & \begin{mat3} 433267157 & 505027703 & 652272231 \end{mat3} \\
\cline{2-2}
          & 3,4,5,6,7,8,9,10,11,11,12,13,13,14,15,15,16,16,17,17,17,18,19,19,19,19,20 \dfreesep 44 \\
\cline{2-2}
          & 5,0,43,0,92,0,244,0,755,0,2161,0,6394,0,18458,0 \\
\cline{2-2}
          & 17,0,245,0,670,0,1948,0,6882,0,21677,0,69920,0,217928,0 \\
\hline
\mlr{29}  & \begin{mat3} 4225313271 & 5167352217 & 6465366243 \end{mat3} \\
\cline{2-2}
          & 3,4,5,6,7,8,9,10,11,11,12,13,13,14,15,15,16,16,17,18,18,19,19,19,19,19,20,21,21,21 \dfreesep 48 \\
\cline{2-2}
          & 7,0,49,0,79,0,277,0,829,0,2234,0,6513,0,19049,0 \\
\cline{2-2}
          & 33,0,321,0,603,0,2281,0,7767,0,23077,0,72815,0,229978,0 \\
\hline
\mlr{31}  & \begin{mat3} 42655051522 & 53764207572 & 70362150606 \end{mat3} \\
\cline{2-2}
          & 3,4,5,6,7,8,9,10,11,11,12,13,13,14,15,15,16,16,17,18,18,19,20,20,20,21,21,21,22,22,23,23 \dfreesep 50 \\
\cline{2-2}
          & 5,0,22,0,80,0,192,0,595,0,1662,0,4746,0,13760,0 \\
\cline{2-2}
          & 22,0,133,0,593,0,1654,0,5523,0,16750,0,52527,0,164589,0 \\
\hline
\end{codearray}
$
\end{table*}
\begin{table*}
\tcfirst
\caption{OBCDF codes, $ R = 1/4 $}
\centering
$
\begin{codearray}
\hline
\mlr{m}   & G(D) \\
\cline{2-2}
          & \hat{d}_{[0,m]} \dfreesep d_{\infty} \\
\cline{2-2}
          & a_{d_{\infty}},\ldots,a_{d_{\infty}+\dsmax} \\
\cline{2-2}
          & c_{d_{\infty}},\ldots,c_{d_{\infty}+\dsmax} \\
\hline
\hline
\mlr{1}   & \begin{mat4} 2 & 4 & 6 & 6 \end{mat4} \\
\cline{2-2}
          & 3,5 \dfreesep 6 \\
\cline{2-2}
          & 1,0,1,0,1,0,1,0,1,0,1,0,1,0,1,0 \\
\cline{2-2}
          & 1,0,2,0,3,0,4,0,5,0,6,0,7,0,8,0 \\
\hline
\mlr{2}   & \begin{mat4} 5 & 5 & 7 & 7 \end{mat4} \\
\cline{2-2}
          & 4,6,6 \dfreesep 10 \\
\cline{2-2}
          & 1,0,2,0,4,0,8,0,16,0,32,0,64,0,128,0 \\
\cline{2-2}
          & 1,0,4,0,12,0,32,0,80,0,192,0,448,0,1024,0 \\
\hline
\mlr{3}   & \begin{mat4} 44 & 54 & 64 & 74 \end{mat4} \\
\cline{2-2}
          & 4,6,8,8 \dfreesep 12 \\
\cline{2-2}
          & 1,0,1,0,3,0,5,0,11,0,21,0,43,0,85,0 \\
\cline{2-2}
          & 1,0,2,0,7,0,16,0,41,0,94,0,219,0,492,0 \\
\hline
\mlr{4}   & \begin{mat4} 46 & 56 & 62 & 72 \end{mat4} \\
\cline{2-2}
          & 4,6,8,8,10 \dfreesep 14 \\
\cline{2-2}
          & 1,0,1,0,3,0,7,0,9,0,25,0,57,0,100,0 \\
\cline{2-2}
          & 1,0,2,0,9,0,22,0,37,0,116,0,291,0,602,0 \\
\hline
\mlr{5}   & \begin{mat4} 45 & 51 & 67 & 73 \end{mat4} \\
\cline{2-2}
          & 4,6,8,8,10,10 \dfreesep 16 \\
\cline{2-2}
          & 1,0,0,0,8,0,0,0,21,0,0,0,122,0,0,0 \\
\cline{2-2}
          & 1,0,0,0,24,0,0,0,91,0,0,0,669,0,0,0 \\
\hline
\mlr{6}   & \begin{mat4} 434 & 564 & 614 & 704 \end{mat4} \\
\cline{2-2}
          & 4,6,8,9,10,11,13 \dfreesep 17 \\
\cline{2-2}
          & 1,1,0,0,1,2,0,4,7,4,11,13,21,30,39,68 \\
\cline{2-2}
          & 1,2,0,0,3,8,0,12,23,14,47,56,105,178,213,396 \\
\hline
\mlr{7}   & \begin{mat4} 406 & 536 & 602 & 752 \end{mat4} \\
\cline{2-2}
          & 4,6,8,9,10,11,12,14 \dfreesep 18 \\
\cline{2-2}
          & 1,0,2,0,1,0,1,0,7,0,17,0,31,0,75,0 \\
\cline{2-2}
          & 1,0,4,0,3,0,4,0,23,0,66,0,129,0,368,0 \\
\hline
\mlr{8}   & \begin{mat4} 471 & 525 & 603 & 727 \end{mat4} \\
\cline{2-2}
          & 4,6,8,9,10,11,13,14,15 \dfreesep 21 \\
\cline{2-2}
          & 1,0,0,2,3,3,0,1,5,11,15,13,29,41,46,68 \\
\cline{2-2}
          & 1,0,0,6,9,8,0,8,27,50,71,66,155,226,274,448 \\
\hline
\mlr{9}   & \begin{mat4} 4314 & 5704 & 6174 & 7024 \end{mat4} \\
\cline{2-2}
          & 4,6,8,9,10,11,13,14,15,16 \dfreesep 22 \\
\cline{2-2}
          & 1,1,0,0,0,0,0,4,3,5,9,17,26,25,43,57 \\
\cline{2-2}
          & 2,1,0,0,0,0,0,14,10,19,36,87,146,129,236,327 \\
\hline
\mlr{10}  & \begin{mat4} 4102 & 5756 & 6106 & 7372 \end{mat4} \\
\cline{2-2}
          & 4,6,8,9,10,11,13,14,15,16,17 \dfreesep 24 \\
\cline{2-2}
          & 1,0,1,0,3,0,3,0,8,0,24,0,44,0,87,0 \\
\cline{2-2}
          & 2,0,1,0,6,0,9,0,32,0,98,0,233,0,485,0 \\
\hline
\mlr{11}  & \begin{mat4} 4633 & 5647 & 6631 & 7135 \end{mat4} \\
\cline{2-2}
          & 4,6,8,9,10,11,13,14,15,16,17,18 \dfreesep 30 \\
\cline{2-2}
          & 4,0,7,0,7,0,27,0,33,0,80,0,197,0,387,0 \\
\cline{2-2}
          & 10,0,22,0,33,0,138,0,169,0,508,0,1329,0,2864,0 \\
\hline
\mlr{12}  & \begin{mat4} 41204 & 52524 & 62074 & 74114 \end{mat4} \\
\cline{2-2}
          & 4,6,8,9,10,11,13,14,15,17,17,17,18 \dfreesep 25 \\
\cline{2-2}
          & 1,0,0,1,0,0,0,3,0,5,1,7,13,9,11,17 \\
\cline{2-2}
          & 1,0,0,2,0,0,0,12,0,22,5,28,45,38,53,88 \\
\hline
\mlr{13}  & \begin{mat4} 47516 & 57666 & 66772 & 71362 \end{mat4} \\
\cline{2-2}
          & 4,6,8,9,10,11,13,14,15,17,17,18,18,18 \dfreesep 32 \\
\cline{2-2}
          & 2,0,0,0,14,0,0,0,46,0,0,0,162,0,0,0 \\
\cline{2-2}
          & 4,0,0,0,55,0,0,0,255,0,0,0,1007,0,0,0 \\
\hline
\end{codearray}
$
\end{table*}
\begin{table*}
\tcnext
\caption{OBCDF codes, $ R = 1/4 $}
\centering
$
\begin{codearray}
\hline
\mlr{m}   & G(D) \\
\cline{2-2}
          & \hat{d}_{[0,m]} \dfreesep d_{\infty} \\
\cline{2-2}
          & a_{d_{\infty}},\ldots,a_{d_{\infty}+\dsmax} \\
\cline{2-2}
          & c_{d_{\infty}},\ldots,c_{d_{\infty}+\dsmax} \\
\hline
\hline
\mlr{14}  & \begin{mat4} 41057 & 52225 & 60503 & 75041 \end{mat4} \\
\cline{2-2}
          & 4,6,8,9,10,11,13,14,15,17,18,19,20,20,20 \dfreesep 27 \\
\cline{2-2}
          & 1,0,0,0,0,1,0,0,0,0,0,4,4,5,12,15 \\
\cline{2-2}
          & 1,0,0,0,0,2,0,0,0,0,0,14,16,20,52,74 \\
\hline
\mlr{15}  & \begin{mat4} 435314 & 503024 & 632704 & 760174 \end{mat4} \\
\cline{2-2}
          & 4,6,8,9,10,11,13,14,15,17,17,18,19,20,21,22 \dfreesep 34 \\
\cline{2-2}
          & 2,0,0,0,2,0,4,0,14,0,35,0,59,0,131,0 \\
\cline{2-2}
          & 3,0,0,0,7,0,14,0,75,0,176,0,328,0,818,0 \\
\hline
\mlr{16}  & \begin{mat4} 467516 & 545232 & 661066 & 713662 \end{mat4} \\
\cline{2-2}
          & 4,6,8,9,10,11,13,14,15,17,18,19,19,20,21,23,23 \dfreesep 38 \\
\cline{2-2}
          & 2,0,3,0,7,0,9,0,25,0,63,0,121,0,292,0 \\
\cline{2-2}
          & 8,0,6,0,34,0,43,0,117,0,379,0,779,0,2109,0 \\
\hline
\mlr{17}  & \begin{mat4} 442753 & 564627 & 657211 & 723135 \end{mat4} \\
\cline{2-2}
          & 4,6,8,9,10,11,13,14,15,17,18,18,20,20,21,21,22,23 \dfreesep 40 \\
\cline{2-2}
          & 4,0,5,0,10,0,16,0,26,0,85,0,136,0,271,0 \\
\cline{2-2}
          & 20,0,13,0,54,0,76,0,128,0,565,0,918,0,1891,0 \\
\hline
\mlr{18}  & \begin{mat4} 4665544 & 5440464 & 6604154 & 7121234 \end{mat4} \\
\cline{2-2}
          & 4,6,8,9,10,11,13,14,15,17,18,19,20,20,21,22,23,23,25 \dfreesep 38 \\
\cline{2-2}
          & 1,0,1,0,2,0,4,0,2,0,19,0,38,0,68,0 \\
\cline{2-2}
          & 1,0,2,0,6,0,16,0,9,0,100,0,216,0,423,0 \\
\hline
\mlr{19}  & \begin{mat4} 4733366 & 5746156 & 6755562 & 7306372 \end{mat4} \\
\cline{2-2}
          & 4,6,8,9,10,11,13,14,15,17,18,19,20,20,21,22,23,23,24,25 \dfreesep 38 \\
\cline{2-2}
          & 1,0,0,0,0,0,2,0,3,0,10,0,23,0,35,0 \\
\cline{2-2}
          & 3,0,0,0,0,0,4,0,15,0,48,0,119,0,194,0 \\
\hline
\mlr{20}  & \begin{mat4} 4502051 & 5535655 & 6465507 & 7055313 \end{mat4} \\
\cline{2-2}
          & 4,6,8,9,10,11,13,14,15,17,18,19,20,20,21,22,23,24,25,26,27 \dfreesep 45 \\
\cline{2-2}
          & 2,3,1,1,8,3,2,10,13,18,20,43,59,95,127,177 \\
\cline{2-2}
          & 4,10,7,2,40,10,20,56,65,134,136,272,393,700,933,1472 \\
\hline
\mlr{21}  & \begin{mat4} 41256354 & 52575164 & 62006044 & 74145334 \end{mat4} \\
\cline{2-2}
          & 4,6,8,9,10,11,13,14,15,17,18,19,20,20,21,22,23,24,25,26,27,28 \dfreesep 44 \\
\cline{2-2}
          & 1,0,1,0,6,0,2,0,8,0,13,0,36,0,72,0 \\
\cline{2-2}
          & 3,0,1,0,26,0,7,0,33,0,66,0,199,0,463,0 \\
\hline
\mlr{22}  & \begin{mat4} 40153002 & 51135112 & 63027276 & 76564146 \end{mat4} \\
\cline{2-2}
          & 4,6,8,9,10,11,13,14,15,17,18,19,20,20,21,23,23,24,25,26,27,28,29 \dfreesep 46 \\
\cline{2-2}
          & 1,0,1,0,3,0,9,0,6,0,17,0,45,0,95,0 \\
\cline{2-2}
          & 1,0,6,0,12,0,29,0,33,0,76,0,244,0,574,0 \\
\hline
\mlr{23}  & \begin{mat4} 47015547 & 57277275 & 66036033 & 71554071 \end{mat4} \\
\cline{2-2}
          & 4,6,8,9,10,11,13,14,15,17,18,19,20,20,21,23,23,24,25,26,27,28,28,28 \dfreesep 44 \\
\cline{2-2}
          & 1,0,0,0,1,0,0,0,3,0,8,0,15,0,26,0 \\
\cline{2-2}
          & 3,0,0,0,2,0,0,0,20,0,38,0,70,0,158,0 \\
\hline
\mlr{24}  & \begin{mat4} 433163304 & 525446524 & 616735614 & 746161474 \end{mat4} \\
\cline{2-2}
          & 4,6,8,9,10,11,13,14,15,17,18,19,20,20,21,23,24,25,25,26,26,27,28,30,30 \dfreesep 48 \\
\cline{2-2}
          & 1,0,0,0,1,1,1,3,3,6,5,3,8,13,18,41 \\
\cline{2-2}
          & 2,0,0,0,4,3,4,19,22,26,20,15,52,59,92,255 \\
\hline
\mlr{25}  & \begin{mat4} 422044512 & 512220442 & 605716076 & 760717206 \end{mat4} \\
\cline{2-2}
          & 4,6,8,9,10,11,13,14,15,17,18,19,20,20,21,23,24,25,25,26,27,27,28,28,30,32 \dfreesep 48 \\
\cline{2-2}
          & 1,0,0,0,2,0,0,0,7,0,0,0,18,0,0,0 \\
\cline{2-2}
          & 1,0,0,0,8,0,0,0,32,0,0,0,91,0,0,0 \\
\hline
\mlr{26}  & \begin{mat4} 404671717 & 510050045 & 621451423 & 747473101 \end{mat4} \\
\cline{2-2}
          & 4,6,8,9,10,11,13,14,15,17,18,19,20,20,21,23,24,25,25,27,27,28,28,29,30,31,32 \dfreesep 50 \\
\cline{2-2}
          & 1,0,0,0,1,0,4,0,4,0,7,0,10,0,19,0 \\
\cline{2-2}
          & 1,0,0,0,3,0,22,0,18,0,26,0,54,0,115,0 \\
\hline
\end{codearray}
$
\end{table*}
\begin{table*}
\tcnext
\caption{OBCDF codes, $ R = 1/4 $}
\centering
$
\begin{codearray}
\hline
\mlr{m}   & G(D) \\
\cline{2-2}
          & \hat{d}_{[0,m]} \dfreesep d_{\infty} \\
\cline{2-2}
          & a_{d_{\infty}},\ldots,a_{d_{\infty}+\dsmax} \\
\cline{2-2}
          & c_{d_{\infty}},\ldots,c_{d_{\infty}+\dsmax} \\
\hline
\hline
\mlr{27}  & \begin{mat4} 4666466064 & 5415455544 & 6454006454 & 7011462034 \end{mat4} \\
\cline{2-2}
          & 4,6,8,9,10,11,13,14,15,17,18,19,20,20,21,23,24,25,25,26,27,28,29,30,30,31,32,33 \dfreesep 50 \\
\cline{2-2}
          & 2,0,0,0,1,0,2,0,3,0,3,0,4,0,30,0 \\
\cline{2-2}
          & 5,0,0,0,1,0,10,0,16,0,12,0,15,0,186,0 \\
\hline
\end{codearray}
$
\end{table*}
\begin{table*}
\caption{Improved $ \text{OBDP}^{(0)} $ codes, $ R = 1/4 $}
\centering
$
\begin{codearray}
\hline
\mlr{m}   & G(D) \\
\cline{2-2}
          & \hat{d}_{[0,m]} \dfreesep d_{\infty} \\
\cline{2-2}
          & a_{d_{\infty}},\ldots,a_{d_{\infty}+\dsmax} \\
\cline{2-2}
          & c_{d_{\infty}},\ldots,c_{d_{\infty}+\dsmax} \\
\hline
\hline
\mlr{7}   & \begin{mat4} 446 & 516 & 622 & 712 \end{mat4} \\
\cline{2-2}
          & 4,6,8,9,10,11,12,14 \dfreesep 18 \\
\cline{2-2}
          & 1,0,0,0,2,0,4,0,3,0,15,0,29,0,75,0 \\
\cline{2-2}
          & 1,0,0,0,6,0,12,0,15,0,56,0,135,0,430,0 \\
\hline
\mlr{8}   & \begin{mat4} 451 & 525 & 623 & 727 \end{mat4} \\
\cline{2-2}
          & 4,6,8,9,10,11,13,14,15 \dfreesep 21 \\
\cline{2-2}
          & 1,0,0,2,2,2,1,1,3,11,15,20,20,30,52,68 \\
\cline{2-2}
          & 1,0,0,6,6,6,5,4,13,46,71,94,84,170,304,474 \\
\hline
\mlr{10}  & \begin{mat4} 4422 & 5766 & 6772 & 7356 \end{mat4} \\
\cline{2-2}
          & 4,6,8,9,10,11,13,14,15,16,17 \dfreesep 26 \\
\cline{2-2}
          & 2,0,2,0,0,12,3,0,10,12,24,10,35,86,98,128 \\
\cline{2-2}
          & 4,0,4,0,0,50,12,0,52,52,120,66,214,460,660,974 \\
\hline
\mlr{11}  & \begin{mat4} 4427 & 5633 & 6551 & 7375 \end{mat4} \\
\cline{2-2}
          & 4,6,8,9,10,11,13,14,15,16,17,18 \dfreesep 30 \\
\cline{2-2}
          & 2,4,4,5,2,7,7,6,20,33,39,73,104,134,215,293 \\
\cline{2-2}
          & 6,16,14,17,10,31,26,30,112,197,234,491,690,916,1606,2277 \\
\hline
\mlr{21}  & \begin{mat4} 44736564 & 56573444 & 67431734 & 73630754 \end{mat4} \\
\cline{2-2}
          & 4,6,8,9,10,11,13,14,15,17,18,19,20,20,21,22,23,24,25,26,27,28 \dfreesep 46 \\
\cline{2-2}
          & 3,0,1,0,4,0,8,0,17,0,41,0,80,0,212,0 \\
\cline{2-2}
          & 11,0,2,0,16,0,40,0,101,0,250,0,516,0,1456,0 \\
\hline
\end{codearray}
$
\end{table*}
\begin{table*}
\caption{Improved $ \text{OBDP}^{(1)} $ codes, $ R = 1/4 $}
\centering
$
\begin{codearray}
\hline
\mlr{m}   & G(D) \\
\cline{2-2}
          & \hat{d}_{[0,m]} \dfreesep d_{\infty} \\
\cline{2-2}
          & a_{d_{\infty}},\ldots,a_{d_{\infty}+\dsmax} \\
\cline{2-2}
          & c_{d_{\infty}},\ldots,c_{d_{\infty}+\dsmax} \\
\hline
\hline
\mlr{1}   & \begin{mat4} 2 & 6 & 6 & 6 \end{mat4} \\
\cline{2-2}
          & 3,4 \dfreesep 7 \\
\cline{2-2}
          & 1,1,1,1,1,1,1,1,1,1,1,1,1,1,1,1 \\
\cline{2-2}
          & 1,2,3,4,5,6,7,8,9,10,11,12,13,14,15,16 \\
\hline
\mlr{6}   & \begin{mat4} 444 & 564 & 654 & 734 \end{mat4} \\
\cline{2-2}
          & 4,6,8,9,10,11,11 \dfreesep 19 \\
\cline{2-2}
          & 1,0,3,4,0,3,4,5,9,21,23,20,57,73,71,164 \\
\cline{2-2}
          & 1,0,9,12,0,12,12,26,49,96,141,120,333,472,465,1132 \\
\hline
\mlr{7}   & \begin{mat4} 466 & 562 & 636 & 752 \end{mat4} \\
\cline{2-2}
          & 4,6,8,9,10,11,12,13 \dfreesep 22 \\
\cline{2-2}
          & 2,0,4,0,10,0,13,0,36,0,64,0,150,0,293,0 \\
\cline{2-2}
          & 3,0,13,0,35,0,57,0,196,0,377,0,985,0,2065,0 \\
\hline
\mlr{8}   & \begin{mat4} 471 & 525 & 677 & 773 \end{mat4} \\
\cline{2-2}
          & 4,6,8,9,10,11,13,14,14 \dfreesep 22 \\
\cline{2-2}
          & 1,0,3,0,2,0,9,0,11,0,37,0,76,0,151,0 \\
\cline{2-2}
          & 2,0,10,0,3,0,32,0,50,0,181,0,449,0,916,0 \\
\hline
\mlr{10}  & \begin{mat4} 4116 & 5262 & 6266 & 7572 \end{mat4} \\
\cline{2-2}
          & 4,6,8,9,10,11,13,14,15,16,16 \dfreesep 27 \\
\cline{2-2}
          & 1,0,3,4,2,6,5,7,14,20,27,34,57,92,135,213 \\
\cline{2-2}
          & 1,0,9,12,10,24,19,30,80,102,141,196,351,582,901,1522 \\
\hline
\mlr{18}  & \begin{mat4} 4576664 & 5557644 & 6425054 & 7004034 \end{mat4} \\
\cline{2-2}
          & 4,6,8,9,10,11,13,14,15,17,18,19,20,20,21,22,23,23,23 \dfreesep 40 \\
\cline{2-2}
          & 2,0,2,0,6,0,5,0,8,0,24,0,85,0,132,0 \\
\cline{2-2}
          & 4,0,4,0,26,0,22,0,44,0,139,0,553,0,842,0 \\
\hline
\mlr{20}  & \begin{mat4} 4076427 & 5363235 & 6155333 & 7422331 \end{mat4} \\
\cline{2-2}
          & 4,6,8,9,10,11,13,14,15,17,18,19,20,20,21,22,23,24,25,26,26 \dfreesep 46 \\
\cline{2-2}
          & 1,0,4,0,7,0,19,0,48,0,80,0,160,0,352,0 \\
\cline{2-2}
          & 3,0,10,0,27,0,102,0,283,0,507,0,1060,0,2536,0 \\
\hline
\mlr{21}  & \begin{mat4} 40236644 & 53771134 & 63422764 & 75637154 \end{mat4} \\
\cline{2-2}
          & 4,6,8,9,10,11,13,14,15,17,18,19,20,20,21,22,23,24,25,26,27,27 \dfreesep 46 \\
\cline{2-2}
          & 1,0,2,0,1,0,8,0,17,0,38,0,95,0,167,0 \\
\cline{2-2}
          & 2,0,5,0,7,0,34,0,80,0,196,0,602,0,1154,0 \\
\hline
\end{codearray}
$
\end{table*}
\begin{table*}
\caption{Improved $ \text{OBDP}^{(2)} $ codes, $ R = 1/4 $, $ m \geq 3 $}
\centering
$
\begin{codearray}
\hline
\mlr{m}   & G(D) \\
\cline{2-2}
          & \hat{d}_{[0,m]} \dfreesep d_{\infty} \\
\cline{2-2}
          & a_{d_{\infty}},\ldots,a_{d_{\infty}+\dsmax} \\
\cline{2-2}
          & c_{d_{\infty}},\ldots,c_{d_{\infty}+\dsmax} \\
\hline
\hline
\mlr{8}   & \begin{mat4} 455 & 571 & 647 & 733 \end{mat4} \\
\cline{2-2}
          & 4,6,8,9,10,11,13,13,13 \dfreesep 24 \\
\cline{2-2}
          & 3,0,3,0,9,0,15,0,36,0,75,0,154,0,318,0 \\
\cline{2-2}
          & 6,0,8,0,39,0,68,0,187,0,441,0,1016,0,2293,0 \\
\hline
\mlr{9}   & \begin{mat4} 4744 & 5564 & 6374 & 7514 \end{mat4} \\
\cline{2-2}
          & 4,6,8,9,10,11,13,14,14,15 \dfreesep 24 \\
\cline{2-2}
          & 1,0,1,0,3,0,12,0,15,0,34,0,78,0,166,0 \\
\cline{2-2}
          & 2,0,3,0,7,0,43,0,69,0,176,0,458,0,1093,0 \\
\hline
\mlr{10}  & \begin{mat4} 4546 & 5532 & 6162 & 7736 \end{mat4} \\
\cline{2-2}
          & 4,6,8,9,10,11,13,14,15,15,16 \dfreesep 28 \\
\cline{2-2}
          & 1,4,3,4,4,5,6,8,20,32,45,65,91,125,161,283 \\
\cline{2-2}
          & 2,12,8,20,14,23,28,40,104,182,272,413,592,877,1178,2111 \\
\hline
\mlr{11}  & \begin{mat4} 4563 & 5537 & 6531 & 7235 \end{mat4} \\
\cline{2-2}
          & 4,6,8,9,10,11,13,14,15,16,16,17 \dfreesep 31 \\
\cline{2-2}
          & 4,4,5,4,6,8,12,25,25,42,61,99,147,211,284,431 \\
\cline{2-2}
          & 10,14,21,12,30,42,60,132,143,258,389,650,991,1520,2192,3438 \\
\hline
\mlr{15}  & \begin{mat4} 443644 & 560034 & 672664 & 731054 \end{mat4} \\
\cline{2-2}
          & 4,6,8,9,10,11,13,14,15,17,17,18,19,20,20,21 \dfreesep 35 \\
\cline{2-2}
          & 2,0,1,4,2,3,1,5,9,9,16,31,41,46,81,126 \\
\cline{2-2}
          & 4,0,3,12,6,10,5,26,53,50,88,186,251,284,535,920 \\
\hline
\mlr{21}  & \begin{mat4} 43553134 & 50374764 & 63210644 & 76045154 \end{mat4} \\
\cline{2-2}
          & 4,6,8,9,10,11,13,14,15,17,18,19,20,20,21,22,23,24,25,26,26,27 \dfreesep 49 \\
\cline{2-2}
          & 2,1,0,6,10,12,5,22,31,44,71,87,139,194,284,396 \\
\cline{2-2}
          & 4,2,0,32,42,68,21,120,201,286,467,616,1073,1476,2264,3274 \\
\hline
\end{codearray}
$
\end{table*}
\begin{table*}
\caption{Improved $ \text{OBDP}^{(3)} $ codes, $ R = 1/4 $, $ m \geq 5 $}
\centering
$
\begin{codearray}
\hline
\mlr{m}   & G(D) \\
\cline{2-2}
          & \hat{d}_{[0,m]} \dfreesep d_{\infty} \\
\cline{2-2}
          & a_{d_{\infty}},\ldots,a_{d_{\infty}+\dsmax} \\
\cline{2-2}
          & c_{d_{\infty}},\ldots,c_{d_{\infty}+\dsmax} \\
\hline
\hline
\mlr{6}   & \begin{mat4} 454 & 534 & 664 & 744 \end{mat4} \\
\cline{2-2}
          & 4,6,8,9,9,10,11 \dfreesep 19 \\
\cline{2-2}
          & 1,0,2,3,0,5,7,6,14,18,15,26,53,57,97,172 \\
\cline{2-2}
          & 1,0,6,6,0,20,29,36,74,88,89,152,325,364,655,1228 \\
\hline
\mlr{8}   & \begin{mat4} 457 & 565 & 633 & 771 \end{mat4} \\
\cline{2-2}
          & 4,6,8,9,10,11,12,13,14 \dfreesep 24 \\
\cline{2-2}
          & 1,2,3,3,2,9,6,12,20,20,42,55,87,106,145,274 \\
\cline{2-2}
          & 2,4,8,11,10,39,26,58,100,112,254,347,568,728,1058,2070 \\
\hline
\mlr{9}   & \begin{mat4} 4554 & 5534 & 6164 & 7644 \end{mat4} \\
\cline{2-2}
          & 4,6,8,9,10,11,13,13,14,15 \dfreesep 26 \\
\cline{2-2}
          & 3,0,4,0,10,0,16,0,28,0,89,0,163,0,376,0 \\
\cline{2-2}
          & 8,0,9,0,42,0,75,0,145,0,517,0,1073,0,2635,0 \\
\hline
\mlr{10}  & \begin{mat4} 4532 & 5656 & 6326 & 7722 \end{mat4} \\
\cline{2-2}
          & 4,6,8,9,10,11,13,14,14,15,16 \dfreesep 28 \\
\cline{2-2}
          & 1,2,4,5,2,5,10,17,15,31,47,46,102,115,200,299 \\
\cline{2-2}
          & 2,4,14,15,8,29,54,83,74,169,268,294,674,775,1448,2243 \\
\hline
\mlr{15}  & \begin{mat4} 435424 & 506704 & 615374 & 772614 \end{mat4} \\
\cline{2-2}
          & 4,6,8,9,10,11,13,14,15,17,17,18,19,19,21,21 \dfreesep 36 \\
\cline{2-2}
          & 1,0,3,0,9,0,12,0,25,0,54,0,103,0,256,0 \\
\cline{2-2}
          & 2,0,9,0,36,0,44,0,146,0,322,0,630,0,1802,0 \\
\hline
\end{codearray}
$
\end{table*}
\begin{table*}
\caption{Improved $ \text{OBDP}^{(4)} $ codes, $ R = 1/4 $, $ m \geq 7 $}
\centering
$
\begin{codearray}
\hline
\mlr{m}   & G(D) \\
\cline{2-2}
          & \hat{d}_{[0,m]} \dfreesep d_{\infty} \\
\cline{2-2}
          & a_{d_{\infty}},\ldots,a_{d_{\infty}+\dsmax} \\
\cline{2-2}
          & c_{d_{\infty}},\ldots,c_{d_{\infty}+\dsmax} \\
\hline
\hline
\mlr{8}   & \begin{mat4} 455 & 523 & 671 & 757 \end{mat4} \\
\cline{2-2}
          & 4,6,8,9,10,10,11,12,12 \dfreesep 24 \\
\cline{2-2}
          & 1,0,8,0,7,0,16,0,37,0,81,0,158,0,336,0 \\
\cline{2-2}
          & 1,0,26,0,30,0,72,0,217,0,483,0,1039,0,2482,0 \\
\hline
\mlr{9}   & \begin{mat4} 4624 & 5344 & 6654 & 7574 \end{mat4} \\
\cline{2-2}
          & 4,6,8,9,10,11,11,12,13,14 \dfreesep 27 \\
\cline{2-2}
          & 3,4,1,2,9,10,18,23,19,39,54,81,137,185,258,406 \\
\cline{2-2}
          & 7,10,5,8,39,44,96,124,115,242,344,534,937,1406,1998,3272 \\
\hline
\mlr{10}  & \begin{mat4} 4272 & 5176 & 6662 & 7246 \end{mat4} \\
\cline{2-2}
          & 4,6,8,9,10,11,13,13,14,14,15 \dfreesep 28 \\
\cline{2-2}
          & 1,0,8,0,8,0,16,0,46,0,87,0,179,0,366,0 \\
\cline{2-2}
          & 1,0,22,0,33,0,78,0,258,0,524,0,1207,0,2611,0 \\
\hline
\mlr{11}  & \begin{mat4} 4555 & 5247 & 6371 & 7273 \end{mat4} \\
\cline{2-2}
          & 4,6,8,9,10,11,13,14,14,15,16,16 \dfreesep 31 \\
\cline{2-2}
          & 4,3,3,4,8,13,18,22,25,50,57,86,121,212,336,441 \\
\cline{2-2}
          & 10,10,13,12,38,62,102,112,139,296,357,590,871,1580,2532,3482 \\
\hline
\mlr{12}  & \begin{mat4} 45264 & 57634 & 64554 & 72344 \end{mat4} \\
\cline{2-2}
          & 4,6,8,9,10,11,13,14,15,15,15,16,17 \dfreesep 33 \\
\cline{2-2}
          & 3,7,4,2,9,9,12,22,40,61,70,90,144,215,321,480 \\
\cline{2-2}
          & 7,24,16,8,43,46,58,136,242,380,466,586,1012,1580,2447,3868 \\
\hline
\mlr{19}  & \begin{mat4} 4153362 & 5234356 & 6243572 & 7444166 \end{mat4} \\
\cline{2-2}
          & 4,6,8,9,10,11,13,14,15,17,18,19,20,20,21,22,22,24,24,25 \dfreesep 44 \\
\cline{2-2}
          & 1,0,4,0,13,0,16,0,30,0,75,0,153,0,314,0 \\
\cline{2-2}
          & 3,0,11,0,58,0,79,0,175,0,449,0,1072,0,2306,0 \\
\hline
\mlr{20}  & \begin{mat4} 4533731 & 5502433 & 6441647 & 7074515 \end{mat4} \\
\cline{2-2}
          & 4,6,8,9,10,11,13,14,15,17,18,19,20,20,21,22,23,23,24,25,26 \dfreesep 46 \\
\cline{2-2}
          & 1,0,1,0,11,0,22,0,42,0,77,0,161,0,369,0 \\
\cline{2-2}
          & 1,0,3,0,52,0,117,0,247,0,485,0,1157,0,2801,0 \\
\hline
\mlr{22}  & \begin{mat4} 47153556 & 57650766 & 67612772 & 73353162 \end{mat4} \\
\cline{2-2}
          & 4,6,8,9,10,11,13,14,15,17,18,19,20,20,21,23,23,24,25,25,27,27,28 \dfreesep 50 \\
\cline{2-2}
          & 1,0,8,0,16,0,22,0,46,0,90,0,184,0,404,0 \\
\cline{2-2}
          & 3,0,32,0,76,0,126,0,304,0,618,0,1322,0,3132,0 \\
\hline
\end{codearray}
$
\end{table*}
\begin{table*}
\caption{Improved $ \text{OBDP}^{(5)} $ codes, $ R = 1/4 $, $ m \geq 9 $}
\centering
$
\begin{codearray}
\hline
\mlr{m}   & G(D) \\
\cline{2-2}
          & \hat{d}_{[0,m]} \dfreesep d_{\infty} \\
\cline{2-2}
          & a_{d_{\infty}},\ldots,a_{d_{\infty}+\dsmax} \\
\cline{2-2}
          & c_{d_{\infty}},\ldots,c_{d_{\infty}+\dsmax} \\
\hline
\hline
\mlr{10}  & \begin{mat4} 4566 & 5162 & 6552 & 7636 \end{mat4} \\
\cline{2-2}
          & 4,6,8,9,10,11,11,12,13,13,14 \dfreesep 29 \\
\cline{2-2}
          & 3,3,2,5,10,10,13,29,27,41,60,78,145,198,287,466 \\
\cline{2-2}
          & 7,8,6,24,46,48,65,166,163,252,378,516,1073,1460,2171,3748 \\
\hline
\mlr{11}  & \begin{mat4} 4527 & 5433 & 6471 & 7375 \end{mat4} \\
\cline{2-2}
          & 4,6,8,9,10,11,13,13,15,15,16,16 \dfreesep 31 \\
\cline{2-2}
          & 2,5,4,8,7,7,13,21,23,51,71,80,143,199,311,442 \\
\cline{2-2}
          & 4,16,16,32,31,38,67,122,125,320,465,540,981,1458,2377,3498 \\
\hline
\mlr{12}  & \begin{mat4} 45534 & 50564 & 63754 & 71544 \end{mat4} \\
\cline{2-2}
          & 4,6,8,9,10,11,13,14,14,15,15,16,18 \dfreesep 33 \\
\cline{2-2}
          & 3,4,2,8,8,11,19,20,32,46,66,112,151,205,342,461 \\
\cline{2-2}
          & 7,12,10,32,38,50,95,120,172,298,438,744,1089,1554,2558,3636 \\
\hline
\mlr{13}  & \begin{mat4} 43372 & 52722 & 65546 & 71176 \end{mat4} \\
\cline{2-2}
          & 4,6,8,9,10,11,13,14,15,15,16,17,18,19 \dfreesep 36 \\
\cline{2-2}
          & 9,0,16,0,13,0,54,0,101,0,220,0,465,0,1017,0 \\
\cline{2-2}
          & 25,0,68,0,66,0,299,0,619,0,1503,0,3447,0,8161,0 \\
\hline
\mlr{14}  & \begin{mat4} 44207 & 56151 & 67355 & 73013 \end{mat4} \\
\cline{2-2}
          & 4,6,8,9,10,11,13,14,15,17,17,17,18,19,20 \dfreesep 33 \\
\cline{2-2}
          & 2,0,0,1,1,5,5,5,3,11,23,19,33,58,91,118 \\
\cline{2-2}
          & 4,0,0,2,3,22,21,20,15,56,139,114,195,360,621,824 \\
\hline
\mlr{15}  & \begin{mat4} 423354 & 532544 & 614564 & 774434 \end{mat4} \\
\cline{2-2}
          & 4,6,8,9,10,11,13,14,15,17,17,17,18,19,19,19 \dfreesep 38 \\
\cline{2-2}
          & 3,0,5,0,12,0,31,0,65,0,93,0,249,0,526,0 \\
\cline{2-2}
          & 6,0,18,0,48,0,159,0,359,0,610,0,1705,0,4048,0 \\
\hline
\mlr{21}  & \begin{mat4} 42647164 & 51270354 & 62464044 & 77553334 \end{mat4} \\
\cline{2-2}
          & 4,6,8,9,10,11,13,14,15,17,18,19,20,20,21,22,23,23,23,24,25,26 \dfreesep 50 \\
\cline{2-2}
          & 7,0,4,0,22,0,45,0,88,0,166,0,386,0,830,0 \\
\cline{2-2}
          & 21,0,17,0,106,0,288,0,547,0,1187,0,2980,0,6895,0 \\
\hline
\mlr{23}  & \begin{mat4} 45704327 & 55073553 & 65567055 & 72610751 \end{mat4} \\
\cline{2-2}
          & 4,6,8,9,10,11,13,14,15,17,18,19,20,20,21,23,23,24,25,25,25,26,27,29 \dfreesep 48 \\
\cline{2-2}
          & 1,0,0,0,6,0,0,0,19,0,0,0,88,0,0,0 \\
\cline{2-2}
          & 3,0,0,0,24,0,0,0,99,0,0,0,514,0,0,0 \\
\hline
\mlr{27}  & \begin{mat4} 4126442534 & 5254267444 & 6202134254 & 7416644164 \end{mat4} \\
\cline{2-2}
          & 4,6,8,9,10,11,13,14,15,17,18,19,20,20,21,23,24,25,25,26,27,28,29,29,30,30,31,32 \dfreesep 55 \\
\cline{2-2}
          & 1,1,1,1,1,1,2,2,3,6,6,10,20,22,39,71 \\
\cline{2-2}
          & 1,4,3,2,5,4,6,12,15,26,28,68,130,130,257,480 \\
\hline
\end{codearray}
$
\end{table*}
\begin{table*}
\caption{Improved $ \text{OBDP}^{(6)} $ codes, $ R = 1/4 $, $ m \geq 11 $}
\centering
$
\begin{codearray}
\hline
\mlr{m}   & G(D) \\
\cline{2-2}
          & \hat{d}_{[0,m]} \dfreesep d_{\infty} \\
\cline{2-2}
          & a_{d_{\infty}},\ldots,a_{d_{\infty}+\dsmax} \\
\cline{2-2}
          & c_{d_{\infty}},\ldots,c_{d_{\infty}+\dsmax} \\
\hline
\hline
\mlr{11}  & \begin{mat4} 4557 & 5163 & 6531 & 7735 \end{mat4} \\
\cline{2-2}
          & 4,6,8,9,10,11,11,11,13,14,14,15 \dfreesep 32 \\
\cline{2-2}
          & 8,0,8,0,31,0,46,0,87,0,202,0,435,0,926,0 \\
\cline{2-2}
          & 20,0,38,0,155,0,265,0,550,0,1402,0,3325,0,7489,0 \\
\hline
\mlr{12}  & \begin{mat4} 44514 & 54374 & 65564 & 72744 \end{mat4} \\
\cline{2-2}
          & 4,6,8,9,10,11,13,13,14,14,16,17,18 \dfreesep 33 \\
\cline{2-2}
          & 1,3,8,9,11,7,12,20,34,47,63,113,137,212,341,465 \\
\cline{2-2}
          & 1,6,32,44,49,30,72,106,200,294,407,766,1003,1550,2575,3766 \\
\hline
\mlr{13}  & \begin{mat4} 44562 & 56516 & 61572 & 77266 \end{mat4} \\
\cline{2-2}
          & 4,6,8,9,10,11,13,14,14,14,16,16,18,19 \dfreesep 36 \\
\cline{2-2}
          & 8,0,14,0,23,0,47,0,97,0,220,0,479,0,1036,0 \\
\cline{2-2}
          & 19,0,65,0,121,0,265,0,615,0,1556,0,3610,0,8502,0 \\
\hline
\mlr{14}  & \begin{mat4} 44255 & 56317 & 67523 & 73731 \end{mat4} \\
\cline{2-2}
          & 4,6,8,9,10,11,13,14,15,15,16,17,18,18,19 \dfreesep 38 \\
\cline{2-2}
          & 8,0,14,0,23,0,55,0,111,0,254,0,472,0,1093,0 \\
\cline{2-2}
          & 20,0,57,0,125,0,321,0,717,0,1818,0,3675,0,9136,0 \\
\hline
\mlr{16}  & \begin{mat4} 467662 & 545066 & 661232 & 713516 \end{mat4} \\
\cline{2-2}
          & 4,6,8,9,10,11,13,14,15,17,18,18,18,20,20,21,22 \dfreesep 40 \\
\cline{2-2}
          & 3,0,8,0,16,0,19,0,66,0,135,0,249,0,567,0 \\
\cline{2-2}
          & 6,0,32,0,75,0,95,0,421,0,874,0,1812,0,4402,0 \\
\hline
\mlr{17}  & \begin{mat4} 424271 & 510447 & 627355 & 776753 \end{mat4} \\
\cline{2-2}
          & 4,6,8,9,10,11,13,14,15,17,18,18,19,20,20,21,22,23 \dfreesep 42 \\
\cline{2-2}
          & 3,2,5,5,3,10,15,30,33,48,58,93,146,206,318,442 \\
\cline{2-2}
          & 6,4,22,25,16,50,76,174,200,300,394,655,1048,1502,2528,3596 \\
\hline
\mlr{20}  & \begin{mat4} 4347511 & 5027147 & 6336553 & 7610515 \end{mat4} \\
\cline{2-2}
          & 4,6,8,9,10,11,13,14,15,17,18,19,20,20,21,21,22,22,23,24,25 \dfreesep 47 \\
\cline{2-2}
          & 2,2,5,2,6,10,15,23,19,45,57,78,114,178,259,373 \\
\cline{2-2}
          & 4,10,17,4,34,52,69,138,121,294,383,550,862,1376,2057,3044 \\
\hline
\mlr{21}  & \begin{mat4} 44250664 & 56453754 & 65705434 & 72324744 \end{mat4} \\
\cline{2-2}
          & 4,6,8,9,10,11,13,14,15,17,18,19,20,20,21,22,22,23,23,25,26,27 \dfreesep 50 \\
\cline{2-2}
          & 3,4,6,11,11,9,18,23,44,53,88,152,189,283,402,607 \\
\cline{2-2}
          & 14,12,26,57,64,55,100,149,306,337,632,1104,1472,2259,3358,5247 \\
\hline
\mlr{23}  & \begin{mat4} 42457565 & 53454077 & 61055521 & 77164443 \end{mat4} \\
\cline{2-2}
          & 4,6,8,9,10,11,13,14,15,17,18,19,20,20,21,23,23,24,24,26,26,27,28,29 \dfreesep 52 \\
\cline{2-2}
          & 2,1,2,3,5,4,19,18,21,45,50,66,106,148,215,310 \\
\cline{2-2}
          & 6,1,12,15,20,22,104,94,122,303,328,442,770,1076,1666,2484 \\
\hline
\mlr{27}  & \begin{mat4} 4076631774 & 5332055614 & 6332426324 & 7507044104 \end{mat4} \\
\cline{2-2}
          & 4,6,8,9,10,11,13,14,15,17,18,19,20,20,21,23,24,25,25,26,27,28,28,29,30,31,32,33 \dfreesep 60 \\
\cline{2-2}
          & 5,0,7,0,17,0,30,0,58,0,138,0,267,0,585,0 \\
\cline{2-2}
          & 18,0,35,0,90,0,178,0,355,0,961,0,2058,0,4737,0 \\
\hline
\end{codearray}
$
\end{table*}
\begin{table*}
\caption{Improved $ \text{OBDP}^{(7)} $ codes, $ R = 1/4 $, $ m \geq 13 $}
\centering
$
\begin{codearray}
\hline
\mlr{m}   & G(D) \\
\cline{2-2}
          & \hat{d}_{[0,m]} \dfreesep d_{\infty} \\
\cline{2-2}
          & a_{d_{\infty}},\ldots,a_{d_{\infty}+\dsmax} \\
\cline{2-2}
          & c_{d_{\infty}},\ldots,c_{d_{\infty}+\dsmax} \\
\hline
\hline
\mlr{13}  & \begin{mat4} 45362 & 55172 & 64556 & 70766 \end{mat4} \\
\cline{2-2}
          & 4,6,8,9,10,11,13,13,14,15,16,17,18,18 \dfreesep 36 \\
\cline{2-2}
          & 7,0,13,0,29,0,41,0,99,0,230,0,451,0,1041,0 \\
\cline{2-2}
          & 16,0,56,0,142,0,224,0,638,0,1629,0,3352,0,8477,0 \\
\hline
\mlr{14}  & \begin{mat4} 44333 & 56251 & 63675 & 76327 \end{mat4} \\
\cline{2-2}
          & 4,6,8,9,10,11,13,14,14,15,15,17,18,19,20 \dfreesep 38 \\
\cline{2-2}
          & 6,0,22,0,17,0,58,0,102,0,248,0,494,0,1083,0 \\
\cline{2-2}
          & 14,0,100,0,87,0,338,0,662,0,1776,0,3825,0,9026,0 \\
\hline
\mlr{15}  & \begin{mat4} 431764 & 511274 & 665314 & 735744 \end{mat4} \\
\cline{2-2}
          & 4,6,8,9,10,11,13,14,15,15,16,17,19,19,20,20 \dfreesep 40 \\
\cline{2-2}
          & 5,8,5,9,18,22,28,36,51,87,133,183,280,402,583,876 \\
\cline{2-2}
          & 12,24,24,47,98,118,172,212,324,611,920,1355,2198,3250,4878,7468 \\
\hline
\mlr{16}  & \begin{mat4} 422716 & 517466 & 622562 & 772172 \end{mat4} \\
\cline{2-2}
          & 4,6,8,9,10,11,13,14,15,17,17,17,19,20,20,21,22 \dfreesep 40 \\
\cline{2-2}
          & 1,4,4,4,6,7,15,23,37,43,60,91,117,182,277,456 \\
\cline{2-2}
          & 2,12,18,14,28,29,78,125,218,279,396,645,856,1380,2114,3638 \\
\hline
\mlr{18}  & \begin{mat4} 4605744 & 5436554 & 6656764 & 7162234 \end{mat4} \\
\cline{2-2}
          & 4,6,8,9,10,11,13,14,15,17,18,19,19,20,21,21,22,22,23 \dfreesep 43 \\
\cline{2-2}
          & 2,2,2,4,6,10,7,12,21,34,69,65,102,163,235,321 \\
\cline{2-2}
          & 8,6,8,20,20,54,39,60,137,220,437,452,726,1224,1835,2560 \\
\hline
\mlr{20}  & \begin{mat4} 4226347 & 5176075 & 6224271 & 7722633 \end{mat4} \\
\cline{2-2}
          & 4,6,8,9,10,11,13,14,15,17,18,19,20,20,20,21,23,23,24,26,26 \dfreesep 48 \\
\cline{2-2}
          & 7,0,11,0,18,0,35,0,77,0,155,0,380,0,791,0 \\
\cline{2-2}
          & 27,0,46,0,87,0,202,0,464,0,1139,0,2894,0,6515,0 \\
\hline
\mlr{21}  & \begin{mat4} 41560434 & 50313664 & 63452754 & 75133744 \end{mat4} \\
\cline{2-2}
          & 4,6,8,9,10,11,13,14,15,17,18,19,20,20,21,21,22,24,24,25,26,27 \dfreesep 50 \\
\cline{2-2}
          & 3,8,5,5,9,14,19,27,28,51,103,143,196,289,413,620 \\
\cline{2-2}
          & 10,26,18,33,44,64,114,167,174,337,712,1051,1500,2243,3336,5292 \\
\hline
\mlr{23}  & \begin{mat4} 41254251 & 50074613 & 62613167 & 77037375 \end{mat4} \\
\cline{2-2}
          & 4,6,8,9,10,11,13,14,15,17,18,19,20,20,21,23,23,23,24,25,25,26,27,27 \dfreesep 52 \\
\cline{2-2}
          & 2,2,6,3,2,8,9,14,24,29,55,67,101,148,240,334 \\
\cline{2-2}
          & 4,6,30,11,14,38,48,78,158,183,384,487,782,1112,1908,2754 \\
\hline
\mlr{27}  & \begin{mat4} 4607721564 & 5462110354 & 6404333434 & 7052573544 \end{mat4} \\
\cline{2-2}
          & 4,6,8,9,10,11,13,14,15,17,18,19,20,20,21,23,24,25,25,26,27,27,28,29,29,30,31,31 \dfreesep 60 \\
\cline{2-2}
          & 5,0,3,0,16,0,28,0,77,0,126,0,271,0,608,0 \\
\cline{2-2}
          & 17,0,12,0,96,0,166,0,523,0,870,0,2136,0,5212,0 \\
\hline
\end{codearray}
$
\end{table*}
\begin{table*}
\caption{OBCDF codes, $ R = 2/3 $}
\centering
$
\begin{codearray}
\hline
\mlr{m}   & G(D) \\
\cline{2-2}
          & \hat{d}_{[0,m]} \dfreesep d_{\infty} \\
\cline{2-2}
          & a_{d_{\infty}},\ldots,a_{d_{\infty}+\dsmax} \\
\cline{2-2}
          & c_{d_{\infty}},\ldots,c_{d_{\infty}+\dsmax} \\
\hline
\hline
\mlr{1}   & \begin{mat32} 0 & 2 & 6 \\ 6 & 6 & 4 \end{mat32} \\
\cline{2-2}
          & 1,2 \dfreesep 3 \\
\cline{2-2}
          & 1,4,14,40,116,339,991,2897,8468,24752,72350,211479,618153,1806861,5281454,15437688 \\
\cline{2-2}
          & 1,10,54,226,856,3072,10647,35998,119478,390904,1264412,4051896,12884337,40702722,127865250,399739078 \\
\hline
\mlr{2}   & \begin{mat32} 2 & 5 & 7 \\ 7 & 2 & 7 \end{mat32} \\
\cline{2-2}
          & 2,2,3 \dfreesep 5 \\
\cline{2-2}
          & 3,10,37,145,511,1833,6791,24945,91072,333115,1219685,4464371,16338207,59795625,218848574,800968310 \\
\cline{2-2}
          & 6,35,198,990,4284,18288,77975,324610,1326573,5366305,21526325,85675943,338748114,1331977916,5212424593,20312193244 \\
\hline
\mlr{3}   & \begin{mat32} 04 & 60 & 74 \\ 40 & 34 & 54 \end{mat32} \\
\cline{2-2}
          & 2,2,3,3 \dfreesep 6 \\
\cline{2-2}
          & 4,10,28,143,497,1859,7204,27247,103730,393902,1497596,5696182,21646452,82278998,312758462,1188845828 \\
\cline{2-2}
          & 8,40,140,945,4252,18729,84832,368483,1572348,6634224,27756484,115127460,473925208,1939848940,7899760960,32027928816 \\
\hline
\mlr{4}   & \begin{mat32} 00 & 46 & 62 \\ 42 & 34 & 56 \end{mat32} \\
\cline{2-2}
          & 2,2,3,3,4 \dfreesep 6 \\
\cline{2-2}
          & 2,4,14,38,142,582,1999,7330,27660,103298,389776,1474542,5597276,21295326,81165458,309839136 \\
\cline{2-2}
          & 4,12,58,181,871,4317,17880,76023,328108,1379427,5811723,24291839,101005089,418173935,1723751838,7078565276 \\
\hline
\mlr{5}   & \begin{mat32} 05 & 40 & 73 \\ 67 & 01 & 50 \end{mat32} \\
\cline{2-2}
          & 2,2,3,3,4,4 \dfreesep 8 \\
\cline{2-2}
          & 5,0,64,0,886,0,13135,0,194508,0,2866789,0,42389535,0,626003049,0 \\
\cline{2-2}
          & 10,0,404,0,8332,0,161410,0,3026338,0,53817120,0,932876586,0,15798364932,0 \\
\hline
\mlr{6}   & \begin{mat32} 024 & 664 & 770 \\ 420 & 104 & 604 \end{mat32} \\
\cline{2-2}
          & 2,2,3,3,4,4,4 \dfreesep 7 \\
\cline{2-2}
          & 1,1,3,4,24,113,428,1610,6157,23800,90711,349547,1348768,5184284,19936565,76680385 \\
\cline{2-2}
          & 1,4,13,30,190,894,4388,18874,81591,353706,1491999,6315544,26557266,110454370,456925125,1881458892 \\
\hline
\mlrr{7}  & \begin{mat32} 130 & 462 & 642 \\ 462 & 150 & 426 \end{mat32} \\
\cline{2-2}
          & 2,2,3,3,4,4,4,5 \dfreesep 10 \\
\cline{2-2}
          & 2,6,35,114,406,1522,6075,23450,88533,338930,1304843,5011746,19266076,74120106,285151035,1096767162 \\
\cline{2-2}
          & \lss{6,34,336,1050,4566,19206,86044,366486,1533770,6411514,26775912,110882670,457412802,1879140162,7689434220,}{31347068334} \\
\hline
\mlr{8}   & \begin{mat32} 020 & 673 & 757 \\ 505 & 174 & 623 \end{mat32} \\
\cline{2-2}
          & 2,2,3,3,4,4,4,5,5 \dfreesep 10 \\
\cline{2-2}
          & 2,0,27,0,181,0,3098,0,43483,0,647810,0,9565637,0,141617422,0 \\
\cline{2-2}
          & 10,0,254,0,1892,0,41034,0,711971,0,12672782,0,217756500,0,3681136744,0 \\
\hline
\mlrr{9}  & \begin{mat32} 0324 & 6670 & 7614 \\ 7674 & 0444 & 5320 \end{mat32} \\
\cline{2-2}
          & 2,2,3,3,4,4,4,5,5,5 \dfreesep 13 \\
\cline{2-2}
          & 5,27,104,366,1437,5456,21089,81199,311171,1197524,4613955,17753163,68278329,262712131,1010761241,3888541058 \\
\cline{2-2}
          & \lss{30,272,1122,4359,19857,84102,361056,1516953,6317270,26252396,108562577,446419658,1827303793,7455646383,30318297037,}{122921047419} \\
\hline
\end{codearray}
$
\end{table*}
\begin{table*}
\caption{Improved $ \text{OBDP}^{(0)} $ codes, $ R = 2/3 $}
\centering
$
\begin{codearray}
\hline
\mlr{m}   & G(D) \\
\cline{2-2}
          & \hat{d}_{[0,m]} \dfreesep d_{\infty} \\
\cline{2-2}
          & a_{d_{\infty}},\ldots,a_{d_{\infty}+\dsmax} \\
\cline{2-2}
          & c_{d_{\infty}},\ldots,c_{d_{\infty}+\dsmax} \\
\hline
\hline
\mlr{2}   & \begin{mat32} 1 & 5 & 6 \\ 5 & 2 & 7 \end{mat32} \\
\cline{2-2}
          & 2,2,3 \dfreesep 5 \\
\cline{2-2}
          & 3,11,39,135,519,1902,6875,25451,93245,341448,1254173,4599593,16872884,61905141,227081651,833068755 \\
\cline{2-2}
          & 5,45,218,949,4518,19355,81065,339912,1389053,5626034,22621891,90140260,357088182,1406861354,5515712953,21537553805 \\
\hline
\mlrr{3}  & \begin{mat32} 04 & 54 & 60 \\ 50 & 14 & 54 \end{mat32} \\
\cline{2-2}
          & 2,2,3,3 \dfreesep 6 \\
\cline{2-2}
          & 1,11,43,132,550,2029,7648,29405,111370,423078,1609894,6111168,23220495,88252281,335297341,1274111805 \\
\cline{2-2}
          & \lss{1,41,311,1037,5446,23287,100723,434034,1830587,7638513,31735764,130530347,534092763,2175156504,8815646941,}{35593865798} \\
\hline
\mlr{4}   & \begin{mat32} 10 & 52 & 66 \\ 42 & 34 & 76 \end{mat32} \\
\cline{2-2}
          & 2,2,3,3,4 \dfreesep 8 \\
\cline{2-2}
          & 20,0,230,0,3639,0,52367,0,774654,0,11383879,0,167452837,0,2462791462,0 \\
\cline{2-2}
          & 97,0,1809,0,40252,0,748532,0,13587443,0,236695274,0,4025404103,0,67200975888,0 \\
\hline
\mlr{5}   & \begin{mat32} 25 & 55 & 76 \\ 63 & 16 & 45 \end{mat32} \\
\cline{2-2}
          & 2,2,3,3,4,4 \dfreesep 10 \\
\cline{2-2}
          & 69,0,925,0,13189,0,197340,0,2908007,0,42990554,0,635277043,0,9387589600,0 \\
\cline{2-2}
          & 435,0,9445,0,177734,0,3296251,0,57951650,0,995745244,0,16769147724,0,278158489411,0 \\
\hline
\mlrr{6}  & \begin{mat32} 100 & 524 & 744 \\ 514 & 350 & 554 \end{mat32} \\
\cline{2-2}
          & 2,2,3,3,4,4,4 \dfreesep 10 \\
\cline{2-2}
          & 2,40,123,458,1669,6503,25819,97451,374195,1442699,5549498,21351798,82107201,315825085,1214831156,4672708118 \\
\cline{2-2}
          & \lss{3,311,1105,4967,20916,92253,407488,1694295,7112044,29754154,123429228,509427663,2091741605,8556947666,34879451597,}{141715371354} \\
\hline
\mlr{7}   & \begin{mat32} 126 & 444 & 662 \\ 742 & 126 & 700 \end{mat32} \\
\cline{2-2}
          & 2,2,3,3,4,4,4,5 \dfreesep 12 \\
\cline{2-2}
          & 50,0,874,0,12157,0,178870,0,2651012,0,39222356,0,580635078,0,8592930704,0 \\
\cline{2-2}
          & 352,0,9816,0,176037,0,3162293,0,55404270,0,946674598,0,15890264133,0,262933618537,0 \\
\hline
\mlrrr{8} & \begin{mat32} 037 & 473 & 532 \\ 676 & 317 & 633 \end{mat32} \\
\cline{2-2}
          & 2,2,3,3,4,4,4,5,5 \dfreesep 13 \\
\cline{2-2}
          & \lss{33,114,391,1533,5896,22446,86351,333298,1284199,4935269,18991631,73070198,281048717,1081357875,4160286510,}{16005386513} \\
\cline{2-2}
          & \lss{241,1206,4595,20758,88541,371861,1577070,6617859,27595063,114034490,469392670,1924179460,7855412213,31970539122,}{129720754690,524921773895} \\
\hline
\mlrrr{9} & \begin{mat32} 0070 & 5664 & 7244 \\ 4624 & 3344 & 7700 \end{mat32} \\
\cline{2-2}
          & 2,2,3,3,4,4,4,5,5,5 \dfreesep 14 \\
\cline{2-2}
          & \lss{13,120,403,1565,5786,21985,84942,327229,1257881,4835277,18618514,71620166,275537751,1060057014,4078264840,}{15690715855} \\
\cline{2-2}
          & \lss{107,1222,4824,21692,89046,374345,1584861,6641402,27559311,113718038,468041757,1916099476,7816655051,31785506012,}{128875336330,521185864619} \\
\hline
\end{codearray}
$
\end{table*}
\begin{table*}
\caption{Improved $ \text{OBDP}^{(1)} $ codes, $ R = 2/3 $}
\centering
$
\begin{codearray}
\hline
\mlr{m}   & G(D) \\
\cline{2-2}
          & \hat{d}_{[0,m]} \dfreesep d_{\infty} \\
\cline{2-2}
          & a_{d_{\infty}},\ldots,a_{d_{\infty}+\dsmax} \\
\cline{2-2}
          & c_{d_{\infty}},\ldots,c_{d_{\infty}+\dsmax} \\
\hline
\hline
\mlrr{4}  & \begin{mat32} 14 & 56 & 62 \\ 46 & 34 & 76 \end{mat32} \\
\cline{2-2}
          & 2,2,3,3,3 \dfreesep 8 \\
\cline{2-2}
          & 6,42,153,510,1853,7338,28378,108499,416534,1595979,6113526,23435761,89845723,344435753,1320384052,5061473869 \\
\cline{2-2}
          & \lss{20,284,1312,5164,22192,99382,428364,1814957,7638336,31851193,131945966,543738121,2229857002,9105969835,37045065940,}{150200458883} \\
\hline
\mlrrr{7} & \begin{mat32} 156 & 464 & 716 \\ 560 & 352 & 602 \end{mat32} \\
\cline{2-2}
          & 2,2,3,3,4,4,4,4 \dfreesep 12 \\
\cline{2-2}
          & \lss{29,118,445,1677,6306,24378,93443,360315,1385727,5327655,20494712,78855614,303346443,1167006455,4489774950,}{17272149336} \\
\cline{2-2}
          & \lss{206,1071,4962,21582,91978,392819,1659948,6984843,29080564,120434649,496310709,2037084645,8326704384,33918070763,}{137743606362,557802462199} \\
\hline
\end{codearray}
$
\end{table*}
\begin{table*}
\caption{Improved $ \text{OBDP}^{(2)} $ codes, $ R = 2/3 $, $ m \geq 3 $}
\centering
$
\begin{codearray}
\hline
\mlr{m}   & G(D) \\
\cline{2-2}
          & \hat{d}_{[0,m]} \dfreesep d_{\infty} \\
\cline{2-2}
          & a_{d_{\infty}},\ldots,a_{d_{\infty}+\dsmax} \\
\cline{2-2}
          & c_{d_{\infty}},\ldots,c_{d_{\infty}+\dsmax} \\
\hline
\hline
\mlrr{3}  & \begin{mat32} 30 & 54 & 54 \\ 54 & 30 & 74 \end{mat32} \\
\cline{2-2}
          & 2,2,2,2 \dfreesep 7 \\
\cline{2-2}
          & 17,53,133,569,2327,8624,32412,123450,468927,1777969,6744637,25591422,97087053,368332736,1397439369,5301764098 \\
\cline{2-2}
          & \lss{86,360,1148,5767,27277,114524,481710,2036585,8488085,35006717,143606787,585895596,2378004579,9611183620,}{38700634801,155311168770} \\
\hline
\mlrrr{8} & \begin{mat32} 053 & 406 & 755 \\ 421 & 361 & 712 \end{mat32} \\
\cline{2-2}
          & 2,2,3,3,4,4,4,4,4 \dfreesep 13 \\
\cline{2-2}
          & \lss{19,112,418,1596,5790,22640,87213,334667,1287593,4947028,19051545,73288627,281959022,1084774198,4173505248,}{16056499631} \\
\cline{2-2}
          & \lss{109,1041,4888,21160,87667,374609,1587526,6629895,27580039,113957986,469744346,1925349325,7862997951,32003035001,}{129873519658,525601328262} \\
\hline
\end{codearray}
$
\end{table*}
\begin{table*}
\caption{OBCDF codes, $ R = 2/4 $}
\centering
$
\begin{codearray}
\hline
\mlr{m}   & G(D) \\
\cline{2-2}
          & \hat{d}_{[0,m]} \dfreesep d_{\infty} \\
\cline{2-2}
          & a_{d_{\infty}},\ldots,a_{d_{\infty}+\dsmax} \\
\cline{2-2}
          & c_{d_{\infty}},\ldots,c_{d_{\infty}+\dsmax} \\
\hline
\hline
\mlr{1}   & \begin{mat42} 0 & 2 & 6 & 6 \\ 6 & 6 & 0 & 4 \end{mat42} \\
\cline{2-2}
          & 2,3 \dfreesep 5 \\
\cline{2-2}
          & 2,4,8,16,32,65,132,268,544,1104,2240,4545,9222,18712,37968,77040 \\
\cline{2-2}
          & 2,8,24,64,160,386,908,2096,4768,10720,23872,52740,115742,252560,548400,1185664 \\
\hline
\mlr{2}   & \begin{mat42} 1 & 3 & 4 & 6 \\ 4 & 6 & 3 & 1 \end{mat42} \\
\cline{2-2}
          & 2,4,4 \dfreesep 6 \\
\cline{2-2}
          & 2,0,10,0,57,0,325,0,1826,0,10266,0,57719,0,324507,0 \\
\cline{2-2}
          & 2,0,30,0,286,0,2286,0,16554,0,114048,0,758868,0,4928302,0 \\
\hline
\mlr{3}   & \begin{mat42} 14 & 34 & 54 & 74 \\ 70 & 60 & 74 & 64 \end{mat42} \\
\cline{2-2}
          & 2,4,4,4 \dfreesep 8 \\
\cline{2-2}
          & 3,0,14,0,88,0,506,0,3181,0,18652,0,113778,0,679038,0 \\
\cline{2-2}
          & 6,0,54,0,500,0,3702,0,30672,0,212178,0,1540550,0,10467134,0 \\
\hline
\mlr{4}   & \begin{mat42} 00 & 24 & 46 & 72 \\ 76 & 52 & 04 & 30 \end{mat42} \\
\cline{2-2}
          & 2,4,5,5,6 \dfreesep 9 \\
\cline{2-2}
          & 1,0,6,15,19,45,118,319,735,1696,4262,10199,24194,58187,140827,339809 \\
\cline{2-2}
          & 1,0,18,50,87,240,756,2292,5709,14758,40898,106810,274556,710220,1843589,4735250 \\
\hline
\mlr{5}   & \begin{mat42} 04 & 35 & 52 & 77 \\ 73 & 67 & 67 & 73 \end{mat42} \\
\cline{2-2}
          & 2,4,5,5,6,7 \dfreesep 12 \\
\cline{2-2}
          & 3,0,28,0,168,0,810,0,4868,0,28224,0,165989,0,968494,0 \\
\cline{2-2}
          & 9,0,148,0,1178,0,6735,0,49288,0,336004,0,2236726,0,14819291,0 \\
\hline
\mlr{6}   & \begin{mat42} 064 & 210 & 540 & 654 \\ 540 & 654 & 064 & 210 \end{mat42}, \begin{mat42} 064 & 250 & 540 & 614 \\ 540 & 614 & 064 & 250 \end{mat42} \\
\cline{2-2}
          & 2,4,5,5,6,7,7 \dfreesep 13 \\
\cline{2-2}
          & 2,1,4,26,48,134,276,610,1494,3500,8598,21083,49836,120051,289182,700811 \\
\cline{2-2}
          & 2,2,12,152,300,912,2104,5264,13766,35456,93754,247068,626784,1610326,4127214,10612592 \\
\hline
\mlr{7}   & \begin{mat42} 024 & 226 & 540 & 632 \\ 546 & 644 & 032 & 240 \end{mat42} \\
\cline{2-2}
          & 2,4,5,5,6,7,7,8 \dfreesep 14 \\
\cline{2-2}
          & 2,0,11,0,53,0,315,0,1754,0,10430,0,59425,0,350928,0 \\
\cline{2-2}
          & 2,0,54,0,302,0,2288,0,16268,0,113456,0,744578,0,4999208,0 \\
\hline
\end{codearray}
$
\end{table*}
\begin{table*}
\caption{Improved $ \text{OBDP}^{(0)} $ codes, $ R = 2/4 $}
\centering
$
\begin{codearray}
\hline
\mlr{m}   & G(D) \\
\cline{2-2}
          & \hat{d}_{[0,m]} \dfreesep d_{\infty} \\
\cline{2-2}
          & a_{d_{\infty}},\ldots,a_{d_{\infty}+\dsmax} \\
\cline{2-2}
          & c_{d_{\infty}},\ldots,c_{d_{\infty}+\dsmax} \\
\hline
\hline
\mlr{3}   & \begin{mat42} 04 & 30 & 54 & 60 \\ 74 & 44 & 34 & 04 \end{mat42} \\
\cline{2-2}
          & 2,4,4,4 \dfreesep 8 \\
\cline{2-2}
          & 1,0,16,0,99,0,549,0,3300,0,20026,0,120863,0,728478,0 \\
\cline{2-2}
          & 1,0,53,0,627,0,4517,0,33307,0,241829,0,1701059,0,11692535,0 \\
\hline
\mlr{4}   & \begin{mat42} 06 & 34 & 52 & 70 \\ 70 & 52 & 34 & 06 \end{mat42}, \begin{mat42} 10 & 22 & 54 & 76 \\ 76 & 54 & 22 & 10 \end{mat42} \\
\cline{2-2}
          & 2,4,5,5,6 \dfreesep 11 \\
\cline{2-2}
          & 2,13,30,53,106,319,698,1750,4162,10103,24042,58171,140412,338686,810984,1956296 \\
\cline{2-2}
          & 2,48,150,314,686,2504,5874,16412,42334,111732,284422,743242,1908884,4901296,12425504,31668176 \\
\hline
\mlr{5}   & \begin{mat42} 05 & 34 & 47 & 62 \\ 65 & 47 & 25 & 17 \end{mat42} \\
\cline{2-2}
          & 2,4,5,5,6,7 \dfreesep 12 \\
\cline{2-2}
          & 1,7,14,36,68,171,419,1031,2495,6007,14636,34613,84110,202840,488371,1181814 \\
\cline{2-2}
          & 1,26,66,193,431,1278,3380,9352,24408,64670,169567,429129,1120242,2867619,7320571,18732286 \\
\hline
\mlr{6}   & \begin{mat42} 014 & 354 & 544 & 704 \\ 770 & 640 & 664 & 734 \end{mat42} \\
\cline{2-2}
          & 2,4,5,5,6,7,7 \dfreesep 15 \\
\cline{2-2}
          & 8,24,55,100,254,628,1450,3625,8588,20659,50428,121925,293767,708327,1709851,4125697 \\
\cline{2-2}
          & 23,129,376,720,2041,5638,14455,39176,99203,256897,671607,1726238,4408373,11237339,28598966,72500580 \\
\hline
\mlr{7}   & \begin{mat42} 052 & 276 & 444 & 730 \\ 674 & 460 & 062 & 326 \end{mat42} \\
\cline{2-2}
          & 2,4,5,5,6,7,7,8 \dfreesep 17 \\
\cline{2-2}
          & 12,31,90,168,343,907,2218,5176,12785,30683,73542,179456,431879,1043088,2517345,6075911 \\
\cline{2-2}
          & 44,160,612,1230,2915,8446,22382,56776,151201,387722,993878,2577130,6568691,16763154,42591943,107997042 \\
\hline
\end{codearray}
$
\end{table*}
\begin{table*}
\caption{Improved $ \text{OBDP}^{(1)} $ codes, $ R = 2/4 $}
\centering
$
\begin{codearray}
\hline
\mlr{m}   & G(D) \\
\cline{2-2}
          & \hat{d}_{[0,m]} \dfreesep d_{\infty} \\
\cline{2-2}
          & a_{d_{\infty}},\ldots,a_{d_{\infty}+\dsmax} \\
\cline{2-2}
          & c_{d_{\infty}},\ldots,c_{d_{\infty}+\dsmax} \\
\hline
\hline
\mlr{5}   & \begin{mat42} 04 & 36 & 55 & 73 \\ 71 & 73 & 42 & 44 \end{mat42} \\
\cline{2-2}
          & 2,4,5,5,6,6 \dfreesep 13 \\
\cline{2-2}
          & 5,17,41,71,177,440,1044,2483,5888,14377,34900,84237,203629,490955,1183571,2856816 \\
\cline{2-2}
          & 12,68,239,453,1272,3596,9282,24281,62949,164950,429519,1109116,2856440,7310302,18642253,47428692 \\
\hline
\mlr{7}   & \begin{mat42} 130 & 252 & 466 & 654 \\ 662 & 446 & 226 & 162 \end{mat42} \\
\cline{2-2}
          & 2,4,5,5,6,7,7,7 \dfreesep 17 \\
\cline{2-2}
          & 8,40,77,164,413,875,2099,5202,12714,30649,74009,178921,432210,1042340,2514627,6071322 \\
\cline{2-2}
          & 30,258,571,1300,3713,8648,22381,59866,158002,405318,1042141,2673744,6827610,17355802,44032753,111469348 \\
\hline
\end{codearray}
$
\end{table*}
\begin{table*}
\caption{Improved $ \text{OBDP}^{(2)} $ codes, $ R = 2/4 $, $ m \geq 3 $}
\centering
$
\begin{codearray}
\hline
\mlr{m}   & G(D) \\
\cline{2-2}
          & \hat{d}_{[0,m]} \dfreesep d_{\infty} \\
\cline{2-2}
          & a_{d_{\infty}},\ldots,a_{d_{\infty}+\dsmax} \\
\cline{2-2}
          & c_{d_{\infty}},\ldots,c_{d_{\infty}+\dsmax} \\
\hline
\hline
\mlr{5}   & \begin{mat42} 12 & 34 & 47 & 75 \\ 63 & 55 & 34 & 16 \end{mat42} \\
\cline{2-2}
          & 2,4,5,5,5,6 \dfreesep 14 \\
\cline{2-2}
          & 34,0,153,0,916,0,5174,0,29696,0,175547,0,1017113,0,5922855,0 \\
\cline{2-2}
          & 141,0,1039,0,7567,0,52357,0,350011,0,2368491,0,15458710,0,100135144,0 \\
\hline
\mlr{6}   & \begin{mat42} 044 & 360 & 534 & 730 \\ 730 & 534 & 360 & 044 \end{mat42}, \begin{mat42} 130 & 214 & 560 & 764 \\ 764 & 560 & 214 & 130 \end{mat42} \\
\cline{2-2}
          & 2,4,5,5,6,6,7 \dfreesep 16 \\
\cline{2-2}
          & 50,0,230,0,1331,0,7340,0,43933,0,254030,0,1485854,0,8648308,0 \\
\cline{2-2}
          & 228,0,1710,0,12040,0,78248,0,542752,0,3567864,0,23422324,0,151104076,0 \\
\hline
\end{codearray}
$
\end{table*}
\begin{table*}
\caption{OBCDF codes, $ R = 3/4 $}
\centering
$
\begin{codearray}
\hline
\mlr{m}   & G(D) \\
\cline{2-2}
          & \hat{d}_{[0,m]} \dfreesep d_{\infty} \\
\cline{2-2}
          & a_{d_{\infty}},\ldots,a_{d_{\infty}+\dsmax} \\
\cline{2-2}
          & c_{d_{\infty}},\ldots,c_{d_{\infty}+\dsmax} \\
\hline
\hline
\mlrr{1}  & \begin{mat43} 0 & 0 & 2 & 6 \\ 2 & 6 & 4 & 4 \\ 6 & 4 & 0 & 2 \end{mat43} \\
\cline{2-2}
          & 1,2 \dfreesep 3 \\
\cline{2-2}
          & 1,6,37,167,774,3680,17368,81799,385602,1817837,8569379,40396837,190434700,897727223,4231969910,19949903507 \\
\cline{2-2}
          & \lss{1,14,171,1253,7860,46660,265052,1460682,7882936,41862053,219503480,1139231902,5862941817,29960126731,152179388275,}{768981534621} \\
\hline
\mlrrr{2} & \begin{mat43} 0 & 2 & 5 & 5 \\ 0 & 5 & 2 & 7 \\ 7 & 2 & 0 & 7 \end{mat43} \\
\cline{2-2}
          & 2,2,3 \dfreesep 5 \\
\cline{2-2}
          & \lss{4,25,101,508,2701,13690,71015,370347,1926355,10029423,52254459,272276980,1418909555,7394896070,38540793300,}{200871624999} \\
\cline{2-2}
          & \lss{8,113,619,4637,30029,184582,1129173,6762847,39762974,230961274,1328081084,7569663711,42833547022,240881766082,}{1347405710159,7501957225754} \\
\hline
\mlrr{3}  & \begin{mat43} 04 & 34 & 40 & 70 \\ 34 & 44 & 34 & 64 \\ 70 & 54 & 70 & 44 \end{mat43} \\
\cline{2-2}
          & 2,2,3,3 \dfreesep 6 \\
\cline{2-2}
          & 12,0,173,0,3148,3376,70240,150226,1670006,5210196,42063376,162163022,1098609758,4801447648,29390452930,138434181474 \\
\cline{2-2}
          & \lss{54,0,1353,0,36921,46254,1121711,2807728,34293244,121755596,1060971971,4547552906,32926918736,157075553226,}{1021028781509,5176910557412} \\
\hline
\end{codearray}
$
\end{table*}
\begin{table*}
\caption{Improved $ \text{OBDP}^{(0)} $ codes, $ R = 3/4 $}
\centering
$
\begin{codearray}
\hline
\mlr{m}   & G(D) \\
\cline{2-2}
          & \hat{d}_{[0,m]} \dfreesep d_{\infty} \\
\cline{2-2}
          & a_{d_{\infty}},\ldots,a_{d_{\infty}+\dsmax} \\
\cline{2-2}
          & c_{d_{\infty}},\ldots,c_{d_{\infty}+\dsmax} \\
\hline
\hline
\mlrrr{1} & \begin{mat43} 0 & 2 & 4 & 6 \\ 2 & 4 & 2 & 6 \\ 6 & 4 & 4 & 4 \end{mat43} \\
\cline{2-2}
          & 1,2 \dfreesep 4 \\
\cline{2-2}
          & \lss{10,46,202,949,4444,20840,97528,456549,2137538,10007099,46849485,219333106,1026840600,4807305254,22506108713,}{105365669809} \\
\cline{2-2}
          & \lss{31,237,1565,9389,53751,297976,1610542,8549416,44753289,231649971,1188111461,6047384125,30582628361,153808653817,}{769851144342,3837192524065} \\
\hline
\mlrrr{2} & \begin{mat43} 2 & 2 & 5 & 7 \\ 2 & 7 & 2 & 5 \\ 7 & 0 & 5 & 0 \end{mat43} \\
\cline{2-2}
          & 2,2,3 \dfreesep 5 \\
\cline{2-2}
          & \lss{1,26,98,498,2562,13741,70610,369270,1920960,10038480,52311060,272924883,1423268651,7424177153,38720885502,}{201963795116} \\
\cline{2-2}
          & \lss{1,144,702,5024,31152,201794,1202186,7182592,41936692,243340712,1393221490,7925492164,44743884419,251230446502,}{1403232677010,7804079023580} \\
\hline
\mlr{3}   & \begin{mat43} 10 & 14 & 44 & 70 \\ 10 & 70 & 14 & 44 \\ 44 & 10 & 34 & 60 \end{mat43} \\
\cline{2-2}
          & 2,2,3,3 \dfreesep 8 \\
\cline{2-2}
          & 131,0,3574,0,97035,0,2712619,0,75442601,0,2098975723,0,58400768397,0,1624878461858,0 \\
\cline{2-2}
          & 1015,0,46147,0,1714524,0,60678502,0,2043490555,0,66759042190,0,2132996507148,0,67012325819048,0 \\
\hline
\end{codearray}
$
\end{table*}
\begin{table*}
\caption{Improved $ \text{OBDP}^{(1)} $ codes, $ R = 3/4 $}
\label{code-last}
\centering
$
\begin{codearray}
\hline
\mlr{m}   & G(D) \\
\cline{2-2}
          & \hat{d}_{[0,m]} \dfreesep d_{\infty} \\
\cline{2-2}
          & a_{d_{\infty}},\ldots,a_{d_{\infty}+\dsmax} \\
\cline{2-2}
          & c_{d_{\infty}},\ldots,c_{d_{\infty}+\dsmax} \\
\hline
\hline
\mlrrr{2} & \begin{mat43} 1 & 3 & 4 & 6 \\ 3 & 4 & 2 & 7 \\ 6 & 5 & 1 & 2 \end{mat43} \\
\cline{2-2}
          & 2,2,2 \dfreesep 6 \\
\cline{2-2}
          & \lss{27,118,529,2978,15201,79518,414729,2160437,11264545,58694027,305964561,1594717595,8311866917,43322929148,}{225804554547,1176932834242} \\
\cline{2-2}
          & \lss{151,898,5436,37195,222112,1341095,7920780,46054230,265200292,1512561844,8565315496,48192500186,269683904075,}{1502047764635,8331381355477,46043767073516} \\
\hline
\end{codearray}
$
\end{table*}
\begin{table}
\caption{Free distances, $ d_{\infty}^{(s)} $, $ R = 1/2 $}
\label{dfree-first}
\centering
$
\begin{array}{|c|>{\arrayrule}c|c|c|c|c|c|c|c|c|c|c|}
\hline
\mlrf{m} & \multicolumn{10}{c|}{s}                              & \mlrf{d_{\infty}^{\text{G}}} \\
\cline{2-11}
         & -\infty & 0  & 1  & 2  & 3  & 4  & 5  & 6  & 7  & 8  &                              \\
\hline
\hline
 1       & 3       &    &    & -  & -  & -  & -  & -  & -  & -  & 4                            \\
\hline
 2       & 5       &    &    & -  & -  & -  & -  & -  & -  & -  & 5                            \\
\hline
 3       & 6       &    &    &    & -  & -  & -  & -  & -  & -  & 6                            \\
\hline
 4       & 6       &    &    & 7  & -  & -  & -  & -  & -  & -  & 8                            \\
\hline
 5       & 8       &    &    &    & 8  & -  & -  & -  & -  & -  & 8                            \\
\hline
 6       & 8       & 8  &    & 9  &    & -  & -  & -  & -  & -  & 10                           \\
\hline
 7       & 10      &    &    &    & 10 &    & -  & -  & -  & -  & 11                           \\
\hline
 8       & 10      &    &    & 12 &    &    & -  & -  & -  & -  & 12                           \\
\hline
 9       & 10      & 12 &    &    & 12 &    &    & -  & -  & -  & 13                           \\
\hline
10       & 9       &    & 12 &    &    & 14 &    & -  & -  & -  & 14                           \\
\hline
11       & 12      & 13 &    & 14 &    &    & 15 &    & -  & -  & 16                           \\
\hline
12       & 11      & 15 &    &    & 15 &    &    & 16 & -  & -  & 16                           \\
\hline
13       & 12      &    &    & 16 &    & 16 &    &    & 16 & -  & 17                           \\
\hline
14       & 12      &    &    &    & 17 &    & 17 &    &    & -  & 18                           \\
\hline
15       & 16      & 17 &    &    &    & 18 &    &    &    &    & 20                           \\
\hline
16       & 14      & 17 & 18 &    &    &    & 19 &    & 19 &    & 20                           \\
\hline
17       & 16      & 18 &    & 19 &    &    &    & 20 &    & 20 & 22                           \\
\hline
18       & 18      & 20 &    &    & 20 &    &    &    & 21 &    & 23                           \\
\hline
19       & 20      & 20 &    &    &    & 22 &    &    &    & 22 & 24                           \\
\hline
20       & 18      & 20 &    & 22 &    &    & 22 &    &    &    & 24                           \\
\hline
21       & 21      & 21 &    &    & 23 &    &    & 24 &    &    & 26                           \\
\hline
22       & 21      & 23 &    &    &    & 24 &    &    & 24 &    & 27                           \\
\hline
23       & 22      & 24 & 24 &    &    &    & 24 &    &    & 25 & 28                           \\
\hline
24       & 23      & 24 &    & 25 &    &    &    & 26 &    &    & 29                           \\
\hline
25       & 21      & 25 &    &    & 26 &    &    &    & 26 &    & 30                           \\
\hline
26       & 24      & 27 &    &    &    & 27 &    &    &    & 28 & 32                           \\
\hline
27       & 26      & 27 & 28 &    &    &    & 28 &    &    &    & 32                           \\
\hline
28       & 22      & 28 &    & 28 &    &    &    & 29 &    &    & 32                           \\
\hline
29       & 24      & 29 &    &    & 30 &    &    &    & 30 &    & 34                           \\
\hline
30       & 26      & 28 & 30 &    &    & 30 &    &    &    & 31 & 35                           \\
\hline
31       & 26      & 30 &    & 31 &    &    & 32 &    &    &    & 36                           \\
\hline
\end{array}
$
\end{table}
\begin{table}
\caption{Free distances, $ d_{\infty}^{(s)} $, $ R = 1/3 $}
\centering
$
\begin{array}{|c|>{\arrayrule}c|c|c|c|c|c|c|c|c|c|c|}
\hline
\mlrf{m} & \multicolumn{10}{c|}{s}                              & \mlrf{d_{\infty}^{\text{G}}} \\
\cline{2-11}
         & -\infty & 0  & 1  & 2  & 3  & 4  & 5  & 6  & 7  & 8  &                              \\
\hline
\hline
 1       & 4       &    & 5  & -  & -  & -  & -  & -  & -  & -  & 6                            \\
\hline
 2       & 8       &    &    & -  & -  & -  & -  & -  & -  & -  & 8                            \\
\hline
 3       & 9       & 10 &    &    & -  & -  & -  & -  & -  & -  & 10                           \\
\hline
 4       & 12      &    &    &    & -  & -  & -  & -  & -  & -  & 12                           \\
\hline
 5       & 12      & 12 & 12 & 12 & 13 & -  & -  & -  & -  & -  & 13                           \\
\hline
 6       & 13      & 14 &    & 15 & 15 & -  & -  & -  & -  & -  & 15                           \\
\hline
 7       & 14      & 16 &    & 16 & 16 &    & -  & -  & -  & -  & 16                           \\
\hline
 8       & 15      & 18 &    &    &    &    & -  & -  & -  & -  & 18                           \\
\hline
 9       & 18      &    & 18 &    & 20 & 20 & 20 & -  & -  & -  & 20                           \\
\hline
10       & 18      &    & 20 & 21 &    & 22 & 22 & -  & -  & -  & 22                           \\
\hline
11       & 21      &    &    &    &    & 22 & 24 & 24 & -  & -  & 24                           \\
\hline
12       & 22      &    &    &    & 23 &    & 24 & 24 & -  & -  & 24                           \\
\hline
13       & 20      & 22 &    &    &    & 24 &    & 26 & 26 & -  & 26                           \\
\hline
14       & 22      &    &    &    &    & 27 &    &    & 28 & -  & 28                           \\
\hline
15       & 26      &    &    & 28 &    &    & 28 & 28 &    & 30 & 30                           \\
\hline
16       & 24      &    &    & 28 &    & 28 &    & 30 & 30 &    & 32                           \\
\hline
17       & 29      &    &    &    & 29 &    & 32 &    & 32 & 32 & 32                           \\
\hline
18       & 23      &    &    &    & 26 &    & 31 & 32 &    & 33 & 34                           \\
\hline
19       & 26      &    &    &    &    & 32 &    & 34 & 34 &    & 36                           \\
\hline
20       & 31      &    &    &    & 33 &    & 34 &    & 36 & 36 & 38                           \\
\hline
21       & 32      &    &    & 34 &    & 35 &    & 36 &    & 37 & 40                           \\
\hline
22       & 34      &    &    &    &    & 36 & 37 &    & 38 &    & 40                           \\
\hline
23       & 37      &    &    &    &    &    & 38 & 39 &    & 40 & 42                           \\
\hline
24       & 34      &    &    & 35 &    & 39 &    & 40 & 41 &    & 44                           \\
\hline
25       & 36      & 41 &    &    &    &    &    &    & 42 & 42 & 46                           \\
\hline
26       & 38      &    &    &    & 41 & 42 &    & 43 &    & 44 & 48                           \\
\hline
27       & 38      &    & 40 &    &    &    &    & 44 & 45 &    & 48                           \\
\hline
28       & 41      & 41 &    & 44 &    &    &    &    & 46 &    & 50                           \\
\hline
29       & 38      & 44 &    &    &    & 44 & 45 &    &    & 48 & 52                           \\
\hline
30       & 41      &    &    &    &    &    & 46 &    & 47 &    & 53                           \\
\hline
31       & 45      &    &    &    &    & 48 &    & 49 &    & 50 & 55                           \\
\hline
\end{array}
$
\end{table}
\begin{table}
\caption{Free distances, $ d_{\infty}^{(s)} $, $ R = 1/4 $}
\centering
$
\begin{array}{|c|>{\arrayrule}c|c|c|c|c|c|c|c|c|c|}
\hline
\mlrf{m} & \multicolumn{9}{c|}{s}                          & \mlrf{d_{\infty}^{\text{G}}} \\
\cline{2-10}
         & -\infty & 0  & 1  & 2  & 3  & 4  & 5  & 6  & 7  &                              \\
\hline
\hline
 1       & 6       &    & 7  & -  & -  & -  & -  & -  & -  & 8                            \\
\hline
 2       & 10      &    &    & -  & -  & -  & -  & -  & -  & 10                           \\
\hline
 3       & 12      &    &    &    & -  & -  & -  & -  & -  & 13                           \\
\hline
 4       & 14      &    &    &    & -  & -  & -  & -  & -  & 16                           \\
\hline
 5       & 16      &    &    &    &    & -  & -  & -  & -  & 18                           \\
\hline
 6       & 17      &    & 19 &    & 19 & -  & -  & -  & -  & 20                           \\
\hline
 7       & 18      & 18 & 22 &    &    &    & -  & -  & -  & 22                           \\
\hline
 8       & 21      & 21 & 22 & 24 & 24 & 24 & -  & -  & -  & 24                           \\
\hline
 9       & 22      &    &    & 24 & 26 & 27 &    & -  & -  & 27                           \\
\hline
10       & 24      & 26 & 27 & 28 & 28 & 28 & 29 & -  & -  & 29                           \\
\hline
11       & 30      & 30 &    & 31 &    & 31 & 31 & 32 & -  & 32                           \\
\hline
12       & 25      &    &    &    &    & 33 & 33 & 33 & -  & 33                           \\
\hline
13       & 32      &    &    &    &    &    & 36 & 36 & 36 & 36                           \\
\hline
14       & 27      &    &    &    &    &    & 33 & 38 & 38 & 38                           \\
\hline
15       & 34      &    &    & 35 & 36 &    & 38 &    & 40 & 40                           \\
\hline
16       & 38      &    &    &    &    &    &    & 40 & 40 & 42                           \\
\hline
17       & 40      &    &    &    &    &    &    & 42 &    & 44                           \\
\hline
18       & 38      &    & 40 &    &    &    &    &    & 43 & 46                           \\
\hline
19       & 38      &    &    &    &    & 44 &    &    &    & 48                           \\
\hline
20       & 45      &    & 46 &    &    & 46 &    & 47 & 48 & 50                           \\
\hline
21       & 44      & 46 & 46 & 49 &    &    & 50 & 50 & 50 & 52                           \\
\hline
22       & 46      &    &    &    &    & 50 &    &    &    & 55                           \\
\hline
23       & 44      &    &    &    &    &    & 48 & 52 & 52 & 56                           \\
\hline
24       & 48      &    &    &    &    &    &    &    &    & 59                           \\
\hline
25       & 48      &    &    &    &    &    &    &    &    & 61                           \\
\hline
26       & 50      &    &    &    &    &    &    &    &    & 64                           \\
\hline
27       & 50      &    &    &    &    &    & 55 & 60 & 60 & 64                           \\
\hline
\end{array}
$
\end{table}
\begin{table}
\caption{Free distances, $ d_{\infty}^{(s)} $, $ R = 2/3 $}
\centering
$
\begin{array}{|c|>{\arrayrule}c|c|c|c|c|}
\hline
\mlrf{m} & \multicolumn{4}{c|}{s} & \mlrf{d_{\infty}^{\text{G}}} \\
\cline{2-5}
         & -\infty & 0  & 1  & 2  &                              \\
\hline
\hline
 1       & 3       &    &    & -  & 4                            \\
\hline
 2       & 5       & 5  &    & -  & 6                            \\
\hline
 3       & 6       & 6  &    & 7  & 8                            \\
\hline
 4       & 6       & 8  & 8  &    & 8                            \\
\hline
 5       & 8       & 10 &    &    & 10                           \\
\hline
 6       & 7       & 10 &    &    & 12                           \\
\hline
 7       & 10      & 12 & 12 &    & 14                           \\
\hline
 8       & 10      & 13 &    & 13 & 16                           \\
\hline
 9       & 13      & 14 &    &    & 16                           \\
\hline
\end{array}
$
\end{table}
\begin{table}
\caption{Free distances, $ d_{\infty}^{(s)} $, $ R = 2/4 $}
\centering
$
\begin{array}{|c|>{\arrayrule}c|c|c|c|c|}
\hline
\mlrf{m} & \multicolumn{4}{c|}{s} & \mlrf{d_{\infty}^{\text{G}}} \\
\cline{2-5}
         & -\infty & 0  & 1  & 2  &                              \\
\hline
\hline
 1       & 5       &    &    & -  & 5                            \\
\hline
 2       & 6       &    &    & -  & 8                            \\
\hline
 3       & 8       & 8  &    &    & 10                           \\
\hline
 4       & 9       & 11 &    &    & 12                           \\
\hline
 5       & 12      & 12 & 13 & 14 & 14                           \\
\hline
 6       & 13      & 15 &    & 16 & 16                           \\
\hline
 7       & 14      & 17 & 17 &    & 18                           \\
\hline
\end{array}
$
\end{table}
\begin{table}
\caption{Free distances, $ d_{\infty}^{(s)} $, $ R = 3/4 $}
\label{dfree-last}
\centering
$
\begin{array}{|c|>{\arrayrule}c|c|c|c|c|}
\hline
\mlrf{m} & \multicolumn{4}{c|}{s} & \mlrf{d_{\infty}^{\text{G}}} \\
\cline{2-5}
         & -\infty & 0  & 1  & 2  &                              \\
\hline
\hline
 1       & 3       & 4  &    & -  & 4                            \\
\hline
 2       & 5       & 5  & 6  & -  & 6                            \\
\hline
 3       & 6       & 8  &    &    & 8                            \\
\hline
\end{array}
$
\end{table}
\begin{table*}
\caption{Codes used in simulations}
\label{simulation-codes}
\centering
\newcommand{\rulea}{\rule[-1.5ex]{0em}{4.2ex}}
\newcommand{\ruleb}{\rule{0em}{2.3ex}}
\renewcommand{\tabularxcolumn}[1]{>{\raggedright\arraybackslash}m{#1}}
\begin{tabularx}{\textwidth}{|c|c|>{\rulea}c|l|c|X|}
\hline
$ R $                      & $ m $  & Code        & $ G(D) $                                                                      & $ d_{\infty} $ & Description                                                                                              \\
\hline
\hline
\multirow{6.2}{*}{$ 1/2 $} & $ 6 $  & $ \codeaa $ & $ \begin{mat2} 554 & 744 \end{mat2} $                                         & $ 10 $         & ODS, denoted $ \begin{mat2} 133 & 171 \end{mat2} $ in \cite{optimum-spectrum}                            \\
\cline{2-6}
                           & $ 12 $ & $ \codeab $ & $ \begin{mat2} 53734 & 72304 \end{mat2} $                                     & $ 16 $         & ODS, denoted $ \begin{mat2} 12767 & 16461 \end{mat2} $ in \cite{optimum-spectrum}                        \\
\cline{2-6}
                           & $ 31 $ & $ \codeac $ & $ \begin{mat2} 51703207732 & 66455246536 \end{mat2} $                         & $ 32 $         & ODP \cite{fundamentals}                                                                                  \\
\cline{2-6}
                           & $ 31 $ & $ \codead $ & $ \begin{mat2} 42523570626 & 64546507642 \end{mat2} $                         & $ 32 $         & $ \text{OBDP}^{(5)} $                                                                                    \\
\hline
\multirow{7.8}{*}{$ 1/3 $} & $ 6 $  & $ \codeba $ & $ \begin{mat3} 554 & 724 & 744 \end{mat3} $                                   & $ 15 $         & ODS, denoted $ \begin{mat3} 133 & 165 & 171 \end{mat3} $ in \cite{optimum-spectrum}                      \\
\cline{2-6}
                           & $ 12 $ & $ \codebb $ & $ \begin{mat3} 55304 & 64734 & 76244 \end{mat3} $                             & $ 24 $         & ODS, denoted $ \begin{mat3} 13261 & 15167 & 17451 \end{mat3} $ in \cite{optimum-spectrum}                \\
\cline{2-6}
                           & $ 19 $ & $ \codebc $ & $ \begin{mat3} 5531236 & 6151572 & 7731724 \end{mat3} $                       & $ 35 $         & ODP \cite{fundamentals}                                                                                  \\
\cline{2-6}
                           & $ 19 $ & $ \codebd $ & $ \begin{mat3} 4340472 & 5270746 & 7165336 \end{mat3} $                       & $ 32 $         & $ \text{OBDP}^{(4)} $                                                                                    \\
\cline{2-6}
                           & $ 31 $ & $ \codebe $ & $ \begin{mat3} 41775322742 & 64230711252 & 70124437316 \end{mat3} $           & $ 48 $         & $ \text{OBDP}^{(4)} $                                                                                    \\
\hline
\multirow{8.2}{*}{$ 1/4 $} & $ 6 $  & $ \codeca $ & $ \begin{mat4} 474 & 534 & 664 & 744 \end{mat4} $                             & $ 20 $         & ODS, denoted $ \begin{mat4} 117 & 127 & 155 & 171 \end{mat4} $ in \cite{optimum-spectrum}                \\
\cline{2-6}
                           & $ 12 $ & $ \codecb $ & $ \begin{mat4} 46254 & 56374 & 65044 & 75564 \end{mat4} $                     & $ 33 $         & \ruleb OFD, denoted $ \begin{mat4} 11453 & 13477 & 15211 & 17335 \end{mat4} $ in \cite{optimum-distance} \\
\cline{2-6}
                           & $ 21 $ & $ \codecc $ & $ \begin{mat4} 45724414 & 55057474 & 65556514 & 72624710 \end{mat4} $         & $ 50 $         & ODP \cite{fundamentals}                                                                                  \\
\cline{2-6}
                           & $ 21 $ & $ \codecd $ & $ \begin{mat4} 43553134 & 50374764 & 63210644 & 76045154 \end{mat4} $         & $ 49 $         & $ \text{OBDP}^{(2)} $                                                                                    \\
\cline{2-6}
                           & $ 27 $ & $ \codece $ & $ \begin{mat4} 4126442534 & 5254267444 & 6202134254 & 7416644164 \end{mat4} $ & $ 55 $         & $ \text{OBDP}^{(5)} $                                                                                    \\
\hline
\multirow{8.1}{*}{$ 2/3 $} & $ 3 $  & $ \codeda $ & $ \begin{mat32} 64 & 64 & 20 \\ 54 & 30 & 54 \end{mat32} $                    & $ 7  $         & OFD, defined by $ H(D) = \begin{mat3} 634 & 514 & 504 \end{mat3} $ in \cite{codes-on-graphs}             \\
\cline{2-6}
                           & $ 6 $  & $ \codedb $ & $ \begin{mat32} 614 & 544 & 340 \\ 140 & 664 & 454 \end{mat32} $              & $ 11 $         & OFD, defined by $ H(D) = \begin{mat3} 70754 & 62364 & 42074 \end{mat3} $ in \cite{codes-on-graphs}       \\
\cline{2-6}
                           & $ 9 $  & $ \codedc $ & $ \begin{mat32} 6734 & 1734 & 4330 \\ 1574 & 5140 & 7014 \end{mat32} $        & $ 14 $         & ODP \cite{further-results}                                                                               \\
\cline{2-6}
                           & $ 9 $  & $ \codedd $ & $ \begin{mat32} 0070 & 5664 & 7244 \\ 4624 & 3344 & 7700 \end{mat32} $        & $ 14 $         & $ \text{OBDP}^{(0)} $                                                                                    \\
\hline
\end{tabularx}
\end{table*}
\clearpage
\begin{figure}
\centering
\begin{tikzpicture}
\begin{semilogyaxis}
  [ xlabel={SNR [dB]},
    ylabel={FER},
    xmin=1,
    xmax=9,
    ymin=1e-7,
    ymax=1 ]
\addplot [ mark=none, black, dashed ] table {fer-viterbi-r12-m6.txt} ;
\addplot [ mark=none, black, dashed ] table {fer-viterbi-r12-m12.txt} ;
\label{fer-viterbi-r12}
\addplot [ mark=none, black ] table {fer-bistack-r12-obdp-m31-2.txt} ;
\addplot [ mark=none, black ] table {fer-bistack-r12-obdp-m31-3.txt} ;
\addplot [ mark=none, black ] table {fer-bistack-r12-obdp-m31-4.txt} ;
\addplot [ mark=none, black ] table {fer-bistack-r12-obdp-m31-5.txt} ;
\addplot [ mark=none, black ] table {fer-bistack-r12-obdp-m31-6.txt} ;
\addplot [ mark=none, black ] table {fer-bistack-r12-obdp-m31-7.txt} ;
\addplot [ mark=none, black ] table {fer-bistack-r12-obdp-m31-8.txt} ;
\addplot [ mark=none, black ] table {fer-bistack-r12-obdp-m31-9.txt} ;
\addplot [ mark=none, black ] table {fer-bistack-r12-obdp-m31-10.txt} ;
\addplot [ mark=none, black ] table {fer-bistack-r12-obdp-m31-11.txt} ;
\addplot [ mark=none, black ] table {fer-bistack-r12-obdp-m31-12.txt} ;
\addplot [ mark=none, black ] table {fer-bistack-r12-obdp-m31-13.txt} ;
\addplot [ mark=none, black ] table {fer-bistack-r12-obdp-m31-14.txt} ;
\addplot [ mark=none, black ] table {fer-bistack-r12-obdp-m31-15.txt} ;
\addplot [ mark=none, black ] table {fer-bistack-r12-obdp-m31-16.txt} ;
\label{fer-bistack-r12-obdp-m31-mlc}
\end{semilogyaxis}
\end{tikzpicture}
\caption
  {FER, $ R = 1/2 $:
   \begin{captionaligned}
   \ref{fer-viterbi-r12} VA($ \codeaa $), VA($ \codeab $) \\
   \ref{fer-bistack-r12-obdp-m31-mlc} BSA($ \codead $,$ \{ 2^{10}, 2^{11}, \ldots, 2^{24} \} $)
   \end{captionaligned}}
\label{fer-mlc-first}
\end{figure}
\begin{figure}
\centering
\begin{tikzpicture}
\begin{semilogyaxis}
  [ xlabel={SNR [dB]},
    ylabel={FER},
    xmin=1,
    xmax=9,
    ymin=1e-7,
    ymax=1 ]
\addplot [ mark=none, black, dashed ] table {fer-viterbi-r13-m6.txt} ;
\addplot [ mark=none, black, dashed ] table {fer-viterbi-r13-m12.txt} ;
\label{fer-viterbi-r13}
\addplot [ mark=none, black ] table {fer-bistack-r13-obdp-m31-2.txt} ;
\addplot [ mark=none, black ] table {fer-bistack-r13-obdp-m31-3.txt} ;
\addplot [ mark=none, black ] table {fer-bistack-r13-obdp-m31-4.txt} ;
\addplot [ mark=none, black ] table {fer-bistack-r13-obdp-m31-5.txt} ;
\addplot [ mark=none, black ] table {fer-bistack-r13-obdp-m31-6.txt} ;
\addplot [ mark=none, black ] table {fer-bistack-r13-obdp-m31-7.txt} ;
\addplot [ mark=none, black ] table {fer-bistack-r13-obdp-m31-8.txt} ;
\addplot [ mark=none, black ] table {fer-bistack-r13-obdp-m31-9.txt} ;
\addplot [ mark=none, black ] table {fer-bistack-r13-obdp-m31-10.txt} ;
\addplot [ mark=none, black ] table {fer-bistack-r13-obdp-m31-11.txt} ;
\addplot [ mark=none, black ] table {fer-bistack-r13-obdp-m31-12.txt} ;
\addplot [ mark=none, black ] table {fer-bistack-r13-obdp-m31-13.txt} ;
\addplot [ mark=none, black ] table {fer-bistack-r13-obdp-m31-14.txt} ;
\addplot [ mark=none, black ] table {fer-bistack-r13-obdp-m31-15.txt} ;
\addplot [ mark=none, black ] table {fer-bistack-r13-obdp-m31-16.txt} ;
\label{fer-bistack-r13-obdp-m31-mlc}
\end{semilogyaxis}
\end{tikzpicture}
\caption
  {FER, $ R = 1/3 $:
   \begin{captionaligned}
   \ref{fer-viterbi-r13} VA($ \codeba $), VA($ \codebb $) \\
   \ref{fer-bistack-r13-obdp-m31-mlc} BSA($ \codebe $,$ \{ 2^{10}, 2^{11}, \ldots, 2^{24} \} $)
   \end{captionaligned}}
\end{figure}
\begin{figure}
\centering
\begin{tikzpicture}
\begin{semilogyaxis}
  [ xlabel={SNR [dB]},
    ylabel={FER},
    xmin=1,
    xmax=9,
    ymin=1e-7,
    ymax=1 ]
\addplot [ mark=none, black, dashed ] table {fer-viterbi-r14-m6.txt} ;
\addplot [ mark=none, black, dashed ] table {fer-viterbi-r14-m12.txt} ;
\label{fer-viterbi-r14}
\addplot [ mark=none, black ] table {fer-bistack-r14-obdp-m27-2.txt} ;
\addplot [ mark=none, black ] table {fer-bistack-r14-obdp-m27-3.txt} ;
\addplot [ mark=none, black ] table {fer-bistack-r14-obdp-m27-4.txt} ;
\addplot [ mark=none, black ] table {fer-bistack-r14-obdp-m27-5.txt} ;
\addplot [ mark=none, black ] table {fer-bistack-r14-obdp-m27-6.txt} ;
\addplot [ mark=none, black ] table {fer-bistack-r14-obdp-m27-7.txt} ;
\addplot [ mark=none, black ] table {fer-bistack-r14-obdp-m27-8.txt} ;
\addplot [ mark=none, black ] table {fer-bistack-r14-obdp-m27-9.txt} ;
\addplot [ mark=none, black ] table {fer-bistack-r14-obdp-m27-10.txt} ;
\addplot [ mark=none, black ] table {fer-bistack-r14-obdp-m27-11.txt} ;
\addplot [ mark=none, black ] table {fer-bistack-r14-obdp-m27-12.txt} ;
\addplot [ mark=none, black ] table {fer-bistack-r14-obdp-m27-13.txt} ;
\addplot [ mark=none, black ] table {fer-bistack-r14-obdp-m27-14.txt} ;
\addplot [ mark=none, black ] table {fer-bistack-r14-obdp-m27-15.txt} ;
\addplot [ mark=none, black ] table {fer-bistack-r14-obdp-m27-16.txt} ;
\label{fer-bistack-r14-obdp-m27-mlc}
\end{semilogyaxis}
\end{tikzpicture}
\caption
  {FER, $ R = 1/4 $:
   \begin{captionaligned}
   \ref{fer-viterbi-r14} VA($ \codeca $), VA($ \codecb $) \\
   \ref{fer-bistack-r14-obdp-m27-mlc} BSA($ \codece $,$ \{ 2^{10}, 2^{11}, \ldots, 2^{24} \} $)
   \end{captionaligned}}
\end{figure}
\begin{figure}
\centering
\begin{tikzpicture}
\begin{semilogyaxis}
  [ xlabel={SNR [dB]},
    ylabel={FER},
    xmin=1,
    xmax=9,
    ymin=1e-7,
    ymax=1 ]
\addplot [ mark=none, black, dashed ] table {fer-viterbi-r23-m3.txt} ;
\addplot [ mark=none, black, dashed ] table {fer-viterbi-r23-m6.txt} ;
\label{fer-viterbi-r23}
\addplot [ mark=none, black ] table {fer-bistack-r23-obdp-m9-2.txt} ;
\addplot [ mark=none, black ] table {fer-bistack-r23-obdp-m9-3.txt} ;
\addplot [ mark=none, black ] table {fer-bistack-r23-obdp-m9-4.txt} ;
\addplot [ mark=none, black ] table {fer-bistack-r23-obdp-m9-5.txt} ;
\addplot [ mark=none, black ] table {fer-bistack-r23-obdp-m9-6.txt} ;
\addplot [ mark=none, black ] table {fer-bistack-r23-obdp-m9-7.txt} ;
\addplot [ mark=none, black ] table {fer-bistack-r23-obdp-m9-8.txt} ;
\addplot [ mark=none, black ] table {fer-bistack-r23-obdp-m9-9.txt} ;
\addplot [ mark=none, black ] table {fer-bistack-r23-obdp-m9-10.txt} ;
\addplot [ mark=none, black ] table {fer-bistack-r23-obdp-m9-11.txt} ;
\addplot [ mark=none, black ] table {fer-bistack-r23-obdp-m9-12.txt} ;
\addplot [ mark=none, black ] table {fer-bistack-r23-obdp-m9-13.txt} ;
\addplot [ mark=none, black ] table {fer-bistack-r23-obdp-m9-14.txt} ;
\addplot [ mark=none, black ] table {fer-bistack-r23-obdp-m9-15.txt} ;
\addplot [ mark=none, black ] table {fer-bistack-r23-obdp-m9-16.txt} ;
\label{fer-bistack-r23-obdp-m9-mlc}
\end{semilogyaxis}
\end{tikzpicture}
\caption
  {FER, $ R = 2/3 $:
   \begin{captionaligned}
   \ref{fer-viterbi-r23} VA($ \codeda $), VA($ \codedb $) \\
   \ref{fer-bistack-r23-obdp-m9-mlc} BSA($ \codedd $,$ \{ 2^{10}, 2^{11}, \ldots, 2^{24} \} $)
   \end{captionaligned}}
\label{fer-mlc-last}
\end{figure}
\flushcolsend
\clearpage
\begin{table}
\caption{VA($ \codeaa $) and BSA($ \codead $, $ M $) equal FERs}
\centering
$
\begin{array}{|>{\arrayrule}c|c|c|}
\hline
\text{SNR [dB]} & \text{FER}          & M      \\
\hline
\hline
5.37            & 9.97 \times 10^{-2} & 2^{11} \\
\hline
7.20            & 9.15 \times 10^{-4} & 2^{10} \\
\hline
\end{array}
$
\label{fer-cross-mlc-first}
\end{table}
\begin{table}
\caption{VA($ \codeab $) and BSA($ \codead $, $ M $) equal FERs}
\centering
$
\begin{array}{|>{\arrayrule}c|c|c|}
\hline
\text{SNR [dB]} & \text{FER}          & M      \\
\hline
\hline
4.66            & 5.37 \times 10^{-2} & 2^{15} \\
\hline
5.29            & 6.92 \times 10^{-3} & 2^{14} \\
\hline
5.98            & 4.52 \times 10^{-4} & 2^{13} \\
\hline
6.78            & 1.28 \times 10^{-5} & 2^{12} \\
\hline
\end{array}
$
\end{table}
\begin{table}
\caption{VA($ \codeba $) and BSA($ \codebe $, $ M $) equal FERs}
\centering
$
\begin{array}{|>{\arrayrule}c|c|c|}
\hline
\text{SNR [dB]} & \text{FER}          & M      \\
\hline
\hline
5.05            & 1.02 \times 10^{-1} & 2^{11} \\
\hline
7.02            & 7.48 \times 10^{-4} & 2^{10} \\
\hline
\end{array}
$
\end{table}
\begin{table}
\caption{VA($ \codebb $) and BSA($ \codebe $, $ M $) equal FERs}
\centering
$
\begin{array}{|>{\arrayrule}c|c|c|}
\hline
\text{SNR [dB]} & \text{FER}          & M      \\
\hline
\hline
4.18            & 6.54 \times 10^{-2} & 2^{15} \\
\hline
4.85            & 8.74 \times 10^{-3} & 2^{14} \\
\hline
5.53            & 7.68 \times 10^{-4} & 2^{13} \\
\hline
6.43            & 1.66 \times 10^{-5} & 2^{12} \\
\hline
\end{array}
$
\end{table}
\begin{table}
\caption{VA($ \codeca $) and BSA($ \codece $, $ M $) equal FERs}
\centering
$
\begin{array}{|>{\arrayrule}c|c|c|}
\hline
\text{SNR [dB]} & \text{FER}          & M      \\
\hline
\hline
4.68            & 1.51 \times 10^{-1} & 2^{11} \\
\hline
6.64            & 1.63 \times 10^{-3} & 2^{10} \\
\hline
\end{array}
$
\end{table}
\begin{table}
\caption{VA($ \codecb $) and BSA($ \codece $, $ M $) equal FERs}
\centering
$
\begin{array}{|>{\arrayrule}c|c|c|}
\hline
\text{SNR [dB]} & \text{FER}          & M      \\
\hline
\hline
3.67            & 1.45 \times 10^{-1} & 2^{15} \\
\hline
4.38            & 2.23 \times 10^{-2} & 2^{14} \\
\hline
5.12            & 1.93 \times 10^{-3} & 2^{13} \\
\hline
6.03            & 5.50 \times 10^{-5} & 2^{12} \\
\hline
\end{array}
$
\end{table}
\begin{table}
\caption{VA($ \codeda $) and BSA($ \codedd $, $ M $) equal FERs}
\centering
$
\begin{array}{|>{\arrayrule}c|c|c|}
\hline
\text{SNR [dB]} & \text{FER}          & M      \\
\hline
\hline
4.98            & 3.93 \times 10^{-1} & 2^{11} \\
\hline
6.88            & 6.43 \times 10^{-3} & 2^{10} \\
\hline
\end{array}
$
\end{table}
\begin{table}
\caption{VA($ \codedb $) and BSA($ \codedd $, $ M $) equal FERs}
\centering
$
\begin{array}{|>{\arrayrule}c|c|c|}
\hline
\text{SNR [dB]} & \text{FER}          & M      \\
\hline
\hline
4.77            & 1.83 \times 10^{-1} & 2^{14} \\
\hline
5.80            & 6.95 \times 10^{-3} & 2^{13} \\
\hline
6.68            & 1.58 \times 10^{-4} & 2^{12} \\
\hline
7.69            & 1.19 \times 10^{-6} & 2^{11} \\
\hline
\end{array}
$
\label{fer-cross-mlc-last}
\end{table}
\flushcolsend
\clearpage
\begin{figure}
\centering
\begin{tikzpicture}
\begin{semilogyaxis}
  [ xlabel={SNR [dB]},
    ylabel={FER},
    xmin=1,
    xmax=9,
    ymin=1e-7,
    ymax=1 ]
\addplot [ mark=none, black, dashdotted ] table {fer-stack-r12-m31-2.txt} ;
\addplot [ mark=none, black, dashdotted ] table {fer-stack-r12-m31-5.txt} ;
\addplot [ mark=none, black, dashdotted ] table {fer-stack-r12-m31-16.txt} ;
\label{fer-stack-r12-m31}
\addplot [ mark=none, black, dashed ] table {fer-bistack-r12-odp-m31-2.txt} ;
\addplot [ mark=none, black, dashed ] table {fer-bistack-r12-odp-m31-5.txt} ;
\addplot [ mark=none, black, dashed ] table {fer-bistack-r12-odp-m31-16.txt} ;
\label{fer-bistack-r12-odp-m31}
\addplot [ mark=none, black ] table {fer-bistack-r12-obdp-m31-2.txt} ;
\addplot [ mark=none, black ] table {fer-bistack-r12-obdp-m31-5.txt} ;
\addplot [ mark=none, black ] table {fer-bistack-r12-obdp-m31-16.txt} ;
\label{fer-bistack-r12-obdp-m31}
\end{semilogyaxis}
\end{tikzpicture}
\caption
  {FER, $ R = 1/2 $:
   \begin{captionaligned}
   \ref{fer-stack-r12-m31} SA($ \codeac $,$ \{ 2^{10}, 2^{13}, 2^{24} \} $) \\
   \ref{fer-bistack-r12-odp-m31} BSA($ \codeac $,$ \{ 2^{10}, 2^{13}, 2^{24} \} $) \\
   \ref{fer-bistack-r12-obdp-m31} BSA($ \codead $,$ \{ 2^{10}, 2^{13}, 2^{24} \} $)
   \end{captionaligned}}
\label{fer-sc-first}
\end{figure}
\begin{figure}
\centering
\begin{tikzpicture}
\begin{semilogyaxis}
  [ xlabel={SNR [dB]},
    ylabel={FER},
    xmin=1,
    xmax=9,
    ymin=1e-7,
    ymax=1 ]
\addplot [ mark=none, black, dashdotted ] table {fer-stack-r13-m19-2.txt} ;
\addplot [ mark=none, black, dashdotted ] table {fer-stack-r13-m19-5.txt} ;
\addplot [ mark=none, black, dashdotted ] table {fer-stack-r13-m19-16.txt} ;
\label{fer-stack-r13-m19}
\addplot [ mark=none, black, dashed ] table {fer-bistack-r13-odp-m19-2.txt} ;
\addplot [ mark=none, black, dashed ] table {fer-bistack-r13-odp-m19-5.txt} ;
\addplot [ mark=none, black, dashed ] table {fer-bistack-r13-odp-m19-16.txt} ;
\label{fer-bistack-r13-odp-m19}
\addplot [ mark=none, black ] table {fer-bistack-r13-obdp-m19-2.txt} ;
\addplot [ mark=none, black ] table {fer-bistack-r13-obdp-m19-5.txt} ;
\addplot [ mark=none, black ] table {fer-bistack-r13-obdp-m19-16.txt} ;
\label{fer-bistack-r13-obdp-m19}
\end{semilogyaxis}
\end{tikzpicture}
\caption
  {FER, $ R = 1/3 $:
   \begin{captionaligned}
   \ref{fer-stack-r13-m19} SA($ \codebc $,$ \{ 2^{10}, 2^{13}, 2^{24} \} $) \\
   \ref{fer-bistack-r13-odp-m19} BSA($ \codebc $,$ \{ 2^{10}, 2^{13}, 2^{24} \} $) \\
   \ref{fer-bistack-r13-obdp-m19} BSA($ \codebd $,$ \{ 2^{10}, 2^{13}, 2^{24} \} $)
   \end{captionaligned}}
\end{figure}
\begin{figure}
\centering
\begin{tikzpicture}
\begin{semilogyaxis}
  [ xlabel={SNR [dB]},
    ylabel={FER},
    xmin=1,
    xmax=9,
    ymin=1e-7,
    ymax=1 ]
\addplot [ mark=none, black, dashdotted ] table {fer-stack-r14-m21-2.txt} ;
\addplot [ mark=none, black, dashdotted ] table {fer-stack-r14-m21-5.txt} ;
\addplot [ mark=none, black, dashdotted ] table {fer-stack-r14-m21-16.txt} ;
\label{fer-stack-r14-m21}
\addplot [ mark=none, black, dashed ] table {fer-bistack-r14-odp-m21-2.txt} ;
\addplot [ mark=none, black, dashed ] table {fer-bistack-r14-odp-m21-5.txt} ;
\addplot [ mark=none, black, dashed ] table {fer-bistack-r14-odp-m21-16.txt} ;
\label{fer-bistack-r14-odp-m21}
\addplot [ mark=none, black ] table {fer-bistack-r14-obdp-m21-2.txt} ;
\addplot [ mark=none, black ] table {fer-bistack-r14-obdp-m21-5.txt} ;
\addplot [ mark=none, black ] table {fer-bistack-r14-obdp-m21-16.txt} ;
\label{fer-bistack-r14-obdp-m21}
\end{semilogyaxis}
\end{tikzpicture}
\caption
  {FER, $ R = 1/4 $:
   \begin{captionaligned}
   \ref{fer-stack-r14-m21} SA($ \codecc $,$ \{ 2^{10}, 2^{13}, 2^{24} \} $) \\
   \ref{fer-bistack-r14-odp-m21} BSA($ \codecc $,$ \{ 2^{10}, 2^{13}, 2^{24} \} $) \\
   \ref{fer-bistack-r14-obdp-m21} BSA($ \codecd $,$ \{ 2^{10}, 2^{13}, 2^{24} \} $)
   \end{captionaligned}}
\end{figure}
\begin{figure}
\centering
\begin{tikzpicture}
\begin{semilogyaxis}
  [ xlabel={SNR [dB]},
    ylabel={FER},
    xmin=1,
    xmax=9,
    ymin=1e-7,
    ymax=1 ]
\addplot [ mark=none, black, dashdotted ] table {fer-stack-r23-m9-2.txt} ;
\addplot [ mark=none, black, dashdotted ] table {fer-stack-r23-m9-5.txt} ;
\addplot [ mark=none, black, dashdotted ] table {fer-stack-r23-m9-16.txt} ;
\label{fer-stack-r23-m9}
\addplot [ mark=none, black, dashed ] table {fer-bistack-r23-odp-m9-2.txt} ;
\addplot [ mark=none, black, dashed ] table {fer-bistack-r23-odp-m9-5.txt} ;
\addplot [ mark=none, black, dashed ] table {fer-bistack-r23-odp-m9-16.txt} ;
\label{fer-bistack-r23-odp-m9}
\addplot [ mark=none, black ] table {fer-bistack-r23-obdp-m9-2.txt} ;
\addplot [ mark=none, black ] table {fer-bistack-r23-obdp-m9-5.txt} ;
\addplot [ mark=none, black ] table {fer-bistack-r23-obdp-m9-16.txt} ;
\label{fer-bistack-r23-obdp-m9}
\end{semilogyaxis}
\end{tikzpicture}
\caption
  {FER, $ R = 2/3 $:
   \begin{captionaligned}
   \ref{fer-stack-r23-m9} SA($ \codedc $,$ \{ 2^{10}, 2^{13}, 2^{24} \} $) \\
   \ref{fer-bistack-r23-odp-m9} BSA($ \codedc $,$ \{ 2^{10}, 2^{13}, 2^{24} \} $) \\
   \ref{fer-bistack-r23-obdp-m9} BSA($ \codedd $,$ \{ 2^{10}, 2^{13}, 2^{24} \} $)
   \end{captionaligned}}
\label{fer-sc-last}
\end{figure}
\flushcolsend
\clearpage
\begin{figure}
\centering
\begin{tikzpicture}
\begin{loglogaxis}
  [ xlabel={$ x $},
    ylabel={$ P [ X \geq x ] $},
    xmin=1e2,
    xmax=1e7,
    ymin=1e-7,
    ymax=1 ]
\addplot [ mark=none, black, dashdotted ] table {comp-stack-r12-4db-ub.txt} ;
\addplot [ mark=none, black, dashdotted ] table {comp-stack-r12-5db-ub.txt} ;
\addplot [ mark=none, black, dashdotted ] table {comp-stack-r12-6db-ub.txt} ;
\addplot [ mark=none, black, dashdotted ] table {comp-stack-r12-7db-ub.txt} ;
\label{comp-stack-r12-ub}
\addplot [ mark=none, black, dashdotted ] table {comp-stack-r12-4db-lb.txt} ;
\addplot [ mark=none, black, dashdotted ] table {comp-stack-r12-5db-lb.txt} ;
\addplot [ mark=none, black, dashdotted ] table {comp-stack-r12-6db-lb.txt} ;
\addplot [ mark=none, black, dashdotted ] table {comp-stack-r12-7db-lb.txt} ;
\label{comp-stack-r12-lb}
\addplot [ mark=none, black, dashed ] table {comp-bistack-r12-odp-4db.txt} ;
\addplot [ mark=none, black, dashed ] table {comp-bistack-r12-odp-5db.txt} ;
\addplot [ mark=none, black, dashed ] table {comp-bistack-r12-odp-6db.txt} ;
\addplot [ mark=none, black, dashed ] table {comp-bistack-r12-odp-7db.txt} ;
\label{comp-bistack-r12-odp}
\addplot [ mark=none, black ] table {comp-bistack-r12-obdp-4db.txt} ;
\addplot [ mark=none, black ] table {comp-bistack-r12-obdp-5db.txt} ;
\addplot [ mark=none, black ] table {comp-bistack-r12-obdp-6db.txt} ;
\addplot [ mark=none, black ] table {comp-bistack-r12-obdp-7db.txt} ;
\label{comp-bistack-r12-obdp}
\end{loglogaxis}
\end{tikzpicture}
\caption
  {\begin{captionaligned}
   Decoding complexity, $ R = 1/2 $, \\
   from top to bottom $ \text{SNR [dB]} = 4,5,6,7 $: \\
   \ref{comp-stack-r12-ub} SA($ \codeac $) (bounds) \\
   \ref{comp-bistack-r12-odp} BSA($ \codeac $) \\
   \ref{comp-bistack-r12-obdp} BSA($ \codead $)
   \end{captionaligned}}
\label{complexity-first}
\end{figure}
\begin{figure}
\centering
\begin{tikzpicture}
\begin{loglogaxis}
  [ xlabel={$ x $},
    ylabel={$ P [ X \geq x ] $},
    xmin=1e2,
    xmax=1e7,
    ymin=1e-7,
    ymax=1 ]
\addplot [ mark=none, black, dashdotted ] table {comp-stack-r13-4db-ub.txt} ;
\addplot [ mark=none, black, dashdotted ] table {comp-stack-r13-5db-ub.txt} ;
\addplot [ mark=none, black, dashdotted ] table {comp-stack-r13-6db-ub.txt} ;
\addplot [ mark=none, black, dashdotted ] table {comp-stack-r13-7db-ub.txt} ;
\label{comp-stack-r13-ub}
\addplot [ mark=none, black, dashdotted ] table {comp-stack-r13-4db-lb.txt} ;
\addplot [ mark=none, black, dashdotted ] table {comp-stack-r13-5db-lb.txt} ;
\addplot [ mark=none, black, dashdotted ] table {comp-stack-r13-6db-lb.txt} ;
\addplot [ mark=none, black, dashdotted ] table {comp-stack-r13-7db-lb.txt} ;
\label{comp-stack-r13-lb}
\addplot [ mark=none, black, dashed ] table {comp-bistack-r13-odp-4db.txt} ;
\addplot [ mark=none, black, dashed ] table {comp-bistack-r13-odp-5db.txt} ;
\addplot [ mark=none, black, dashed ] table {comp-bistack-r13-odp-6db.txt} ;
\addplot [ mark=none, black, dashed ] table {comp-bistack-r13-odp-7db.txt} ;
\label{comp-bistack-r13-odp}
\addplot [ mark=none, black ] table {comp-bistack-r13-obdp-4db.txt} ;
\addplot [ mark=none, black ] table {comp-bistack-r13-obdp-5db.txt} ;
\addplot [ mark=none, black ] table {comp-bistack-r13-obdp-6db.txt} ;
\addplot [ mark=none, black ] table {comp-bistack-r13-obdp-7db.txt} ;
\label{comp-bistack-r13-obdp}
\end{loglogaxis}
\end{tikzpicture}
\caption
  {\begin{captionaligned}
   Decoding complexity, $ R = 1/3 $, \\
   from top to bottom $ \text{SNR [dB]} = 4,5,6,7 $: \\
   \ref{comp-stack-r13-ub} SA($ \codebc $) (bounds) \\
   \ref{comp-bistack-r13-odp} BSA($ \codebc $) \\
   \ref{comp-bistack-r13-obdp} BSA($ \codebd $)
   \end{captionaligned}}
\end{figure}
\begin{figure}
\centering
\begin{tikzpicture}
\begin{loglogaxis}
  [ xlabel={$ x $},
    ylabel={$ P [ X \geq x ] $},
    xmin=1e2,
    xmax=1e7,
    ymin=1e-7,
    ymax=1 ]
\addplot [ mark=none, black, dashdotted ] table {comp-stack-r14-4db-ub.txt} ;
\addplot [ mark=none, black, dashdotted ] table {comp-stack-r14-5db-ub.txt} ;
\addplot [ mark=none, black, dashdotted ] table {comp-stack-r14-6db-ub.txt} ;
\addplot [ mark=none, black, dashdotted ] table {comp-stack-r14-7db-ub.txt} ;
\label{comp-stack-r14-ub}
\addplot [ mark=none, black, dashdotted ] table {comp-stack-r14-4db-lb.txt} ;
\addplot [ mark=none, black, dashdotted ] table {comp-stack-r14-5db-lb.txt} ;
\addplot [ mark=none, black, dashdotted ] table {comp-stack-r14-6db-lb.txt} ;
\addplot [ mark=none, black, dashdotted ] table {comp-stack-r14-7db-lb.txt} ;
\label{comp-stack-r14-lb}
\addplot [ mark=none, black, dashed ] table {comp-bistack-r14-odp-4db.txt} ;
\addplot [ mark=none, black, dashed ] table {comp-bistack-r14-odp-5db.txt} ;
\addplot [ mark=none, black, dashed ] table {comp-bistack-r14-odp-6db.txt} ;
\addplot [ mark=none, black, dashed ] table {comp-bistack-r14-odp-7db.txt} ;
\label{comp-bistack-r14-odp}
\addplot [ mark=none, black ] table {comp-bistack-r14-obdp-4db.txt} ;
\addplot [ mark=none, black ] table {comp-bistack-r14-obdp-5db.txt} ;
\addplot [ mark=none, black ] table {comp-bistack-r14-obdp-6db.txt} ;
\addplot [ mark=none, black ] table {comp-bistack-r14-obdp-7db.txt} ;
\label{comp-bistack-r14-obdp}
\end{loglogaxis}
\end{tikzpicture}
\caption
  {\begin{captionaligned}
   Decoding complexity, $ R = 1/4 $, \\
   from top to bottom $ \text{SNR [dB]} = 4,5,6,7 $: \\
   \ref{comp-stack-r14-ub} SA($ \codecc $) (bounds) \\
   \ref{comp-bistack-r14-odp} BSA($ \codecc $) \\
   \ref{comp-bistack-r14-obdp} BSA($ \codecd $)
   \end{captionaligned}}
\end{figure}
\begin{figure}
\centering
\begin{tikzpicture}
\begin{loglogaxis}
  [ xlabel={$ x $},
    ylabel={$ P [ X \geq x ] $},
    xmin=1e2,
    xmax=1e7,
    ymin=1e-7,
    ymax=1 ]
\addplot [ mark=none, black, dashdotted ] table {comp-stack-r23-4db-ub.txt} ;
\addplot [ mark=none, black, dashdotted ] table {comp-stack-r23-5db-ub.txt} ;
\addplot [ mark=none, black, dashdotted ] table {comp-stack-r23-6db-ub.txt} ;
\addplot [ mark=none, black, dashdotted ] table {comp-stack-r23-7db-ub.txt} ;
\label{comp-stack-r23-ub}
\addplot [ mark=none, black, dashdotted ] table {comp-stack-r23-4db-lb.txt} ;
\addplot [ mark=none, black, dashdotted ] table {comp-stack-r23-5db-lb.txt} ;
\addplot [ mark=none, black, dashdotted ] table {comp-stack-r23-6db-lb.txt} ;
\addplot [ mark=none, black, dashdotted ] table {comp-stack-r23-7db-lb.txt} ;
\label{comp-stack-r23-lb}
\addplot [ mark=none, black, dashed ] table {comp-bistack-r23-odp-4db.txt} ;
\addplot [ mark=none, black, dashed ] table {comp-bistack-r23-odp-5db.txt} ;
\addplot [ mark=none, black, dashed ] table {comp-bistack-r23-odp-6db.txt} ;
\addplot [ mark=none, black, dashed ] table {comp-bistack-r23-odp-7db.txt} ;
\label{comp-bistack-r23-odp}
\addplot [ mark=none, black ] table {comp-bistack-r23-obdp-4db.txt} ;
\addplot [ mark=none, black ] table {comp-bistack-r23-obdp-5db.txt} ;
\addplot [ mark=none, black ] table {comp-bistack-r23-obdp-6db.txt} ;
\addplot [ mark=none, black ] table {comp-bistack-r23-obdp-7db.txt} ;
\label{comp-bistack-r23-obdp}
\end{loglogaxis}
\end{tikzpicture}
\caption
  {\begin{captionaligned}
   Decoding complexity, $ R = 2/3 $, \\
   from top to bottom $ \text{SNR [dB]} = 4,5,6,7 $: \\
   \ref{comp-stack-r23-ub} SA($ \codedc $) (bounds) \\
   \ref{comp-bistack-r23-odp} BSA($ \codedc $) \\
   \ref{comp-bistack-r23-obdp} BSA($ \codedd $)
   \end{captionaligned}}
\label{complexity-last}
\end{figure}
\flushcolsend
\clearpage
\begin{table}
\caption{Average number of extended nodes, $ R = 1/2 $}
\centering
$
\begin{array}{|>{\arrayrule}c|c|c|c|}
\hline
\text{SNR [dB]} & \text{SA($ \codeac $), LB} & \text{BSA($ \codeac $)} & \text{BSA($ \codead $)} \\
\hline
\hline
4               & 1370470                    & 39916.6                 & 39230.4                 \\
\hline
5               & 53787.6                    & 1444.59                 & 1400.58                 \\
\hline
6               & 690.758                    & 366.085                 & 362.433                 \\
\hline
7               & 285.136                    & 282.817                 & 282.126                 \\
\hline
\end{array}
$
\label{average-complexity-first}
\end{table}
\begin{table}
\caption{Average number of extended nodes, $ R = 1/3 $}
\centering
$
\begin{array}{|>{\arrayrule}c|c|c|c|}
\hline
\text{SNR [dB]} & \text{SA($ \codebc $), LB} & \text{BSA($ \codebc $)} & \text{BSA($ \codebd $)} \\
\hline
\hline
4               & 81000                      & 2221.87                 & 1950.36                 \\
\hline
5               & 2053.29                    & 538.329                 & 480.53                  \\
\hline
6               & 328.403                    & 319.052                 & 300.176                 \\
\hline
7               & 260.336                    & 268.519                 & 260.128                 \\
\hline
\end{array}
$
\end{table}
\begin{table}
\caption{Average number of extended nodes, $ R = 1/4 $}
\centering
$
\begin{array}{|>{\arrayrule}c|c|c|c|}
\hline
\text{SNR [dB]} & \text{SA($ \codecc $), LB} & \text{BSA($ \codecc $)} & \text{BSA($ \codecd $)} \\
\hline
\hline
4               & 61758.3                    & 2582.11                 & 1831.05                 \\
\hline
5               & 1460.96                    & 590.712                 & 452.307                 \\
\hline
6               & 316.926                    & 347.656                 & 298.378                 \\
\hline
7               & 261.146                    & 284.164                 & 261.333                 \\
\hline
\end{array}
$
\end{table}
\begin{table}
\caption{Average number of extended nodes, $ R = 2/3 $}
\centering
$
\begin{array}{|>{\arrayrule}c|c|c|c|}
\hline
\text{SNR [dB]} & \text{SA($ \codedc $), LB} & \text{BSA($ \codedc $)} & \text{BSA($ \codedd $)} \\
\hline
\hline
4               & 636974                     & 1870.24                 & 1871.56                 \\
\hline
5               & 52076.1                    & 589.551                 & 590.943                 \\
\hline
6               & 817.769                    & 186.364                 & 186.402                 \\
\hline
7               & 140.06                     & 134.111                 & 134.036                 \\
\hline
\end{array}
$
\label{average-complexity-last}
\end{table}
\end{document}